\documentstyle[preprint,aps,eqsecnum,epsf]{revtex}
\tighten
\newdimen\rotdimen
\def\vspec#1{\special{ps:#1}}
\def\rotstart#1{\vspec{gsave currentpoint currentpoint translate
   #1 neg exch neg exch translate}}
\def\rotfinish{\vspec{currentpoint grestore moveto}}
\def\rotr#1{\rotdimen=\ht#1\advance\rotdimen by\dp#1%
   \hbox to\rotdimen{\hskip\ht#1\vbox to\wd#1{\rotstart{90 rotate}%
   \box#1\vss}\hss}\rotfinish}
\def\RFig#1#2#3{\begin{center} \epsfxsize=#1in
  \setbox0=\hbox to\hsize{\hfil\epsfbox{#2}\hfil} \rotr0
  {\sl #3} \end{center}}
\begin{document}
\preprint{PITT-96-;  CMU-HEP-96-; LPTHE-96/32} 
\draft
\title{\bf ANALYTIC AND NUMERICAL STUDY OF PREHEATING DYNAMICS}
\author{{\bf D. Boyanovsky$^{(a)}$, 
H.J. de Vega$^{(b)}$, R. Holman$^{(c)}$ and J. F. J. Salgado$^{(b)}$ }}   
\address
{ (a)  Department of Physics and Astronomy, University of
Pittsburgh, Pittsburgh, PA. 15260, U.S.A. \\
 (b)  LPTHE, \footnote{Laboratoire Associ\'{e} au CNRS UA280.}
Universit\'e Pierre et Marie Curie (Paris VI) 
et Denis Diderot  (Paris VII), Tour 16, 1er. \'etage, 4, Place Jussieu
75252 Paris, Cedex 05, France \\
 (c) Department of Physics, Carnegie Mellon University, Pittsburgh,
PA. 15213, U. S. A. }
\date{July 1996}
\maketitle
\begin{abstract}
We analyze the phenomenon of preheating, i.e. explosive particle production due
to parametric amplification of quantum fluctuations in the unbroken
symmetry case, 
or spinodal instabilities in the broken symmetry phase, using the
Minkowski space 
$O(N)$ vector model in the large $N$ limit to study the non-perturbative issues
involved. We give analytic results for weak couplings and times short compared
to the time at which the fluctuations become of the same order as the tree
level terms, as well as numerical results including the full backreaction.
 In the case where the symmetry is
unbroken, the analytical results agree spectacularly well with the numerical
ones in their common domain of validity. In the broken symmetry case,
interesting situations, corresponding to slow roll initial conditions from the
unstable minimum at the origin, give rise to a new and unexpected
phenomenon: the dynamical relaxation of the vacuum energy. That is, particles
are abundantly produced at the expense of the quantum vacuum energy while 
the zero mode  comes back to almost its initial value. In both cases
we obtain analytically 
and numerically the equation of state which in both 
cases can be written in terms of an effective polytropic index that
interpolates between vacuum and radiation-like domination.

We find that simplified analysis based 
on harmonic behavior of the zero mode, giving rise to a Mathieu equation for
the non-zero modes miss important physics. Furthermore, such  analysis that
do not include the full 
backreaction  and do not conserve energy,   results in unbound particle
production. Our results rule out  the possibility of symmetry
restoration  by non-equilibrium fluctuations in the cases relevant for
new inflationnary scenarios. Finally, estimates of the
reheating  
temperature are given, as well as a discussion of the inconsistency of
a kinetic approach 
 to thermalization when a non-perturbatively large number of particles
are created. 
  
\end{abstract}

\section{Introduction}

It has recently been realized\cite{branden,lindekov,dis} that as the zero
momentum mode of a quantum field evolves, it can drive a large
amplification of quantum fluctuations. This, in turn, gives rise to copious
particle production for bosonic fields, creating quanta in a highly
non-equilibrium distribution, radically changing the standard picture of
reheating the post-inflationary universe\cite{linde,rocky,dolgov}. This process
has other possible applications, such as in understanding the hadronization stage of the
quark-gluon plasma\cite{muller} as well as trying to understand out of
equilibrium particle production in strong electromagnetic fields and in heavy
ion collisions \cite{strongfields,mottola,habib,boyveg}.

The actual processes giving rise to preheating can be different depending on
the potential for the scalar field involved as well as the initial
conditions. For example, in new inflationary scenarios, where the inflaton
field's zero mode evolves down the flat portion of a potential admitting
spontaneous symmetry breaking, particle production occurs due to the existence
of unstable field modes which get amplified until the zero mode leaves the
instability region. These are the instabilities that give rise to spinodal
decomposition and phase separation. 
In contrast, if we start with chaotic initial conditions, so
that the field has large initial amplitude, particles are created from the
parametric amplification of the quantum fluctuations due to the oscillations of
the zero mode. 

In this paper we analyze the details of this so-called {\bf preheating} process
both analytically as well as numerically. Preheating is a non-perturbative
process, with typically $1\slash \lambda$ particles being produced, where
$\lambda$ is the self coupling of the field. Due to this fact, any attempts at
analyzing the detailed dynamics of preheating must also be non-perturbative in
nature. This leads us to consider the $O(N)$ vector model in the large $N$
limit. This is a non-perturbative approximation  that has many
important features that 
justify its use:  unlike the Hartree or
mean-field approximation\cite{dis}, it can be systematically improved
in the $ 1\slash N $ 
expansion. It conserves energy, satisfies the Ward identities of the
underlying symmetry,  
and again unlike the Hartree approximation it predicts the correct
order of the transition in equilibrium. 

This
approximation has also been used in other non-equilibrium
contexts\cite{strongfields,mottola,habib,boyveg}. In this work, we consider this model in
Minkowski space, saving the discussion of the effects of the
expansion of the universe for later work.

Our findings are summarized  as follows.

We are able to provide consistent non-perturbative analytic estimates of the
non-equilibrium processes occurring during the preheating stage taking into
account the {\bf exact} evolution of the inflaton zero mode for large
amplitudes when the quantum back-reaction due to the produced particles is
negligible i.e. at early and intermediate times. We also compute the momentum
space distribution of the particle number as well as the effective equation of
state during this stage.  Explicit expressions for the growth of quantum
fluctuations, the preheating time scale, as well as the effective 
(time dependent) polytropic index defining the equation of state are
given in sec. III. 
 
We then go beyond the early/intermediate time regime and evolve the equations
of motion numerically, taking into account back-reaction effects. 
(That is, the non-linear quantum field interaction). These results
confirm the analytic results in their domain of validity and show how, when
back-reaction effects are large enough to compete with tree level effects,
dissipational effects arise in the zero mode. Energy 
conservation is guaranteed in the full backreaction problem, leading
to the eventual  shut-off of particle production. This is an important
ingredient in the dynamics that  determines the relevant time scales.

We also find a novel 
dynamical relaxation of the vacuum energy
in this regime when the theory is in the broken phase. Namely, 
particles are produced at the expense of the quantum vacuum energy while 
the zero mode contributes very little.
We find a radiation type equation of state for late times ($ p \approx
\frac13 \; \varepsilon $) despite the lack of thermal equilibrium. 

Finally we discuss the
calculation of the reheating temperature in a class of models, paying
particular attention to when the kinetic approach to thermalization and
equilibration is applicable.

There have been a number of papers (see refs.\cite{branden,lindekov,yoshimura}
-\cite{tkachev}) dedicated to the analysis of the preheating process
where particle production and back-reaction are estimated in different
approximations \cite{paris}.


The layout of the paper is as follows. Section II presents the model, the
evolution equations, the renormalization of the equations of motion and
introduces the relevant definitions of particle number, energy and pressure and
the details of their renormalization. The unbroken and broken symmetry cases
are presented in detail and the differences in their treatment are clearly
explained.

In sections III through V we present a detailed analytic and numerical
treatment of both the unbroken and broken symmetry phases emphasizing the
description of particle production, energy, pressure and the equation of
state. In the broken symmetry case, when the inflaton zero mode begins very
close to the top of the potential, we find that there is a novel phenomenon of
relaxation of the vacuum energy that explicitly shows where the energy used
to produced the particles comes from. We also discuss why the phenomenon of
symmetry restoration at preheating, discussed by various
authors\cite{lindekov,tkachev,kolbriotto,kolblinde} is {\bf not} seen
to occur in the cases treated by us here  and  relevant for
new inflationnary scenarios \cite{paris}.  

In section VI we provide estimates, under suitably specified assumptions, of
the reheating temperature in the $ O(N) $ model as well as other
models in which 
the inflaton couples to lighter scalars. In this section we argue that
thermalization cannot be studied with a kinetic approach because of the
non-perturbatively large occupation number of long-wavelength modes.

Finally, we summarize our results and discuss future avenues of study in the
conclusions. We also include an appendix where we gather many important technical
details on the evaluation of the Floquet mode functions and Floquet indices
used in the main text.

\section{Scalar Field Dynamics in the Large $N$ limit}

As mentioned above, preheating is a non-perturbative phenomenon so that a
non-perturbative treatment of the field theory is necessary. This leads us to
consider the $O(N)$ vector model in the large $N$ limit.  

In this section we introduce this model, obtain the non-equilibrium evolution
equations, the energy momentum tensor and analyze the issue of renormalization.
We will then be poised to study each particular case in detail in the later
sections.

The Lagrangian density is the following:
\begin{eqnarray}
 {\cal{L}} &=& \frac{1}{2} \partial_{\mu} \vec{\phi} \cdot \partial^{\mu} 
    \vec{\phi} - V(\vec{\phi}\cdot \vec{\phi} ) \nonumber  \\
   V(\sigma, \vec{\pi} ) &=& \frac{1}{2} m^2 \vec{\phi} \cdot \vec{\phi} + 
   \frac{ \lambda}{8N} (  \vec{\phi} \cdot \vec{\phi} )^2   \label{onLagrangian}
\end{eqnarray}
for $\lambda$ fixed in the large $N$ limit. Here $\vec{\phi}$ is an $O(N)$
vector, $\vec{\phi} = (\sigma, \vec{\pi} )$ and $\vec{\pi}$ represents the
$N-1$ ``pions''. In what follows, we will consider two different cases of the
potential $ V(\sigma, \vec{\pi} ) $, with ($m^2 < 0$) or without ($m^2 > 0$)
symmetry breaking.

We can decompose the field $\sigma$ into its zero mode and fluctuations
$\chi( \vec{x},t )$ about the zero mode:
\begin{equation}
\sigma  (\vec{x},t ) = \sigma_0(t)+ \chi ( \vec{x},t)
\end{equation}

The generating functional of real time non-equilibrium Green's functions can be
written in terms of a path integral along a complex contour in time,
corresponding to forward and backward time evolution and at finite temperature
a branch down the imaginary time axis. This requires doubling the
number of fields 
which now carry a label $\pm$ corresponding to forward ($+$), and backward
($-$) time evolution. The reader is referred to the literature for more
details\cite{noneq,hu}. This generating functional along the complex contour
requires the Lagrangian density along the contour, which for zero temperature
is given by\cite{dis}

\begin{eqnarray}
&& {\cal{L}} [ \sigma_0 +\chi^+, \vec{\pi}^+ ] - {\cal{L}} [\sigma_0+\chi^-,
 \vec{\pi}^- ] = \left\{ {\cal{L}} [ \sigma_0,\vec{\pi}^+ ] + \frac{\delta
 {\cal{L}} } {\delta \sigma_0} \chi^+ + \frac{1}{2} \left( \partial_{\mu}
 \chi^+ \right)^2 \right.  \nonumber \\ && + \frac{1}{2} \left( \partial_{\mu}
 \vec{\pi}^+ \right)^2 \left. -\left( \frac{1}{2!} V^{\prime \prime}
 (\sigma_0,\vec{\pi}^+ ) \chi^{+2} + \frac{1}{3!} V^{[3]} (\sigma_0,\vec{\pi}^+
 ) (\chi^+)^3 + \frac{1}{4!} V^{[4]} (\sigma_0, \vec{\pi}^+ ) (\chi^+)^4
 \right) \right\} \nonumber \\ &&- \left\{ \left( \chi^+ \rightarrow \chi^-
 \right), \left( \vec{\pi}^+ \rightarrow \vec{\pi}^- \right) \right\}
\end{eqnarray}
The tadpole condition $\langle \chi^{\pm}(\vec{x},t) \rangle =0$ will lead to
the equations of motion as discussed in \cite{dis} and references therein.

A consistent and elegant version of the large $ N $ limit for non-equilibrium
problems can be obtained by introducing an auxiliary field and is presented
very thoroughly in reference\cite{mottola}. This formulation has the advantage
that it can incorporate the $ O(1/N) $ corrections in a systematic
fashion. Alternatively, the large $ N $ limit can be implemented via a
Hartree-like 
factorization\cite{dis} in which i) there are no cross correlations between the
pions and sigma field and ii) the two point correlation functions of the pion
field are diagonal in the $ O(N-1) $ space of the remaining unbroken symmetry
group. To leading order in large $ N $ both methods are completely
equivalent and 
for simplicity of presentation we chose the factorization method.

The factorization of the non-linear terms in the Lagrangian is (again for
both $\pm$ components):
\begin{eqnarray}
\chi^4 & \rightarrow & 6 \langle \chi^2 \rangle \chi^2 +\text{constant}
\label{larg1} \\ \chi^3 & \rightarrow & 3 \langle \chi^2 \rangle \chi
   \label{larg2} \\ \left( \vec{\pi} \cdot \vec{\pi} \right)^2 & \rightarrow &
   2 \langle \vec{\pi}^2 \rangle \vec{\pi}^2 - \langle \vec{\pi}^2 \rangle^2+
   {\cal{O}}(1/N) \label{larg3} \\ \vec{\pi}^2 \chi^2 & \rightarrow & \langle
   \vec{\pi}^2 \rangle \chi^2 +\vec{\pi}^2 \langle \chi^2 \rangle \label{larg4}
   \\ \vec{\pi}^2 \chi & \rightarrow & \langle \vec{\pi}^2 \rangle \chi
   \label{larg5}
\end{eqnarray}

To obtain a large N limit, we define
\begin{equation} 
\vec{\pi}(\vec x, t) = \psi(\vec x, t)
\overbrace{\left(1,1,\cdots,1\right)}^{N-1} \; \; ; \; \; \sigma_0(t) = \phi(t)
\sqrt{N} \label{filargeN}
\end{equation} 
where the large N limit is implemented by the requirement that
\begin{equation}
 \langle
\psi^2 \rangle \approx {\cal{O}} (1) \; , \;  \langle
\chi^2 \rangle \approx {\cal{O}} (1) \; , \;  \phi \approx  {\cal{O}} (1).
\label{order1}
\end{equation}
The leading contribution is obtained by neglecting the $ {\cal{O}}
({1}\slash {N})$ terms in the formal large $N$ limit. The resulting Lagrangian
density is quadratic, with a linear term in $\chi$ :
\begin{eqnarray}
{\cal{L}} [\sigma_0+\chi^+, \vec{\pi}^+ ] &-& {\cal{L}} [\sigma_0+\chi^-,
\vec{\pi}^- ] = \left\{ \frac{1}{2} \left( \partial_{\mu} \chi^+ \right)^2
+\frac{1}{2} \left( \partial_{\mu} \vec{\pi}^+ \right)^2 - \chi^+ V^{\prime}
(t) \right. \nonumber \\ &-& \frac{1}{2}{\cal{M}}^2_{\chi} (t) (\chi^+)^2
-\left.\frac{1}{2}{\cal{M}}^2_{\vec{\pi}} (t) (\vec{\pi}^+)^2 \right\} -
\left\{ \left( \chi^+ \rightarrow \chi^- \right),\left( \vec{\pi}^+ \rightarrow
\vec{\pi}^- \right)\right\} \nonumber \\
\end{eqnarray}
where,
\begin{eqnarray}
V^{'} (\phi(t),t)         &=&{\sqrt{N}}  
\phi (t) \left[ m^2 +\frac{\lambda}{2} \phi^2 (t) + \frac{\lambda}{2} \langle \psi^2 (t) \rangle \right] 
\label{vprime} \\
{\cal{M}}^2_{\vec{\pi}}(t) &=& m^2 + \frac{\lambda}{2} \phi^2 (t) + 
  \frac{\lambda}{2}\langle \psi^2 (t) \rangle  \\
  {\cal{M}}^2_{\chi}(t)      &=& m^2 + \frac{3 \lambda}{2} \phi^2 (t) + 
  \frac{\lambda}{2} \langle \psi^2 (t) \rangle.
\end{eqnarray}
Note that we  have used spatial translational invariance to write
\begin{equation}
\langle \psi^2(\vec x, t) \rangle \equiv \langle \psi^2(t) \rangle \label{psi2}
\end{equation}
The necessary (zero temperature ) non-equilibrium Green's functions are
constructed from the following ingredients
\begin{eqnarray}
G^>_k(t,t') = \frac{i}{2W_k} V_k(t)V^*_k(t') \label{greater} \\
G^<_k(t,t') = \frac{i}{2W_k} V_k(t')V^*_k(t) \label{lesser}
\end{eqnarray}
while the Heisenberg field operator $\psi(\vec x,t)$ can be written as
\begin{equation}
\psi(\vec x , t) = \frac{1}{\sqrt{{\cal{V}}}}\sum_k \frac{1}{\sqrt{2 W_k}}\left[
a_k V_k(t) e^{i \vec k \cdot \vec x} + a^{\dagger}_k V^*_k(t)
 e^{-i \vec k \cdot \vec x}  \right],
\end{equation}
where $a_k \, , a^{\dagger}_k$ are the canonical destruction and annihilation
operators and ${\cal{V}}$ the quantization volume.

The evolution equations for the expectation value $\phi(t)$ and the mode
functions $V_k(t)$ can be obtained by using the tadpole method\cite{dis} and
are given by:
\begin{eqnarray}
&& \ddot{\phi}(t) +  \phi (t) \left[ m^2 +\frac{\lambda}{2} \phi^2 (t) + 
\frac{\lambda}{2} \langle \psi^2 (t) \rangle  \right]     
   =0    \label{zeromodeeqn}  \\
&& \left[ \frac{d^2}{d t^2} + k^2 + m^2 + \frac{\lambda}{2} \phi^2 (t) + 
  \frac{\lambda}{2}\langle \psi^2 (t) \rangle \right] 
     V_{k} (t)  =0 \; ; \;  V_k(0)=1 \; ; \; \dot{V}_k(0) = -i W_k \label{pionmodes} \\
&& \langle \psi^2 ( t ) \rangle =   
   \int \frac{d^3 k}{( 2 \pi)^3} \frac{\left| V_k (t) \right|^2 }{2W_{k}} 
 \label{fluctuation} \\
&& {W_{ k}} =  \sqrt{k^2+m^2_0}. \label{initialfreqs}
\end{eqnarray}

The initial state $|i\rangle$ is chosen to be the vacuum for these modes, i.e.
 $a_k |i\rangle =0$. 
The frequencies $W_k$ (i.e. $m^2_0$) will determine the initial state and will be 
discussed for each particular case below. 

The fluctuations $\chi(\vec x, t)$ obey an independent equation, that does not
enter in the dynamics of the evolution of the expectation value or the
$\vec{\pi}$ fields to this order and decouples in the leading order in the
large $N$ limit\cite{dis}.

It is clear from the above equations that the Ward identities of Goldstone's
theorem are fulfilled. Because $V^{'}(\phi
(t),t)=\sqrt{N}\phi(t){\cal{M}}_{\vec{\pi}}^2(t)$, whenever $V'(\phi(t),t)$
vanishes for $\phi \neq 0$ then ${\cal{M}}_{\vec{\pi}}=0$ and the ``pions'' are
the Goldstone bosons. This observation will be important in the discussions of
symmetry breaking in a later section.

Since in this approximation, the dynamics for the $\vec{\pi}$ and $\chi$ fields
decouple, and the dynamics of $\chi$ does not influence that of $\phi$, the
mode functions or $\langle \psi^2 \rangle$, we will only concentrate on the
solution for the $\vec{\pi}$ fields. We note however, that if the dynamics is
such that the asymptotic value of $\phi \neq 0$ the masses for $\chi$ and the
``pion'' multiplet $\vec{\pi}$ are different, and the original $O(N)$ symmetry
is broken down to the $O(N-1)$ subgroup.

\subsection{Renormalization of the $O(N)$ Model}

We briefly review the most relevant features of the renormalization program in
the large $N$ limit that will be used frequently in our analysis. For more
details the reader is referred to\cite{mottola,dis,boyveg}.

In this approximation, the Lagrangian is quadratic, and there are no
counterterms. This implies that the equations for the mode functions must be
finite. This requires that
\begin{equation}
m^2+\frac{\lambda}{2}\phi^2(t)+\frac{\lambda}{2}\langle \psi^2(t) \rangle =  
m^2_R+\frac{\lambda_R}{2}\phi^2(t)+\frac{\lambda_R}{2}\langle
\psi^2(t) \rangle_R 
= - v(t)
\label{renorm}
\end{equation}
Defining
\begin{equation}
 \frac{V_k(t)} { \sqrt{W_k}} =\varphi_k(t) \; \; , \; \; 
 \varphi_k(0)= \frac{1} { \sqrt{W_k}} \; \; , \; \; \dot{\varphi_k}(0)= -i 
{ \sqrt{W_k}} \label{varphi}
\end{equation}

The function $\varphi_k(t)$ is written as linear combinations of WKB solutions
of the form
\begin{equation}
\varphi_k(t) = A_k \exp{\int^t_0 R_k(t') dt' } + B_k
 \exp{\int^t_0 R^*_k(t') dt' } \label{combo1}
\end{equation}
with $R_k(t)$ obeying a Riccati equation\cite{boyveg} and the coefficients $A_k
\; , B_k$ are fixed by the initial conditions. After some algebra we find
\begin{eqnarray}
 \mid \varphi_k(t) \mid^2 &\buildrel{k \to \infty}\over=& \frac1k +
{{v(t)}\over {2\; k^3}} + {{3\; v(t)^2 - {\ddot v}(t)}\over{8
\; k^5}} + O(k^{-7}) +{\text{oscillatory terms}}  \; ,\cr \cr
 \mid {\dot\varphi}_k(t) \mid^2 &\buildrel{k \to \infty}\over=& k -
{{v(t)} \over {2 \; k}} +{{{\ddot v}(t) - v(t)^2 }\over{ 8
\; k^3}} +  O(k^{-5})+{\text{oscillatory terms}}  \; , \label{asymptotics} 
\end{eqnarray}

Using this asymptotic forms, we obtain\cite{dis,boyveg} the following
renormalized quantities
\begin{eqnarray}
&&
m^2+\frac{\lambda}{16\pi^2}\left[\Lambda^2-m^2_R
\ln\left(\frac{\Lambda}{\kappa}\right)\right] = m^2_R \label{renormass} \\
&&
\lambda\left[1-\frac{\lambda_R}{16\pi^2}\ln\left(\frac{\Lambda}{\kappa}\right)
\right]= \lambda_R \label{renorcoup} 
\end{eqnarray} 
and  
\begin{equation}
\langle \psi^2(t) \rangle_R = \int \frac{k^2dk}{4\pi^2} \left\{
|\varphi_k(t)|^2 - 
\frac{1}{k} + \frac{\Theta(k-\kappa)}{2k^3} 
\left( m^2_R+\frac{\lambda_R}{2}\phi^2(t)+\frac{\lambda_R}{2}\langle \psi^2(t)
\rangle_R \right) \right\}, \label{psiren}
\end{equation}
where we have introduced the (arbitrary) renormalization scale
$\kappa$. Eqs. \ref{renorm} and \ref{psiren} lead to the renormalization
conditions valid in the large $N$ limit.

At this point it is convenient to absorb a further {\em finite} renormalization
in the definition of the mass and introduce the following quantities:
\begin{eqnarray}
&& M^2_R =  m^2_R + \frac{\lambda_R}{2}\langle \psi^2(0) \rangle_R  \label{submass} \\
&& \tau = |M_R| t \; \; , \; \;  q = \frac{k}{|M_R| }
\; \; , \; \; \Omega_q= \frac{W_k}{|M_R|}  \label{dimensionlessvars} \\
&&\eta^2(\tau) = \frac{\lambda_R }{2 |M_R|^2 }\; \phi^2(t)  \\
&& g \Sigma(\tau) = 
\frac{\lambda_R}{2|M_R|^2 } \left[ \langle \psi^2(t) \rangle_R- \langle \psi^2(0) \rangle_R 
\right]  \; \; , \; \; \left( \Sigma(0) = 0 \right) \label{sigma} \\
&& g = \frac{\lambda_R}{8\pi^2} \label{gren} \\
&&\varphi_q(\tau) \equiv |M_R| \; \varphi_k(t) \label{dimensionlessmodes}
\end{eqnarray}

For simplicity in our numerical calculations later, we will chose the
renormalization scale $\kappa = |M_R|$.  The evolution equations are now
written in terms of these dimensionless variables, in which dots now stand for
derivatives with respect to $\tau$.

\subsection{Unbroken Symmetry}

In this case $M^2_R = |M_R|^2$, and in terms of the dimensionless variables
introduced above we find the following equations of motion:

\begin{eqnarray}
& & \ddot{\eta}+ \eta+
\eta^3+ g \;\eta(\tau)\, \Sigma(\tau)  = 0 \label{modo0} \\
& & \left[\;\frac{d^2}{d\tau^2}+q^2+1+
\;\eta(\tau)^2 + g\;  \Sigma(\tau)\;\right]
 \varphi_q(\tau) =0 \; , \label{modok} \\
&& \varphi_q(0) = {1 \over {\sqrt{ \Omega_q}}} \quad , \quad 
{\dot \varphi}_q(0) = - i \; \sqrt{ \Omega_q} \label{conds1} \\
&&\eta(0) = \eta_0  \quad , \quad {\dot\eta}(0) = 0 \label{conds2}
\end{eqnarray}

As mentioned above, the choice of $\Omega_q$ determines the initial state. We
will choose these such that at $t=0$ the quantum fluctuations are in the ground
state of the oscillators at the initial time. Recalling that by definition
$g\Sigma(0)=0$, we choose the dimensionless frequencies to be

\begin{equation}
{\Omega_q}= \sqrt{q^2+1 +  \eta^2_0}. \label{initialfreqs2}
\end{equation} 

The Wronskian of two solutions of (\ref{modok}) is given by
\begin{equation}\label{wrfi} 
{\cal{W}}\left[ \varphi_q, {\bar  \varphi_q}\right] = 2 i \; ,
\end{equation}
while $g \Sigma(\tau)$ is given by
\begin{eqnarray}
g \Sigma(\tau) & = & g \int_0^{\infty} q^2
dq \left\{ 
\mid \varphi_q(\tau) \mid^2 - \frac{1}{\Omega_q}
  \right. \nonumber \\
 &   & \left. + \frac{\theta(q-1)}{2q^3}\left[ 
 -\eta^2_0 + \eta^2(\tau) + g \; \Sigma(\tau) \right] \right\} \;
. \label{sigmafin} 
\end{eqnarray}

\subsection{Broken Symmetry}

In the case of broken symmetry $M^2_R=-|M^2_R|$ and the field equations in the
$N = \infty$ limit become:

\begin{eqnarray}\label{modo0R}
& & \ddot{\eta}- \eta+
\eta^3+ g \;\eta(\tau)\, \Sigma(\tau)  = 0  \\
& & \left[\;\frac{d^2}{d\tau^2}+q^2-1+
\;\eta(\tau)^2 + g\;  \Sigma(\tau)\;\right]
 \varphi_q(\tau) =0 \label{modokR}
\end{eqnarray}
where $ \Sigma(\tau)$ is given in terms of the mode functions $\varphi_q(\tau)$
by the same expression of the previous case, (\ref{sigmafin}). Now the choice
of boundary conditions is more subtle. The situation of interest is when
$0<\eta^2_0 < 1$, corresponding to the situation where the expectation value
rolls down the potential hill from the origin. The modes with $q^2 < 1-
\eta^2_0$ are unstable and thus do not represent simple harmonic oscillator
quantum states. Therefore one {\em must} chose a different set of boundary
conditions for these modes. Our choice will be that corresponding to the ground
state of an {\it upright} harmonic oscillator. This particular initial
condition corresponds to a quench type of situation in which the initial
state is evolved in time in an inverted parabolic potential (for early times
$t>0$). Thus we shall use the following initial conditions for the mode
functions:
\begin{eqnarray}
&& \varphi_q(0) = {1 \over {\sqrt{ \Omega_q}}} \quad , \quad 
{\dot \varphi}_q(0) = - i \; \sqrt{ \Omega_q} \label{conds12} \\
&&\Omega_q= \sqrt{q^2 +1 +\eta^2_0} \quad {\rm for} \; \; q^2 <
1-\eta^2_0 \label{unsfrequ} \\ 
&&\Omega_q=  \sqrt{q^2 -1 + \eta^2_0}  \quad {\rm for} \;  \; q^2
>1-\eta^2_0 \quad  ; 
 \quad 0 \leq \eta^2_0 <1 \label{stafrequ} \; .
\end{eqnarray}
along with the initial conditions for the zero mode given by eq.(\ref{conds2}).

\subsection{Particle Number}
Although the notion of particle number is ambiguous in a time dependent
non-equilibrium situation, a suitable definition can be given with respect to
some particular pointer state. We consider two particular definitions that
are physically motivated and relevant as we will see later. The first is when
we define particles with respect to the initial Fock vacuum state, while the
second corresponds to defining particles with respect to the adiabatic vacuum
state.

In the former case we write the spatial Fourier transform of the
fluctuating field 
$\psi(\vec x,t)$ in (\ref{filargeN}) and its canonical momentum $ \Pi(\vec x,
t)$ as
\begin{eqnarray}
\psi_k(t) & = & \frac{1}{\sqrt{2}}\left[a_k \varphi_k(t) +
a^{\dagger}_{-k} \varphi^*_k(t) \right]  
\label{psioft} \\
\Pi_k(t)  & = &   \frac{1}{\sqrt{2}}\left[a_k \dot{\varphi}_k(t) +
a^{\dagger}_{-k}  
\dot{\varphi}^*_k(t) \right] 
\label{pioft}
\end{eqnarray}
with the {\em time independent} creation and annihilation operators, such that
$a_k$ annihilates the initial Fock vacuum state. Using the initial conditions
on the mode functions, the Heisenberg field operators are written as
\begin{eqnarray}
\psi_k(t) & = & {\cal{U}}^{-1}(t) \; \psi_k(0) \; {\cal{U}}(t) = 
 \frac{1}{\sqrt{2 W_k}}\left[ \tilde{a}_k(t)+ \tilde{a}^{\dagger}_{-k}(t)
\right] \label{psiheis} \\
\Pi_k(t)  & = & {\cal{U}}^{-1}(t) \; \Pi_k(0) \; {\cal{U}}(t) = 
-i \sqrt{\frac{W_k}{2}}
\left[ \tilde{a}_k(t)- \tilde{a}^{\dagger}_{-k}(t)
\right] \label{piheis} \\
\tilde{a}_k(t)    & = & {\cal{U}}^{-1}(t) \; a_k \; {\cal{U}}(t) \label{akoft}
\end{eqnarray}
with ${\cal{U}}(t)$ the time evolution operator with the boundary condition
${\cal{U}}(0)=1$.  The Heisenberg operators $\tilde{a}_k(t)\ ,
\tilde{a}^{\dagger}_k(t)$ are related to $a_k, a^{\dagger}_k$ by a Bogoliubov
(canonical) transformation (see reference\cite{dis} for details).

The particle number defined with respect to the initial Fock vacuum state is
defined in term of the dimensionless variables introduced above as
\begin{equation}
N_q(\tau) = \langle \tilde{a}^{\dagger}_k(t) \tilde{a}_k(t) \rangle =
\frac{\Omega_q}{4}\left[|\varphi_q(\tau)|^2+\frac{|\dot{\varphi}_q(\tau)|^2}{\Omega^2_q}
\right] -\frac{1}{2}. \label{partnumber}
\end{equation}
It is this definition of particle number that will be used for the numerical
study.

In order to define the particle number with respect to the adiabatic vacuum
state we note that the mode equations (\ref{modok},\ref{modokR}) are those of
harmonic oscillators with time dependent squared frequencies
\begin{equation}
\omega^2_q(\tau)= q^2 \pm1+\eta^2(\tau)+g\Sigma(\tau) \label{timedepfreqs2}
\end{equation}
with $+$ for the unbroken symmetry case and $-$ for the broken symmetry case,
respectively. When the frequencies are real, the adiabatic
modes can be introduced in the following manner:
\begin{eqnarray}
\psi_k(t) & = & 
 \frac{1}{\sqrt{2\omega_k(t)}}\left[ \alpha_k(t) e^{-i\int_0^t\omega_k(t') dt'}
+\alpha^{\dagger}_{-k}(t) e^{i\int_0^t\omega_k(t') dt'} 
\right] \label{psiad} \\
\Pi_k(t)  & = &-i \sqrt{\frac{\omega_k(t)}{2}}
\left[ \alpha_k(t)  e^{-i\int_0^t \omega_k(t') dt'}
- \alpha^{\dagger}_{-k}(t)  e^{i\int_0^t \omega_k(t') dt'}
\right] \label{piad} \\ \nonumber
\end{eqnarray}
where now $\alpha_k(t)$ is a canonical operator that destroys the adiabatic
vacuum state, and is related to $a_k, \; a^{\dagger}_k$ by a Bogoliubov
transformation. This expansion diagonalizes the instantaneous Hamiltonian in
terms of the canonical operators $\alpha(t) \; , \alpha^{\dagger}(t)$.  The
adiabatic particle number is
\begin{equation}
{N}^{ad}_q(\tau) = \langle \alpha^{\dagger}_k(t) \alpha_k(t) \rangle =  
\frac{\omega_q(\tau)}{4}\left[|\varphi_q(\tau)|^2+\frac{|\dot{\varphi}_q(\tau)|^2}{\omega^2_q(\tau)}
\right] -\frac{1}{2} \label{adpartnumber}
\end{equation}

As mentioned above, the adiabatic particle number can {\em only} be defined
when the frequencies $\omega_k(t)$ are real. Thus, in the broken symmetry state
they can only be defined for wave-vectors larger than the maximum unstable
wave-vector, $k> k_u=|M_R|\sqrt{1-\eta^2_0}$.  These adiabatic modes and the
corresponding adiabatic particle number have been used previously within the
non-equilibrium context\cite{strongfields,mottola,habib} and will be very
useful in the analysis of the energy below. Both definitions coincide at $t=0$
because $\omega_q(0)= \Omega_q$. Notice that $ {N}^{ad}_q(0) =
{N}_q(0) = 0 $ due to the fact that we are choosing zero initial temperature.
(We considered a non-zero initial temperature in refs.\cite{dis,boyveg}).

\subsection{Energy and Pressure}
The energy momentum tensor for this theory is given by
\begin{equation}
T^{\mu \nu} = \partial^{\mu}\vec{\phi}\cdot \partial^{\nu}\vec{\phi} - g^{\mu
\nu} \left[ \partial_{\alpha}\vec{\phi}\cdot \partial^{\alpha}\vec{\phi}
-V(\vec{\phi} \cdot \vec{ \phi})\right] \label{tmunu}
\end{equation}

Using the large $ N $ factorization (\ref{larg1}-\ref{larg4},
\ref{filargeN}) we find the energy density operator to be
\begin{eqnarray}
\frac{E}{N {\cal{V}}} & = & \frac{1}{2}{\dot{\phi}}^2(t)+\frac{1}{2}m^2
\phi^2(t)+\frac{\lambda}{8} \phi^4(t) +\frac{1}{2
{\cal{V}}}\sum_k\left[\dot{\psi}_k(t)\dot{\psi}_{-k}(t)+\omega^2_k(t)\psi_k(t)\psi_{-k}(t)
\right] -\frac{\lambda}{8}\langle \psi^2(t) \rangle^2 \nonumber \\ & + &
{\rm linear ~terms ~in}\; \psi + {\cal{O}}(1/N) \label{ener1} \\
\omega^2_k(t) & = &k^2+ m^2+\frac{\lambda}{2}\phi^2(t)+\frac{\lambda}{2}\langle
\psi^2(t) \rangle \label{freqs3} .
\end{eqnarray}
Taking the expectation value in the initial state and the infinite volume
limit, defining $\varepsilon = <E>/N{\cal{V}}$, and recalling that the tadpole
condition requires that the expectation value of $\psi$ vanishes, we find the
expectation value of the energy to be
\begin{equation}
\varepsilon   =   \frac{1}{2}{\dot{\phi}}^2(t)+\frac{1}{2}m^2 \phi^2(t)+\frac{\lambda}{8}
\phi^4(t) 
+\frac{1}{8\pi^2}\int k^2 dk \left[|\dot{\varphi}_k(t)|^2+\omega^2_k(t)|\varphi_k(t)|^2 \right]
-\frac{\lambda}{8}\langle \psi^2(t) \rangle^2  \label{ener2}
\end{equation}
It is now straightforward to prove that this bare energy is conserved using the
equations of motion (\ref{zeromodeeqn})-(\ref{fluctuation}). It is important to
account for the last term when taking the time derivative because this term
cancels a similar term in the time derivative of $\omega^2_k(t)$. 

Since we consider translationally as well as  rotationally invariant
states, the expectation value of $ T^{\mu \nu} $ takes the fluid form
\begin{equation}\label{fluido}
T^{00} = <E> \quad , \quad T^{ij} = <P> \; \delta^{ij}  \quad , \quad
T^{i0} = 0 \; .
\end{equation}

We want to emphasize that the full evolution of the zero mode plus the
back-reaction with quantum fluctuations conserves energy. Such is obviously {\em
not} the case in most treatments of reheating in the literature in which
back-reaction effects on the zero mode are neglected. Without energy
conservation, the quantum fluctuations grow without bound. In cosmological
scenarios energy is not conserved but its time dependence is not arbitrary; in
a fixed space-time background metric it is determined by the covariant
conservation of the energy momentum tensor. There again only a full account of
the quantum back-reaction will maintain covariant conservation of the energy
momentum tensor.

We can write the integral in eq.(\ref{ener2}) as
\begin{eqnarray} 
& &\frac{1}{8\pi^2}\int_0^{\Lambda} k^2 dk
\left[|\dot{\varphi}_k(t)|^2+\omega^2_k(t)|\varphi_k(t)|^2 \right] =
\varepsilon_U+ \frac{1}{2\pi^2}\int_{k_u}^{\Lambda}k^2\,dk\;
\omega_k(t)\left({N}^{ad}_k(t)+\frac{1}{2}\right)
\label{enefluc} \\
& & \varepsilon_U= \frac{1}{8\pi^2}\int^{k_u}_0 k^2 dk
\left[|\dot{\varphi}_k(t)|^2+\omega^2_k(t) \;
|\varphi_k(t)|^2 \right] \label{eneunst}
\end{eqnarray}
where $\Lambda$ is a spatial upper momentum cutoff, taken to infinity after
renormalization. In the broken symmetry case, $\varepsilon_U$ is the
contribution to the energy-momentum tensor from the unstable modes with
negative squared frequencies, $k^2_u=|M_R|^2[1-\eta^2_0]$ and
${N}^{ad}_k(t)$ is the adiabatic particle number given by
eq.(\ref{adpartnumber}). 
For the unbroken symmetry case $\varepsilon_U=0$ and $ k_u=0 $. 

This representation is particularly useful in dealing with renormalization of
the energy. Since the energy is conserved, a subtraction at $t=0$ suffices to
render it finite in terms of the renormalized coupling and mass. Using energy
conservation and the renormalization conditions in the large $N$ limit, we find
that the contribution $ \int_{k_u}^{\infty}k^2dk \;\omega_k(t) \;
{N}^{ad}_k(t) $ is finite. This also follows from the asymptotic
behaviors (\ref{asymptotics}).

In terms of dimensionless quantities, the renormalized energy density
is, after taking $\Lambda \rightarrow \infty$:
\begin{eqnarray}
\varepsilon & = & \frac{2 |M_R|^4}{\lambda_R} \left\{
\frac{\dot{\eta}^2}{2}+\frac{\eta^2}{2}{\cal{M}}^2(\tau)+\varepsilon_F-
\frac{{\cal{M}}^4(\tau)}{4}\pm
\frac{1}{2}\left(1-\frac{\lambda_R}{|M_R|^2} <\psi^2(0)>_R\right)
{\cal{M}}^2(\tau) \right.  \nonumber \\ 
& + & \left. 
\frac{g}{8} \left[
\frac{{\cal{M}}^4(\tau)}{4}+ {\cal{M}}^2(\tau) q_u 
\sqrt{q^2_u+{\cal{M}}^2(\tau)}
-2 q_u \left(q^2_u+{\cal{M}}^2(\tau)\right)^{3/2} \right.\right.
\nonumber \\ 
&+ &  \left. \left. {\cal{M}}^4(\tau)
\ln\left[q_u+\sqrt{q^2_u+{\cal{M}}^2(\tau)}\right] \right] \right\} +
 {\cal{C}} \label{renorenergy}\\ \varepsilon_F & = & \frac{g}{2}
\int_0^{q_u} q^2 dq \; \left[|\dot{\varphi}_q|^2+{\omega}^2_q(\tau)
|\varphi_q|^2\right]+ 
2g \int_{q_u}^{\infty} q^2 dq \;{\omega}_q(\tau) \; {N}^{ad}_q(\tau)
\label{flucenergy} \\ 
{\cal{M}}^2(\tau) & = & \pm 1+\eta^2(\tau)+g\Sigma(\tau) \quad , \quad
{\omega}^2_q(\tau)  =  q^2 + {\cal{M}}^2(\tau) \; ,
\label{dimenfreq}
\end{eqnarray}
where the lower sign and $q_u = \sqrt{1- \eta^2_0}$ apply to the broken
symmetry case while the upper sign and $q_u=0$ correspond to the unbroken
symmetry case. The constant ${\cal{C}}$ is chosen such that $\varepsilon$
coincides with the classical energy for the zero mode. The quantity
${\cal{M}}(\tau)$ 
is identified as the effective (dimensionless) mass for the ``pions''.

We find using the renormalized  eqs. (\ref{modo0}),  (\ref{modok}),
 (\ref{sigmafin}),  (\ref{modo0R}) and  (\ref{modokR}), 
 that the renormalized
energy $ \varepsilon $ is indeed {\bf conserved} both for unbroken and for
broken symmetry. 

The pressure is obtained from the spatial components of the energy momentum
tensor (see eq.(\ref{fluido}))
and we find the expectation value of the pressure density $p=<P>/N{\cal{V}}$ to
be given by
\begin{equation}
p= \dot{\phi}^2+ \frac{1}{4\pi^2} \int k^2 dk \left[|\dot{\varphi}_k(t)|^2 +
\frac{k^2}{3}|\varphi_k(t)|^2 \right]- \varepsilon \label{press}
\end{equation}
Using the large-$k$ behavior of the mode functions (\ref{asymptotics}), we find
that aside from the time independent divergence that is present also in the
energy the pressure needs an extra subtraction $\frac{1}{6} \ddot{{\phi}^2}
/k^3$ compared with the energy. Such a term corresponds to an additive
renormalization of the energy-momentum tensor of the form
\begin{equation}
\delta T^{\mu \nu} = A(\eta^{\mu \nu} \partial^2 - \partial^{\mu}
\partial^{\nu}) \phi^2 \label{additive}
\end{equation}
with $A$ a (divergent) constant\cite{zinn}. Performing the integrals with a
spatial ultraviolet cutoff, and in terms of the renormalization scale $\kappa$
introduced before, we find
\begin{equation}
A = -\frac{g}{12} \ln[\frac{\Lambda}{\kappa}] \label{presren}
\end{equation}

In terms of dimensionless quantities and after subtracting a time independent
quartic divergence, we finally find setting $ \Lambda = \infty $,
\begin{eqnarray}
p &=& \frac{2 |M_R|^4}{\lambda_R} \left\{\dot{\eta}^2+ g \int_0^{\infty} q^2 dq
\left[\frac{q^2}{3} |\varphi_q(\tau)|^2+ |\dot{\varphi}_q(\tau)|^2
-\frac43 \, q - \frac1{3q} {\cal{M}}^2(\tau) \right. \right. \cr \cr
&+& \left. \left. {{\theta(q-1)}\over {12 q^3}}
 \;\frac{d^2}{d\tau^2}\left({\eta}^2+g\Sigma(\tau)\right) \right] 
 \right\} -\varepsilon \; . \label{finpress}
\end{eqnarray}

At this stage we can recognize why the effective potential is an
irrelevant quantity to study the dynamics.

The sum of terms {\em without} $\varepsilon_F$ in (\ref{renorenergy})
for $q_u=0$  are identified 
with the effective potential in this approximation for a time independent
 $\eta \; ; g\Sigma$. These arise from the ``zero point'' energy of
the oscillators with time dependent frequency in (\ref{enefluc}).

 In the broken symmetry case the term $\varepsilon_F$ describes the
dynamics of the 
spinodal instabilities\cite{boyveg} since the mode functions will grow in time.
 Ignoring these instabilities and setting $q_u=0$ as is done in a
calculation of the effective  potential 
 results in an imaginary part. In the unbroken symmetry ($q_u=0$) case
the sum of terms 
 without  $\varepsilon_F$ give the effective potential in the large $
N $ limit, but the term $\varepsilon_F$ 
describes the profuse particle production via parametric amplification,
the mode functions in the
unstable bands give a contribution to this term that eventually
becomes non-perturbatively large and 
comparable to the tree level terms as will be described in detail
below. Clearly both in the broken 
and unbroken symmetry cases the effective potential misses {\em all}
of the interesting 
 non-perturbative  dynamics, that is 
the exponential growth of quantum fluctuations and the ensuing
particle production, 
 either associated with unstable bands in the
unbroken symmetry case or spinodal instabilities in the broken symmetry phase. 
 
The expression for the renormalized energy density given by
(\ref{renorenergy}-\ref{dimenfreq}) 
differs from the effective potential in several fundamental aspects:
i) it is always real as opposed 
to the effective potential that becomes complex in the spinodal
region, ii) it accounts for particle 
production and time dependent phenomena.

The effective
potential is a useless tool to study the dynamics precisely because it misses
 the profuse particle production 
associated with these dynamical, non-equilibrium and non-perturbative processes.

\section{The Unbroken Symmetry Case}

\subsection{Analytic Results}

In this section we turn to the analytic treatment of equations
(\ref{modo0},\ref{modok},\ref{sigmafin}) in the unbroken symmetry case. Our
approximations will only be valid in the weak coupling regime and for times
small enough so that the quantum fluctuations, i.e. $g\Sigma(\tau)$ are not
large compared to the ``tree level'' quantities. We will see that this
encompasses the times in which most of the interesting physics occurs. 

Since $\Sigma(0) = 0 $, the back-reaction term $ g \Sigma(\tau) $ is expected to
be small for small $ g $ during an interval say $0 \leq \tau < \tau_1$. This
time $\tau_1$, to be determined below, determines the relevant time scale for
preheating and will be called the preheating time.

During the interval of time in which the back-reaction term $ g \Sigma(\tau) $
can be neglected, we can solve eq.(\ref{modo0}) in terms of elliptic functions,
with the result:
\begin{eqnarray}\label{etac}
\eta(\tau) &=& \eta_0\; \mbox{cn}\left(\tau\sqrt{1+\eta_0^2},k\right)
\cr \cr
k &=& {{\eta_0}\over{\sqrt{2( 1 +  \eta_0^2)}}}\; , 
\end{eqnarray}
where cn stands for the Jacobi cosine.  Notice that $ \eta(\tau) $ has period $
4 \omega \equiv {{ 4 \, K(k)}\slash {\sqrt{1+\eta_0^2}}} $, where $ K(k) $ is
the complete elliptic integral of first kind. In addition we note that since
\begin{equation}
 \eta(\tau + 2 \omega ) =  - \eta(\tau) \; , \label{halfperiod}
\end{equation}
if we neglect the back-reaction in the mode equations, the `potential'
$(-1-\eta^2(\tau))$ is periodic with period $2\omega$.  Inserting this form for
$\eta(\tau)$ in eq.(\ref{modok}) and neglecting $ g \Sigma(\tau) $ yields
\begin{equation}\label{modsn}
 \left[\;\frac{d^2}{d\tau^2}+q^2+1+ \eta_0^2 -  \eta_0^2\;
\mbox{sn}^2\left(\tau\sqrt{1+\eta_0^2},k\right) \;\right]
 \varphi_q(\tau) =0 \; . \label{nobackreaction}
\end{equation}
This is the Lam\'e equation for a particular value of the coefficients that
make it solvable in terms of Jacobi functions \cite{herm}.  We summarize here
the results for the mode functions. The derivations are given in the Appendix.

Since the coefficients of eq.(\ref{modsn}) are periodic with period $ 2 \omega
$, the mode functions can be chosen to be quasi-periodic (Floquet type) with
quasi-period $ 2 \omega $.
\begin{equation}\label{floq}
 U_q(\tau + 2  \omega) =   e^{i F(q)} \; U_q(\tau),
\end{equation}
where the Floquet indices $ F(q) $ are independent of $\tau$.  In the allowed
zones, $ F(q) $ is a real number and the functions are bounded with a constant
maximum amplitude. In the forbidden zones $ F(q) $ has a non-zero imaginary
part and the amplitude of the solutions either grows or decreases
exponentially.

Obviously, the Floquet modes $ U_q(\tau) $ cannot obey in general the initial
conditions given by (\ref{pionmodes}) and the proper mode functions with these
initial conditions will be obtained as linear combinations of the Floquet
solutions.  We normalize the Floquet solutions as
\begin{equation}\label{ucero}
U_q(0) = 1 \; .
\end{equation}
We choose $ U_q(\tau) $ and $ U_q(-\tau) $ as an independent set of solutions
of the second order differential equation (\ref{modsn}).  It follows from
eq.(\ref{floq}) that $ U_q(-\tau) $ has $ -F(q) $ as its Floquet index.

We can now express the modes $\varphi_q(\tau)$ with the proper boundary
conditions (see \ref{pionmodes}) as the following linear combinations of
$U_q(\tau) $ and $ U_q(-\tau) $

\begin{equation}
\varphi_q(\tau)={1 \over {2 \sqrt{\Omega_q}}}\;\left[
\left(1 - {{2i\Omega_q}\over {{\cal{W}}_q}} \right)\;  U_q(-\tau)+
\left(1 + {{2i\Omega_q}\over {{\cal{W}}_q}} \right)\; U_q(\tau)
\right]\; , \label{combi2} 
\end{equation}
where ${\cal{W}}_q$ is the Wronskian of the two Floquet solutions
\begin{equation}
{\cal{W}}_q\equiv W[ U_q(\tau), U_q(-\tau)]= - 2 {\dot U}_q(0)\;
. \label{wronskian} 
\end{equation}

Eq.(\ref{modsn}) corresponds to a Schr\"odinger-like equation with a one-zone
potential\cite{ince}. We find {\em two} allowed bands and {\em two} forbidden
bands.  The allowed bands correspond to
\begin{equation}\label{bandaq}
 -1 - {{\eta_0^2}\over 2} \leq q^2 \leq 0
 \quad \mbox{and} \quad
 {{\eta_0^2}\over 2} \leq q^2 \leq +\infty \; ,
\end{equation}
and the forbidden bands to
\begin{equation}\label{bandap}
-\infty \leq q^2 \leq  -1 - {{\eta_0^2}\over 2}
 \quad \mbox{and} \quad
0  \leq q^2 \leq  {{\eta_0^2}\over 2} \, .
\end{equation}
The last forbidden band is for {\em positive} $q^2$ and hence will contribute
to the exponential growth of the fluctuation function $\Sigma(\tau)$.

The mode functions can be written explicitly in terms of Jacobi
$\vartheta$-functions for each band. We find for the forbidden band,
\begin{equation}\label{uqproh}
 U_q(\tau) = e^{- {\tau\;\sqrt{1+\eta_0^2}\; Z(2  K(k)\,v)}} \;
{{\vartheta_4(0) \; \vartheta_1(v+ {{\tau}\over {2\omega}} )}\over
{\vartheta_1(v)\; \vartheta_4( {{\tau}\over {2\omega}} )}}\; ,
\end{equation}
where $v$ is a function of $q$ in the forbidden band $ 0 \leq q \leq
{{\eta_0}\over {\sqrt2}} $ defined by 
\begin{equation}\label{qprohi}
q =  {{\eta_0}\over {\sqrt2}}\, \mbox{cn}(2  K(k)\,v,k) \; , \; 0 \leq v
\leq \frac{1}{2}. 
\end{equation}
and $ Z(u) $ is the Jacobi zeta function \cite{erd}. 
It can be expanded  in  series as follows
\begin{equation}\label{Zjacob}
 2\, K(k)\; Z(2  K(k)\,v) = 4\pi\; \sum_{n=1}^{\infty} {{{\hat q}^n}\over{1 -
{\hat q}^{2n}}}\; \sin(2n\pi v)
\end{equation}
where ${\hat q }\equiv e^{-\pi K'(k)/  K(k)} $. 
The Jacobi $\vartheta$-functions  can be expanded  in  series as
follows \cite{gr} 
\begin{eqnarray}
\vartheta_{1}(v|{\hat q}) & = &  2 \sum_{n=1}^{\infty} \;  (-1)^{n+1}\;
{\hat q}^{(n-1/2)^2}\; \sin(2n-1)\pi  v  \; , \cr \cr
\vartheta_{4}(v|{\hat q}) & = & 1 +  2 \sum_{n=1}^{\infty} \;  (-1)^{n}\;
{\hat q}^{n^2}\; \cos(2n \pi  v)  \; .
\end{eqnarray}

We explicitly see in eq.(\ref{uqproh}) that $ U_q(\tau) $ factorizes into a
real exponential with an exponent linear in $\tau$ and an antiperiodic function
of $ \tau $ with period $2 \omega$. Recall that
\begin{equation}\label{tetaP}
 \vartheta_1(x+1) = - \vartheta_1(x) \quad , \quad 
 \vartheta_4(x+1) = + \vartheta_4(x) \; .
\end{equation}
We see that the solution $ U_q(\tau) $ decreases with $\tau$. The other
independent solution $U_q(-\tau) $ grows with $\tau$.

The Floquet indices can be read comparing eq.(\ref{floq}), (\ref{uqproh}) and
(\ref{tetaP}),
\begin{equation}
 F(q) = 2 i \, K(k)\; Z(2  K(k)\,v)  \pm  \pi \; .\label{floqindex}
\end{equation}

$ U_q(\tau) $ turns out to be a real function  in the forbidden band. 
It has real zeroes at
\begin{equation}
\tau = 2 \omega ( n - v)  \quad , \quad  n\; \epsilon {\cal Z} \; .\label{realzeros}
\end{equation}
and complex poles at
\begin{equation}\label{polos}
\tau = 2 \omega  n_1 + (2 n_2 + 1 ) \omega'   \quad , \quad  
n_1 , n_2 \; \epsilon {\cal Z} \; .
\end{equation}
where $\omega'$ is the complex period of the Jacobi functions.
Notice that the pole positions are $q$-independent,
and that   $ U_q(\tau) $ becomes  an antiperiodic
function on the borders of this forbidden band,   $ q = 0 $ and  $ q=  {{\eta_0}\over {\sqrt2}} $.
We find using eq.(\ref{uqproh}) and  ref.\cite{erd},
\begin{eqnarray} 
 U_q(\tau)|_{q = 0} &=&  \mbox{cn}(\tau\sqrt{1+\eta_0^2},k)  \label{zeromod} \\
\lim_{q\to {{\eta_0}\over {\sqrt2}} }
\left[ v  U_q(\tau)\right] &=& {1 \over {\pi \vartheta^2_3(0)}}\;
 \mbox{sn}(\tau\sqrt{1+\eta_0^2},k) \; ,
\end{eqnarray} 
respectively. 

The functions $ U_q(\tau) $ transform under complex conjugation in the
forbidden band as
\begin{equation}\label{conjuP}
 [U_q(\tau)]^{*} =  U_q(\tau) \; .
\end{equation}

For the allowed band $  {{\eta_0}\over {\sqrt2}} \leq q \leq \infty $,
we find for the mode functions
\begin{equation}\label{Uqperm}
 U_q(\tau) = e^{ -{{\tau}\over {2\omega}}
{{\vartheta_1'}\over{\vartheta_1}}(i \,{{K'(k)}\over{ K(k)}}\, v)} \;
{{\vartheta_4(0) \; \vartheta_4({{i\,K'(k)}\over{ K(k)}}\, v  +
{{\tau}\over {2\omega}} )}\over {\vartheta_4(i\,{{K'(k)}\over{ K(k)}}\, v)\; 
\vartheta_4( {{\tau}\over {2\omega}} )}}\; , 
\end{equation}
where
\begin{eqnarray}
&&q = \sqrt{\eta_0^2 + 1}\;
{{\mbox{dn}}\over{\mbox{sn}}}\left(2\, K'(k) \, v, k'\right)\; , \\
&& 
0 \leq v \leq \frac12 \quad , \quad \infty \geq q \geq  {{\eta_0}\over
{\sqrt2}}  
\end{eqnarray}
We see that  $ U_q(\tau) $ in this allowed band factorizes into 
a phase proportional to $\tau$ and a complex periodic function with 
period $2 \omega$. This function  $ U_q(\tau) $ has {\em no real zeroes}
 in $\tau$ except when $q$ is at the lower border 
$q = {{\eta_0}\over {\sqrt2}}$ . Its poles in $\tau$ 
are $q$-independent and they are the same as those in the forbidden band
[see eq.(\ref{polos})].

The Floquet indices can be read off by comparing eq.(\ref{floq}), (\ref{tetaP})
and (\ref{Uqperm})
\begin{equation}
 F(q) =  i {{\vartheta_1'}\over{\vartheta_1}}\left(i \,
{{K'(k)}\over{K(k)}}\, v\right)  \; .
\end{equation}
These indices are real in the allowed band.

The functions $U_q(\tau)$ transform under complex conjugation in the allowed
band as
\begin{equation}\label{conjuA}
[U_q(\tau)]^{*} =  U_q(-\tau) \; .
\end{equation}

Obviously these modes will give contributions to the fluctuation $\Sigma(\tau)$
which are always bounded in time and at long times will be subdominant with
respect to the contributions of the modes in the forbidden band that grow
exponentially.

The form of these functions is rather complicated, and it is useful to find
convenient approximations of them for calculational convenience.

The expansion of the $\vartheta$-functions in powers of ${\hat q }= e^{-\pi
K'(k)/ K(k)} $ converges quite rapidly in our case. Since $ 0 \leq
k \leq 1/\sqrt2 $ [see eq.(\ref{etac})], we have
\begin{equation}\label{qsomb}
 0 \leq {\hat q } \leq e^{-\pi} = 0.0432139\ldots \; . 
\end{equation}
 ${ \hat q }$ can be  computed with high precision from the series \cite{gr}
$$
{ \hat q } = \lambda + 2  \, \lambda^5 +  15 \, \lambda^9 + 150 \,
\lambda^{13} + 1707  \, \lambda^{17} + \ldots \; ,
$$
where (not to be confused with the coupling constant) 
\begin{equation}
\lambda \equiv \frac12 \; {{1 - \sqrt{k'}}\over {1 + \sqrt{k'}}}\; .
\end{equation}
We find from eq.(\ref{etac}) 
\begin{equation}
\lambda = \frac12 \; {{ (1+\eta_0^2)^{1/4} -  (1+\eta_0^2/2)^{1/4}}
\over { (1+\eta_0^2)^{1/4} +  (1+\eta_0^2/2)^{1/4}}} \; . \label{lambdaprox}
\end{equation}
The quantity $\lambda$ can be computed and is a small number: for $ 0 \leq
\eta_0 \leq \infty $, we find $ 0 \leq \lambda \leq 0.0432136\ldots $.
Therefore, to very good approximation,  with an error smaller than
$\sim 10^{-7} $, we may use:
\begin{equation}\label{qaprox}
{ \hat q } =  \frac12 \;  {{ (1+\eta_0^2)^{1/4} -  (1+\eta_0^2/2)^{1/4}}
\over { (1+\eta_0^2)^{1/4} +  (1+\eta_0^2/2)^{1/4}}}  \; .
\end{equation}

We find in the forbidden band from  eq.(\ref{uqproh}) and \cite{erd}
\begin{equation}\label{Uapro}
\displaystyle{
 U_q(\tau) = e^{-4 \tau \; \sqrt{1 + \eta_0^2 } \; {\hat q }\; \sin(2\pi v)\;
\left[ 1 + 2 \, {\hat q} \;\left( \cos2\pi v - 2 \right)  + O( {\hat q
 }^2)  \right]} 
\; \; {{1 - 2 {\hat q }}\over{1-  2{{\hat q} \; 
\cos({{\pi \tau}\over{\omega}}) }}}\; 
{{\sin\pi(v + {{\tau}\over{2\omega}})}\over {\sin\pi v}} \; 
\left[ 1 + O( {\hat q }^2) \right]} \; ,
\end{equation}
where now we can relate $v$ to $q$ in the simpler form
\begin{equation}\label{qapro}
q = {{\eta_0}\over {\sqrt2}}\, \cos \pi v \,  \left[ 1 - 4  {\hat q }
\; \sin^2 \pi v +   4  {\hat q }^2
\; \sin^2 \pi v \; ( 1 + 4 \cos^2 \pi v )+ O( {\hat q }^3)\right]  \; ,
\end{equation}
which makes it more convenient to write $q(v)$ in the integrals, and
\begin{equation}\label{omegapro}
{{\pi}\over{2\omega}} =  \sqrt{1+\eta_0^2}\;  \left[ 1 - 4 {\hat q}
+ 12   {\hat q }^2 +  O( {\hat q }^4) \right] \; , 
\end{equation}
where $ 0  \leq v  \leq \frac12 $. 

The Floquet indices can now be written in a very compact form amenable for
analytical estimates
\begin{equation}
F(q) = 4i\, \pi  \; {\hat q }\; \sin(2\pi v)\;\left[ 1 + 2 \, {\hat q}
\; \cos2\pi v   + O( {\hat q }^2)\right] +\pi \; .\label{easyfloquet}
\end{equation}

In this approximation the zero mode
(\ref{etac}) becomes
\begin{equation}
\eta(\tau) = \eta_0  \; \cos \left({{\pi\tau}\over{2\omega}} \right)\;
\left[ 1 -  4  {\hat q }  \, \sin^2({{\pi\tau}\over{2\omega}})\;  
+ O( {\hat q }^2) \right] \; . \label{zeromodeappx}
\end{equation}

This expression is very illuminating, because we find that a Mathieu equation
approximation, based on the first term of eq.(\ref{zeromodeappx}) to the
evolution of the mode functions is {\em never} a good approximation. The reason
for this is that the second and higher order terms are of the same order as the
secular terms in the solution which after resummation lead to the
identification of the unstable bands. In fact, whereas the Mathieu equation has
{\em infinitely many} forbidden bands, the exact equation has only {\em one}
forbidden band. Even for small $ {\hat q} $, the Mathieu equation is
not a good approximation to the Lam\'e equation \cite{paris}.


From eq.(\ref{Uqperm}) analogous formulae can be obtained for the allowed band

\begin{equation}
 U_q(\tau) = e^{ -{{i\pi \tau}\over {2\omega}}\;\coth\left[
{{\pi \, K'(k)}\over{ K(k)}}\, v\right]} \;
\;  {{1 - 2 {\hat q }}\over{1-  2{{\hat q} \; 
\cos({{\pi \tau}\over{\omega}}) }}}\; 
{{{1 -  2 {\hat q } \, \cos\left[{{\pi\,\tau}\over{ \omega}}-2iv\,
\log{\hat q} \right]}}\over{1 -  2 {\hat q } \, 
\cosh(2 \, v \, \log{\hat q})}}
\left[ 1 + O( {\hat q }^2) \right] \; ,  
\end{equation}
where
\begin{equation}
q = {{\sqrt{\eta_0}}\over{ 2^{3/2}\; 
\sinh\left( {{\pi \, K'(k)}\over{ K(k)}}\, v\right)}}\; 
\left( {{ \eta_0^2 +2}\over{  {\hat q }}} \right)^{1/4}\;
\left\{ 1 +  2 {\hat q }\;  \cosh(2 \, v \, \log{\hat q}) 
+  O( {\hat q }^2) \right\} \; .
\end{equation}
Here,
$$
0 \leq v \leq \frac12 \quad , \quad \infty \geq q \geq  {{\eta_0}\over
{\sqrt2}}  \; .
$$
Note that eq.(\ref{omegapro}) holds in all bands.

We can now estimate the size and growth of the quantum fluctuations, at least
for relatively short times and weak couplings.  For small times $ 0 \leq \tau <
\tau_1 $ (to be determined consistently later) and small coupling $g << 1 $, we
can safely neglect the back-reaction term $ g \Sigma(\tau)$ in eq.(\ref{modok})
and express the modes $\varphi_q(\tau)$ in terms of the functions $ U_q(\tau) $
and $ U_q(-\tau) $ for this however, we need the Wronskian, which in the
forbidden band is found to be given by:
\begin{equation}
{\cal{W}}_q = -\frac{1}{\omega}\, {d \over {dv}}\log{{\vartheta_1(v)}\over
{\vartheta_4(v)}} = - 2 \sqrt{1 +\eta_0^2 } \; {{ \mbox{cn} \; \mbox{dn}}\over{
\mbox{sn}}}(2vK(k),k)\; .
\end{equation}
In terms of the variable $q^2$ this becomes, after using eq.(\ref{qprohi}):
\begin{equation}
{\cal{W}}_q = - 2 q \sqrt{{ {{\eta_0^2}\over 2}+1 + q^2 }\over {
{{\eta_0^2}\over 2}- q^2} }\; .
\end{equation}
This Wronskian is regular and non-zero except at the four borders
of the bands.

We find from eq.(\ref{combi2}) that $|\varphi_q(\tau) |^2 $ is given by
\begin{equation}\label{mod2}
 \mid\varphi_q(\tau)  \mid^2 = {1 \over {4  \Omega_q}}\;\left\{ 
\left[  U_q(\tau) +  U_q(-\tau) \right]^2 + {{4  \Omega_q^2} \over {
{\cal{W}}_q^2}} \left[  U_q(\tau) -  U_q(-\tau)\right]^2 \right\} 
\end{equation}
where we took into account eqs. (\ref{conjuP} and (\ref{conjuA}).  Notice that
both terms in the rhs of eq.(\ref{mod2}) are real and positive for real $ q $.
For very weak coupling and after renormalization, the contribution to
$g\Sigma(\tau)$ from the stable bands will always be perturbatively small,
while the contribution from the modes in the unstable band will grow in time
exponentially, eventually yielding a non-perturbatively large
contribution. Thus these are the only important modes for the fluctuations and
the back-reaction.  An estimate of the preheating time scale can be obtained by
looking for the time when $g\Sigma(\tau)$ is of the same order of the classical
contributions to the equations of motion. In order to obtain an estimate for
the latter, we consider the average over a period of the classical zero mode:
\begin{equation}
1+<\eta^2(\tau)> = (1+\eta^2_0) \left[\frac{2E(k)}{K(k)}-1\right]
\end{equation}
which yields for small and large initial amplitudes the following results
\begin{eqnarray}
&& <\eta^2(\tau)> \buildrel{ \eta_0 \to 0 }\over=
\frac{\eta^2_0}{2} \\ 
&& <\eta^2(\tau)> \buildrel{ \eta_0 \to \infty }\over=
0.4569 \ldots \; \eta^2_0 \;
\end{eqnarray} 
Therefore the average over a period of $\eta^2(\tau)$ is to a very good
approximation $\eta^2_0/2$ for all initial amplitudes. This result provides an
estimate for the preheating time scale $\tau_1$; this occurs when
$g\Sigma(\tau_1) \approx (1+\eta^2_0/2)$. Furthermore, at long times (but before
$g\Sigma \approx (1+\eta^2_0/2)$) we need only keep the exponentially growing
modes and $g \Sigma(\tau) $ can be approximated by
\begin{equation}\label{Sest}
g \Sigma_{est}(\tau)  = \frac{g}{4} \int_0^{ {{\eta_0}\over
{\sqrt2}}}q^2 
dq\; { 1 \over {\Omega_q}}\left[ 1 + {{4 \Omega_q^2} \over
{{\cal{W}}_q^2}} \right]\; \mid  U_q(-\tau) \mid^2 \; .
\end{equation}
Moreover, choosing $\tau$ such that the oscillatory factors in $ U_q(-\tau) $
attain the value 1 (the envelope), and using eq.(\ref{uqproh}) we finally
obtain:
\begin{equation}\label{Senv}
 \Sigma_{est-env}(\tau)  =    {1 \over 4} \int_0^{ {{\eta_0}\over
{\sqrt2}}} q^2 
dq \; { 1 \over {\Omega_q}}\left[ 1 + {{4 \Omega_q^2} \over
{{\cal{W}}_q^2}} \right]\; e^{2 \tau\;\sqrt{1+\eta_0^2}\; Z(2  K(k)\,v,k)}
\end{equation}
where $v$ depends on the integration variable through
eq.(\ref{qprohi}). 

The Jacobi $ Z $ function can be accurately represented using eq.(\ref{Zjacob})
\begin{equation}
 Z(2  K(k)\,v,k) =  4\, {\hat q }\; \sin 2 \pi v  \left[ 1 -  2\, {\hat
 q }\left(  2 - \cos2\pi v \right) \right]  + O( {\hat q }^3) \; .
\end{equation}
where we recall that $ \hat q < 0.0433 $.

The integral (\ref{Senv}) will be dominated by the point $ q $ that maximizes
the coefficient of $\tau$ in the exponent. This happens at $ q = q_1, \; v =
v_1$, where
\begin{eqnarray}
q_1 & = &  \frac12 \; \eta_0 \; (1 - {\hat q })  + O( {\hat q }^2) \label{quno} \\
Z(2  K(k)\,v_1,k) &  =  & 4  {\hat q } \; (1 - 4 {\hat q })  + O( {\hat q }^3) \;
\end{eqnarray}

We can compute the integral (\ref{Senv}) by saddle point approximation to find:
\begin{eqnarray}\label{sadd}
 \Sigma_{est-env}(\tau)  &=&    
\displaystyle{
{{q_1^2 \; \left[ 1 + {{4 \Omega_{q_1}^2} \over
{{\cal{W}}_{q_1}^2}} \right] }\over {2  \;\Omega_{q_1}}} \;
e^{ 8\, \tau\, \sqrt{1+\eta_0^2}\; {\hat q } \; (1 - 4 {\hat q })}}
\;\cr \cr
& & \displaystyle{ \int_{-\infty}^{\infty}\, dq \;
e^{ -64 \tau \, (q-q_1)^2 \; {\hat q }
\;\sqrt{1+\eta_0^2}\;\eta_0^{-2} \;(1-2  {\hat q })} \left[ 1 + O
({\hat q })\right] }\cr \cr
& = &  \displaystyle{
{{\eta_0^3 \, \sqrt{\pi}  \,  \left[ 1 + {{4 \Omega_{q_1}^2} \over
{{\cal{W}}_{q_1}^2}} \right] }\over {64 \; (1+\eta_0^2)^{1/4}\;
\sqrt{\tau  {\hat q }}\; \Omega_{q_1}}}\; 
e^{ 8\, \tau\;\sqrt{1+\eta_0^2}\; {\hat q } \; (1 - 4 {\hat q })} 
\left[ 1 + O ({1 \over {\tau}}) \right] }\label{estenvfor} \; .
\end{eqnarray}
We can relate $\hat{q}$ to $\eta_0$ using eq.(\ref{qaprox}), and we have used the
small $\hat{q}$ expansion 
\begin{eqnarray}
Z''(2  K(k)\,v_1,k)      &  = &  
-16  {\hat q } \; (1 - 4 {\hat q })  + O( {\hat q }^3) \\
\frac{dq}{dv}|_{v_1} & = &  - \frac{1}{2\pi}(1 - 9 {\hat q })  + 
O( {\hat q}^2) \; .
\end{eqnarray}

In summary, during the preheating time where parametric resonance is
important, $ \Sigma_{est-env}(\tau) $ can be represented to a very good
approximation by the formula
\begin{equation}\label{polenta}
 \Sigma_{est-env}(\tau) = { 1 \over { N \, \sqrt{\tau}}}\; e^{B\,\tau}\; ,
\end{equation}
where $ B $ and $ N $ are functions of $ \eta_0 $ given by
\begin{eqnarray}\label{ByN}
B &=&  \displaystyle{
8\, \sqrt{1+\eta_0^2}\; {\hat q } \; (1 - 4 {\hat q }) +  O(
{\hat q }^3) }\; , \cr \cr
N &=&  \frac{64}{ \pi^{1/2}} \; {{(1+\eta_0^2)^{1/4}\;
\sqrt{  {\hat q }}\; \Omega_{q_1}} \over { \eta_0^3 \,  \,  \left[ 1 +
{{4 \Omega_{q_1}^2} \over {{\cal{W}}_{q_1}^2}}  \right] } }\cr \cr
 &=& {4 \over {\sqrt{ \pi}}} \; \sqrt{  {\hat q }}\;
{{ ( 4 + 3 \, \eta_0^2) \, \sqrt{  4 + 5 \, \eta_0^2}}\over{
 \eta_0^3 \, (1+\eta_0^2)^{3/4}}} \left[ 1 + O ({\hat q })\right]\; \; .
\end{eqnarray}
and eq.(\ref{qaprox}) gives $ {\hat q } $ as a function of $ \eta_0 $. This is
one of the main results of this work.

We display in Table I below some relevant values of $ {\hat q }, \; B $ and $ N
$ as functions of $ \eta_0 $.

We notice that the limiting values of $ B $ and $ N $ for $ \eta_0 \to \infty $
yield a very good approximation even for $ \eta_0 \sim 1 $. Namely,
\begin{equation}
 \Sigma(\tau) \approx \sqrt{{ \eta_0^3}\over {\tau}}\;{{e^{B_{\infty}\,
  \eta_0 \, \tau}}\over {N_{\infty} }} \; . \label{appxfluc}
\end{equation}
with the asymptotic values given by
\begin{eqnarray}
B_{\infty} & = &  8 e^{-\pi} (1 - 4 e^{-\pi}) [ 1 + O(\eta_0^{-2})]
  = 0.285953 \ldots [ 1 + O(\eta_0^{-2})]  \label{binfty} \\
N_{\infty} &  = &  \frac{12}{\sqrt{\pi}} \sqrt5 \, e^{-\pi/2} [ 1 +
  O(\eta_0^{-2})]= 3.147\ldots [ 1 + O(\eta_0^{-2})] \label{ninfty} .
\end{eqnarray}

These rather simple expressions (\ref{polenta}-\ref{ninfty}) allow us to perform
analytic estimates with great accuracy and constitute one of our main analytic
results. The accuracy of this result will be discussed below in connection with
the full numerical analysis including back-reaction.

Using this estimate for the back-reaction term, we can now estimate the value of
the preheating time scale $\tau_1$ at which the back-reaction becomes comparable
to the classical terms in the differential equations. Such a time is defined by
$ g \Sigma(\tau_1) \sim (1+\eta^2_0/2) $. From the results presented above, we
find
\begin{equation}
\tau_1 \approx {1 \over B} \, \log{{N(1+\eta^2_0/2) \over { g \sqrt B}}}\; .\label{maxtime}
\end{equation}

The time interval from $\tau=0$ to $\tau\sim \tau_1$ is when most of the particle
production takes place. After $\tau \sim \tau_1 $ the quantum fluctuation become
large enough to shut-off the growth of the modes and particle production
essentially stops. We will compare these results to our numerical analysis
below.

We can now use our analytic results to study the different contributions to the
energy and pressure coming from the zero mode and the quantum fluctuations and
begin by analyzing the contribution to the energy $ \epsilon_0 $ and pressure $
p_0$ from the zero mode $\eta(\tau)$.

The dimensionless energy and pressure, 
(normalized by the factor $2 M^4_R / \lambda_R$)
 are given by the following expressions,
\begin{eqnarray}\label{e0p0}
 \epsilon_0(\tau) &=&  \frac12 \left[ {\dot \eta}^2 +  \eta(\tau)^2 +  \frac12
\eta(\tau)^4  \right]  \; , \cr \cr
p_0(\tau) &=& 
 \frac12 \left[ {\dot \eta}^2 -  \eta(\tau)^2 -  \frac12
\eta(\tau)^4 \right] \; .
\end{eqnarray}

When the back-reaction term $g\Sigma(\tau)$ can
be neglected, we can use eq.(\ref{etac}) as a good
approximation to  $\eta(\tau)$. In this approximation 
\begin{eqnarray}\label{e0p0A}
\epsilon_0 &=&
 \frac12  \eta_0^2 \; \left[   1 +  \frac12 \eta_0^2
\right] \; , \cr  \cr
p_0(\tau) + \epsilon_0 &=&  \eta_0^2 \left( 1 + \eta_0^2
\right)\; \mbox{sn}^2 \mbox{dn}^2\left(\tau\sqrt{1+\eta_0^2},k\right) \; .
\end{eqnarray}
The zero mode energy is conserved and the pressure oscillates between plus and
minus $ \epsilon_0 $ with period $2 \omega$.

Averaging $ p_0(\tau) $ over one period yields
\begin{equation}\label{promp}
<p_0> \equiv  {1 \over {2 \omega}}\; \int_0^{2 \omega} d\tau \;
p_0(\tau)\; .
\end{equation}
Inserting eq.(\ref{e0p0A}) into  eq.(\ref{promp}) yields \cite{pbm}
\begin{equation}\label{prompA}
<p_0> = -\frac16  \eta_0^2 \; \left[  1 -  \frac12 \eta_0^2 \right]+
\frac23\, (  1 + \eta_0^2 ) \left[ 1 - {{E(k)}\over {K(k)}}\right]
\end{equation}
where $k$ is given by  eq.(\ref{etac}). 

$<p_0>$ vanishes for small $ \eta_0 $ faster than $ \epsilon_0$,
\begin{equation}\label{pp0}
<p_0> \buildrel{ \eta_0 \to 0 }\over= \frac{1}{24} \eta_0^4 + O(
\eta_0^6),
\end{equation}
so that the zero mode contribution to the equation of state is that
of dust for small $\eta_0 $.
For large  $\eta_0 $ we find from eq. (\ref{prompA}),
\begin{equation}
<p_0>  \buildrel{ \eta_0 \to \infty }\over= \frac{1}{12} \eta_0^4 +
 \eta_0^2  \left[ \frac12 - \frac23\; {{E(1/\sqrt2)}\over
{K(1/\sqrt2)}}\right] +O(1) 
\end{equation}
where $  \frac12 - \frac23\; {{E(1/\sqrt2)}\over
{K(1/\sqrt2)}} = 0.01437\ldots $. The equation of state approaches
that of radiation for $ \eta_0 \to \infty $:
\begin{equation}
<p_0>  \buildrel{ \eta_0 \to \infty }\over=  \epsilon_0 \left[
\frac13 - {{0.6092\ldots}\over {  \eta_0^2}} + O( \eta_0^{-4})\right]
\; . \label{zerorad}
\end{equation}

Thus we see that for small amplitudes the zero mode stress-energy, averaged
over an oscillation period, behaves as dust while for large amplitudes, the
behavior is that of a radiation fluid. The ratio $<p_0>/ \varepsilon_0$ for
zero mode vs.$ \varepsilon_0 $ is shown in figure 1.  

The  contribution from the $k \neq 0$ modes originates in the quantum fluctuations
during the  the stage of parametric amplification.

Since we have fluid behaviour, we can define an effective (time-dependent)
polytropic index $ \gamma(\tau) $ as
\begin{equation}
 \gamma(\tau)\equiv {{ p(\tau)}\over{\varepsilon }} + 1 \; .
\end{equation}
where renormalized quantities are understood throughout. Within a cosmological
setting whenever $\gamma(\tau)$ reaches a constant value such equation of state
implies a scale factor $ R(\tau) = R_0\; \tau^{2\over {3 \gamma}} $.

In the case being studied here, that of Minkowski space, $ \varepsilon $ is
time-independent and hence equal to the initial energy density (divided by $ N
$ and restoring pre-factors) which after a suitable choice of the constant ${\cal{C}}$ is given by:
\begin{equation}
\varepsilon = \frac{2|M_R|^4}{\lambda_R} \left\{
\frac12  \eta_0^2 \; \left[   1 +  \frac12 \eta_0^2
\right]\right\} \; .
\end{equation}

As argued before, for weak coupling the important contribution to the quantum
fluctuations come from the modes in unstable bands, since these grow
exponentially in time and give rise to a non-perturbatively large
contribution. Thus we concentrate only on these modes in calculating the
pressure.

The contribution of the forbidden band to the renormalized
$ p(\tau) + \varepsilon $ can be written as
\begin{equation}
 [p(\tau) + \varepsilon]_{unst}  = \frac{2|M_R|^4}{\lambda_R}\left\{
 g\, \int_0^{\eta_0/\sqrt2}q^2
dq
\; \left[ \; \mid  {\dot\varphi}_q(\tau) \mid^2  
+ \frac{1}{3} \, q^2 \,   \mid {\varphi}_q(\tau) \mid^2   \right] \right\} \; .
\end{equation}

After renormalization, the terms that we have neglected in this approximation
are perturbatively small (of order $g$) whereas the terms inside the bracket
eventually become of order 1 (comparable to the tree -level contribution).  We
now only keep the exponentially growing pieces in the mode functions $
{\varphi}_q(\tau) $ and $ {\dot\varphi}_q(\tau) $ since these will dominate the
contribution to the pressure.  This is simplified considerably by writing to
leading order in $ {\hat q} $
\begin{equation}
 {\dot\varphi}_q(\tau) =  {\varphi}_q(\tau) \left\{ \sqrt{1+\eta_0^2}\;
{\rm cot} \left[ \pi (v - {{\tau}\over{2\omega}}) \right] + O( {\hat
q} ) \right\} \; . \label{fidotappx}
\end{equation}
Averaging over a period of oscillation yields
\begin{eqnarray}\label{estpE}
 [p(\tau) + \varepsilon]_{unst}  &=&  \frac{2|M_R|^4}{\lambda_R} \left\{
 \frac{g}{4} \; 
  \int_0^{ {{\eta_0}\over {\sqrt2}}} \frac{q^2 
dq}{(4\pi)^2}\; { 1 \over {2 \; \Omega_q \, \sin^2{\pi v}}}
\left[ 1 + \frac{4 \Omega_q^2}
{ {\cal{W}}_q^2} \right]  \right. \\
& & \left. e^{2 \tau\;\sqrt{1+\eta_0^2}\; Z(2  K(k)\,v,k)}\;
\left[  1+\eta_0^2  + \frac{1}{3} \, q^2 \right] \right\}\; .
\end{eqnarray}

This integral is similar to the one in eq.(\ref{Senv}) and we find that they
are proportional in the saddle point approximation. In fact,
\begin{equation}
 [p(\tau)+\varepsilon]_{unst}  = \frac{2|M_R|^4}{\lambda_R}\left[
g \Sigma_{est-env}(\tau) \; 
\left( 1 + \frac{13}{12} \; \eta_0^2\right) \right] \; .
\end{equation}
where $  \Sigma_{est-env}(\tau)$ is given by eq.(\ref{polenta}).

The effective polytropic index $ \gamma(\tau) $ is:
\begin{equation}
 \gamma(\tau) =  g \Sigma_{est-env}(\tau) \; {{12 + 13  \, \eta_0^2}\over
{3 \eta_0^2 ( \eta_0^2 + 2 )}} \; .
\end{equation}
When $  g \Sigma_{est-env}(\tau_1)
 \sim  1 + \eta_0^2 / 2 $, i.e. at the end of the preheating phase, 
$\gamma(\tau)$ is given by
\begin{equation}
\gamma_{eff}\propto 
{{12 + 13 \, \eta_0^2}\over
{6 \eta_0^2 }} \label{gammaeff}
\end{equation} 
We note here that for very large $ \eta_0 $ the effective polytropic index is
$\gamma_{eff} \simeq 13/6 \sim{\cal{O}}(1)$.  It is clear then that the physics
can be interpreted in terms of two fluids, one the contribution from the zero
mode and the other from the fluctuations, each with an equation of state that
is neither that of dust nor of radiation, but described in terms of an
effective polytropic index.

We can now use our approximations to obtain an estimate for the number of
particles produced during the preheating stage. In terms of dimensionless
quantities, the particle number, defined with respect to the initial Fock
vacuum state is given by eq.(\ref{partnumber}).

This particle number will only obtain a significant contribution from the
unstable modes in the forbidden band where to leading order in $\hat{q}$ we can
approximate $ \varphi_q(\tau) $ and $ \dot{\varphi}_q(\tau) $ by its
exponentially growing pieces [see eq.(\ref{mod2})], as follows:
\begin{eqnarray}
|\varphi_q(\tau)|^2 & \simeq &  
 { 1 \over {4\, \Omega_q}}\left[ 1 + {{4 \, \Omega_q^2} \over
{{\cal{W}}_q^2}} \right] \; \mid  U_q(-\tau) \mid^2 \\
 |\dot{\varphi}_q(\tau)|^2 & \simeq & (1+\eta_0^2)\;
{\rm cotg}^2 \left[ \pi (v - {{\tau}\over{2\omega}}) \right]\;
|\varphi_q(\tau)|^2 + O( {\hat q} ) \; .
\end{eqnarray}

The total number of produced particles $ N(\tau) $ per volume $|M_R|^3$ is
given by:
\begin{equation}
{\cal{N}}= \frac{N(\tau)}{|M_R|^3} \equiv \int {{d^3 q}\over {(2\pi)^3}} \;  N_q(\tau) \; .
\end{equation}
The asymptotic behaviour (\ref{asymptotics}) ensures that this integral
converges. 

Following the same steps as in eq.(\ref{Senv}) and (\ref{estpE}), we find
\begin{equation}
 N(\tau)_{unst} =  {1 \over {8 \pi^2}}\; \Sigma_{est-env}(\tau) \; \left[ {{1 +
\eta_0^2}\over {\Omega_{q_1}}} + \Omega_{q_1} \right] =\frac{1}{\lambda_R}
{{4 + \frac92 \, \eta_0^2}\over { \sqrt{4 + 5 \, \eta_0^2}}} \; 
  \left(g \Sigma_{est-env}(\tau) \right) \; .
\end{equation}
where we used eq.(\ref{quno}) and $ \Sigma_{est-env}(\tau) $ is given by the
simple formula (\ref{polenta}). Notice that by the end of the preheating stage,
when $g\Sigma(\tau) \approx 1 + \eta_0^2/2 $ the total number of particles
produced is non-perturbatively large, both in the amplitude as well as in the
coupling
\begin{equation}
{\cal{N}}_{tot} \approx \frac{1}{\lambda_R} {{(4 + \frac92 \, \eta_0^2)(1 +
\eta_0^2/2)}\over { \sqrt{4 + 5 \, \eta_0^2}}} \label{totalpart}
\end{equation}

The total number of {\em adiabatic} particles can also be computed in a similar
manner with a very similar result insofar as the non-perturbative form in terms
of coupling and initial amplitude.

\subsection{Numerical Results}

We now evolve our equations for the zero and non-zero modes numerically,
including the effects of back-reaction. We will see that up to the preheating
time, our analytic results agree extremely well with the full numerical evolution.

The procedure used was to solve equations (\ref{modo0}, \ref{modok}) with the
initial conditions (\ref{conds1}, \ref{conds2},\ref{initialfreqs2}) and
(\ref{sigmafin}) using a fourth order Runge-Kutta algorithm for the
differential equation and an 11-point Newton-Cotes integrator to compute the
fluctuation integrals. We tested the cutoff sensitivity by running our code for
cutoffs $\Lambda/|M_R|= 100,70,50,20$ and for very small couplings (which is
the case of interest. We found no appreciable cutoff dependence. The typical
numerical error both in the differential equations and the integrals are less
than one part in $10^9$.

Figure 2.a shows $\eta(\tau)$ vs.$\tau$ for $\eta_0=4.0 \; , g=10^{-12}$.  For
this weak coupling, the effect of back-reaction is negligible for a long time,
allowing several undamped oscillations of the zero mode. Figure 2.b shows
$g\Sigma(\tau)$ vs. $\tau$. It can be seen that the back-reaction becomes
important when $g\Sigma(\tau) \approx 1 + \eta_0^2/2 $ as the evolution of
$\eta(\tau)$ begins to damp out.  This happens for $\tau \approx 25 $ in
excellent agreement with the analytic prediction given by
eq.(\ref{maxtime}) $ \tau_1 = 26.2\ldots $,
the difference between the analytic estimate for $\Sigma(\tau)$ given by
eq.(\ref{appxfluc}) and the numerical result is less than $5 \%$ in the range
$0 < \tau < 30$.  Figure 2.c shows $g{\cal{N}}(\tau)$ vs. $\tau$ and we see
that the analytic expression (\ref{totalpart}) gives an approximate
estimation 
$ \lambda_R {\cal{N}}_{tot} \approx 74.6 \ldots $ for the final number
of produced particles.

Figures 2.d-2.f, show $gN_q(\tau)$ for $\tau = 40,120,200$, we see that the
prediction of the width of the unstable band $0 < q < \eta_0 / \sqrt{2}$ is
excellent and is valid even for very long times beyond the regime of validity
of the small time, weak coupling approximation. However, we see that the peak
becomes higher, narrower and moves towards $q \approx 0.5$ as time
evolves beyond $ \tau_1 $. This feature persists in all numerical
studies of the unbroken phase 
that we have carried out, this changes in the peak width, height and position
are clearly a result of
back-reaction effects.  We have searched for unstable bands for $0<q<20$ and we
only found one band precisely in the region predicted by the analytic estimate.
All throughout the evolution {\em there is only one unstable band}. The band
develops some structure with the height, position and width of the peak varying
at long times but no other unstable bands develop and the width of the band
remains constant.  For values of $q$ outside the unstable band we find
typically $gN_q < 10^{-13}$ at all times. This is a remarkable and unexpected
feature.

Obviously this is very different from the band structure of a Mathieu
equation. The Mathieu equation gives rise to an infinite number of narrowing
bands, so that quantitative estimates of particle production, etc. using the
Mathieu equation approximation would be gross misrepresentations of the actual
dynamics, with discrepancies that are non-perturbatively large when the
back-reaction becomes important \cite{paris}. 
Since particle production essentially happens in the forbidden bands,
the quantitative predictions obtained from  a single forbbiden band
and an infinite number, as predicted by WKB or Mathieu equation
analysis, will yield  different physics. 

We have carried the numerical evolution including only the wave-vectors in the
unstable region and we find that this region of $q$-wavevectors is the most
relevant for the numerics. Even using a cutoff as low as $q_{c} = 4$ in this
case gives results that are numerically indistinguishable from those obtained with
much larger cutoffs $q_{c} =70-100$. The occupation number of modes outside the
unstable bands very quickly becomes negligible small and for $q \approx 4$ it
is already of the same order of magnitude as the numerical error $ \leq
10^{-10}$.  Clearly this is a feature of the weak coupling case under
consideration.  Keeping only the contribution of the modes in the unstable
band, the energy and pressure can be written as
\begin{eqnarray}
\varepsilon & = & \frac{2 |M_R|^4}{\lambda_R}\left\{
\frac{\dot{\eta}^2}{2}+\frac{\eta^2}{2}+\frac{\eta^4}{4}+ 2g
\int_0^{q_{c}}q^2dq \Omega_q
N_q(\tau)+\frac{g}{2}\Sigma(\tau)\left[\eta^2(\tau)-\eta^2_0+\frac{g}{2}\Sigma(\tau)\right]
+ \right.  \nonumber \\ & & \left.  {\cal{O}}(g) \right\} \label{enesu} \\ p &
= & \frac{2 |M_R|^4}{\lambda_R}\left\{g \int_0^{q_{c}}q^2 dq \left[
\frac{q^2}{3} |\varphi_q(\tau)|^2 +|\dot{\varphi}(\tau)|^2\right] +
\dot{\eta}^2 + {\cal{O}}(g) \right\} -\varepsilon
\label{presu}  \\
q_{c} & = & \frac{\eta_0}{\sqrt{2}} \label{qmaxi}
\end{eqnarray}
where we have made explicit that we have neglected terms of order $g$
in eqs.(\ref{enesu}-\ref{presu}).  The terms multiplied by $g$ in
eqs.(\ref{enesu}, 
\ref{presu}) become of order 1 during the preheating stage.  For the parameters
used in figures 2, we have checked numerically that the energy (\ref{enesu}) is
conserved to order g within our numerical error. Figure 2.g shows the pressure
$ {2 |M_R|^4\; p(\tau)}/{\lambda_R} $ versus $ \tau $. Initially $ p(0) =
-\varepsilon $ (vacuum dominated) but at the end of preheating the equation of
state becomes almost that of radiation $ p_{\infty} = \varepsilon /3 $.

For very small coupling ($g \sim 10^{-12} $), the backreaction shuts-off 
suddenly the particle production at the end of the preheating (see
fig. 2c). Later on, ($ \tau $ larger than $ 100 $ for  $g \sim
10^{-12} $) the time evolution is periodic in a very good
approximation. That is, this non-linear system exhibits a limiting
cycle behaviour. The modulus of the $k$-modes do not grow in time and
no particle production takes place. This tells us that no forbidden
bands are present for $ q^2 > 0 $ in the late time regime.

We have numerically studied several different values of $\eta_0 \; ,
g$ finding  
the same qualitative behavior for the evolution of the zero mode, particle
production and pressure. In all cases we have found remarkable agreement (at
most $5 \%$ difference) with the analytical predictions in the time regime for
which $0 < g \Sigma(\tau) \leq 1$. The asymptotic value of the pressure,
however, only becomes consistent with a radiation dominated case for large
initial amplitudes. For smaller amplitudes $\eta_0 =1$ we find that
asymptotically the polytropic index is smaller than $4/3$. This asymptotic
behavior is beyond the regime of validity of the approximations in the analytic
treatment and must be studied numerically.  This polytropic index depends
crucially on the band structure because most of the contribution comes from the
unstable modes. 

\section{The Broken Symmetry Case}

\subsection{Analytic Results}

As in the unbroken case, for $ g << 1 $ we can neglect $ g \Sigma(\tau) $ in
eq. (\ref{modo0R}) until a time $\tau_2$ at which point the fluctuations have
grown to be comparable to the `tree level' terms.

We then find for $ 0 \leq \eta_0 \leq 1 $,
\begin{eqnarray}\label{etacR}
\eta(\tau) &=&  {{  \eta_0}\over 
{\mbox{dn}\left(\tau\sqrt{1-{{\eta_0^2}\over 2}},k\right)}} \cr \cr \cr
k &=& \sqrt{{1-\eta_0^2}\over{ 1 -  {{\eta_0^2}\over 2}}}\; , 
\end{eqnarray}
Notice that $ \eta(\tau) $ has period $ 2 \omega \equiv {{ 2 \, K(k)}\over
{\sqrt{1-{{\eta_0^2}\over 2}}}} $. The elliptic modulus $ k $ is given by
eq.(\ref{etacR}).

For $ 1 \leq \eta_0 \leq \sqrt2 $  we find
\begin{eqnarray}\label{etac2}
\eta(\tau) &=& \eta_0\; \mbox{dn}\left(\tau\eta_0/\sqrt{2},k\right)
\cr \cr
k &=& \sqrt{2(1 - {\eta_0}^{-2})} \; .
\end{eqnarray}
This solution follows by shifting eq.(\ref{etacR}) by a half-period and
changing $ \eta_0^2 \to 2 - \eta_0^2 $. It has a period $ 2 \omega \equiv
{{2\sqrt2 }\over {\eta_0}} \, K(k) $. For $ \eta_0 \to 1 $, $ 2 \omega \to
\pi\sqrt2 $ and the oscillation amplitude vanishes, since $ \eta = 1 $ is a
minimum of the classical potential.

For $ \eta_0 > \sqrt2 $ we obtain
\begin{eqnarray}\label{etac3}
\eta(\tau) &=& \eta_0\; \mbox{cn}\left(\sqrt{ \eta_0^2 - 1}\tau,k\right)
\cr \cr
k &=& {{ \eta_0}\over {\sqrt{2({\eta_0}^2-1)}}} \; . 
\end{eqnarray}
This solution has $ 4 \omega \equiv {{ 4 \, K(k)}\over {\sqrt{\eta_0^2 - 1}}} $
as period.

The solutions for $ \eta_0 < \sqrt2 $ and $ \eta_0 > \sqrt2 $ are qualitatively
different since in the second case $ \eta(\tau) $ oscillates over the two
minima $ \eta = \pm 1 $. In the limiting case $ \eta_0 = \sqrt2 $ these
solutions degenerate into the instanton solution
\begin{equation}
\eta(\tau) = {{\sqrt2}\over{\cosh\tau}} \; , 
\end{equation}
and the period becomes infinite.

Inserting this form for $\eta(\tau)$ in eq.(\ref{modokR}) and neglecting $ g
\Sigma(\tau) $ yields for $ 0 \leq \eta_0 \leq 1 $,
\begin{equation}\label{modsnR}
 \left[\;\frac{d^2}{d\tau^2}+q^2-1+ {{\eta_0^2} \over 
{\mbox{dn}^2\left(\tau\sqrt{1-{{\eta_0^2}\over 2}},k\right)}}
\;\right] \varphi_q(\tau) =0.
\end{equation}
This is again a Lam\'e equation for a one-zone potential and can also
be solved in 
closed form in terms of Jacobi functions. We summarize here the results for the
mode functions, with the derivations again given in the Appendix.

As for unbroken symmetry case, there are {\em two} allowed bands and {\em two}
forbidden bands The allowed bands for $ 0 \leq \eta_0 \leq 1 $ correspond to
\begin{equation}\label{rbandaq}
 0 \leq q^2 \leq  {{\eta_0^2}\over 2}
 \quad \mbox{and} \quad
 1-{{\eta_0^2}\over 2} \leq q^2 \leq +\infty \; ,
\end{equation}
and the forbidden bands to
\begin{equation}\label{rbandap}
-\infty \leq q^2 \leq  0
 \quad \mbox{and} \quad
 {{\eta_0^2}\over 2}  \leq q^2 \leq  1-{{\eta_0^2}\over 2} \, .
\end{equation}
The last forbidden band exists for positive $q^2$ and hence 
contributes to the growth of $\Sigma(\tau)$.

The Floquet solutions obey eqs. (\ref{floq}) and (\ref {ucero}) and the modes
$\varphi_q(\tau)$ can be expressed in terms of $U_q(\tau) $ and $ U_q(-\tau) $
following eq.(\ref{combi2}).

It is useful to write the solution $ U_q(\tau) $ in terms of Jacobi
$\vartheta$-functions. For the forbidden band $ {{\eta_0^2}\over 2}
\leq q^2 \leq 1-{{\eta_0^2}\over 2}$ after some calculation (see the Appendix),
\begin{equation}\label{urotaJ}
 U_q(\tau) = e^{- {\tau\;\sqrt{1-{{\eta_0^2}\over 2} }\; Z(2  K(k)\,v)}} \;
{{\vartheta_3(0) \; \vartheta_2(v+ {{\tau}\over {2\omega}} )}\over
{\vartheta_2(v)\; \vartheta_3( {{\tau}\over {2\omega}} )}}\; ,
\end{equation}
where $ 0 \leq v \leq \frac12 $ is related with $q$ through
\begin{equation}
 q^2 =    1-{{\eta_0^2}\over 2}
-( 1 - \eta_0^2 )\;
\mbox{sn}^2\left( 2\, K(k)\, v ,k\right)\; ,
\end{equation}
and $ k $ is a function of $ \eta_0 $ as defined by eq.(\ref{etacR}).

We see explicitly here that $ U_q(\tau) $ factorizes into a real exponential
with an exponent linear in $\tau$ and an antiperiodic function of $ \tau $ with
period $2 \omega$.

The Floquet indices for this forbidden band are given by
\begin{equation}
F(q) = 2 \,i \,  K(k)\; Z(2  K(k)\,v) \pm \pi\; .
\end{equation}

For the allowed band $ 1- {{\eta_0^2}\over 2} \leq q^2 \leq +\infty \; ,$ we
find for the modes,
\begin{equation}\label{urotaP1}
 U_q(\tau) = e^{ -{{\tau}\over {2\omega}}\;
{{\vartheta_1'}\over{\vartheta_1}}({{i\alpha}\over {2\omega}})} \; 
{{\vartheta_3(0) \; \vartheta_3({{i\alpha+ \tau}\over {2\omega}} )}\over
{\vartheta_3({{i\alpha}\over {2\omega}})\; \vartheta_3( {{\tau}\over
{2\omega}} )}}\; ,
\end{equation}
where $ q $ and $ \alpha $ are related by:
$$
 q = {{\sqrt{1 -{{\eta_0^2}\over 2}}}\over{ \mbox{sn}
\left( \alpha\sqrt{1 -{{\eta_0^2}\over 2}},k'\right)}} \; 
$$
with $ {{K'(k)}\over {\sqrt{1 -\eta_0^2}}} \geq \alpha \geq 0 $.
The Floquet indices for this first allowed band are given by
\begin{equation}
F(q) = i \, {{\vartheta_1'}\over{\vartheta_1}}\left({{i\alpha}\over
{2\omega}}\right) \; . 
\end{equation}

Analogous expressions hold in the other allowed band, $ 0 \leq q^2 \leq
{{\eta_0^2}\over 2} $:
\begin{equation}\label{urotaP2}
 U_q(\tau) = e^{ -{{\tau}\over {2\omega}}\;
{{\vartheta_2'}\over{\vartheta_2}}({{i\alpha}\over {2\omega}})} \; 
{{\vartheta_3(0) \; \vartheta_4({{i\alpha+ \tau}\over {2\omega}} )}\over
{\vartheta_4({{i\alpha}\over {2\omega}})\; \vartheta_3( {{\tau}\over
{2\omega}} )}}\; .
\end{equation}
Here, $ q   = {{\eta_0}\over \sqrt2}\;
\mbox{sn}\left( \alpha\sqrt{1 -{{\eta_0^2}\over 2}},k'\right) $
and  $ {{K'(k)}\over {\sqrt{1 -\eta_0^2}}}  \geq \alpha  \geq 0 $, and 
the Floquet indices for this band are given by
\begin{equation}
F(q) = i \, {{\vartheta_2'}\over{\vartheta_2}}\left({{i\alpha}\over
{2\omega}}\right) \; . 
\end{equation}

For $\eta_0 \approx 1$ the situation is very similar to the unbroken symmetry
case; the zero mode oscillates quasi-periodically around the minimum of the
tree level potential. There are effects from the curvature of the potential,
but the dynamics can be analyzed in the same manner as in the unbroken case,
with similar conclusions and will not be repeated here. 

The case $\eta_0 << 1$ is especially interesting\cite{dis} for broken symmetry
because of new and interesting phenomena\cite{dis} that has been recently
associated with symmetry
restoration\cite{lindekov,tkachev,kolbriotto,kolblinde}.

In this limit, the elliptic modulus $k$ [see eq. (\ref{etacR})] approaches
unity and the (real) period $2\omega$ grows as
\begin{equation}
2\, \omega\simeq 2 \, K(k) +O(\eta_0^2) \simeq
2\, \ln\left({{\sqrt{32}}\over{\eta_0}}\right) +O(\eta_0^2)
\end{equation}
In this limit, both the potential in eq.(\ref{modsnR}) and the mode
functions (\ref{urotaJ}) can be approximated by hyperbolic functions\cite{erd}:
\begin{eqnarray}
&& {1 \over{{\mbox{dn}}(\tau\sqrt{1 -{{\eta_0^2}\over 2}},k)}}= 
\cosh\tau +O(\eta_0^2) \\ 
&& Z(u) = \tanh u -{u \over {\Lambda}} +O(\eta_0^2)
\end{eqnarray}
where
\begin{eqnarray}\label{varu}
\cosh u &=& { 4 \over {\eta_0^2}}\left(q - \sqrt{q^2 - {{\eta_0^2}\over
2}}\right) \left[1 +O(\eta_0^2)\right] \; ,\cr \cr
\Lambda &\equiv&  \log\left({{\sqrt{32}}\over{\eta_0}}\right) 
\quad , \quad  0 \leq u \leq  \Lambda
\end{eqnarray}

Using the imaginary Jacobi transformation \cite{erd},
\begin{equation}
\vartheta_{2,3}(v|{\hat q}) = \sqrt{{K(k)}\over{K(k')}}\; e^{- {{\pi
K(k)}\over{K(k')}} \; v^2}\; \vartheta_{4,3}(-i {{K(k)}\over{K(k')}} v|{\dot
q})  \; , 
\end{equation}
where $ {\hat q} = e^{- {{\pi K(k')}\over{K(k)}}} \; , \; {\dot q} = e^{- {{\pi
K(k)}\over{K(k')}}} $ and the series expansions\cite{erd}
\begin{eqnarray}
&&\vartheta_{3}(v|{\hat q}) = 1 + 2 \sum_{n=1}^{\infty} \;  {\hat
q}^{n^2}\; \cos(2\pi n v)  \\
&& \vartheta_{4}(v|{\hat q}) =\vartheta_{3}(v+ \frac12\; |{\hat q}) \; ,
\end{eqnarray}
we can derive expressions for the mode functions $ U_q(\tau) $ valid for small
$\eta_0$:
\begin{equation}
U_q(\tau)= e^{-\tau \; \tanh u\left(1 +  {{\eta_0^2}\over8}\right) }\; 
{{1 -  {{\eta_0^2}\over 8}\; \cosh u \; \cosh( u + 2 \tau)}\over
{1  -  {{\eta_0^2}\over8}\; \cosh u}} \left[ 1+O(\eta_0^2)\right] \label{Uq}\; .
\end{equation}
Here $u$ is related with $q$ through eq.(\ref{varu}).  We see that the function
$ U_q(-\tau) $ grows with $\tau$ almost as $e^{\tau}$ for $q$ near the lower
border of the forbidden band $u \simeq \Lambda$. This fast growth can be
interpreted as the joint effect of the non-periodic exponential factor in
eq.(\ref{urotaJ}) and the growth of the periodic $\vartheta$-functions. Since
the real period is here of the order $ \Lambda$, the two effects cannot be
separated. The unstable growth for $\tau \leq \omega$ also reflects the spinodal
instabilities associated with phase separation\cite{boyveg}.

In this case, there is a range of parameters for which the quantum fluctuations
grow to become comparable to the tree level contribution within just one or
very few periods. The expression (\ref{Uq}) determines that $\Sigma(\tau)
\approx e^{2\tau}$ from the contributions of modes near the lower edge of the
band. The condition for the quantum fluctuations to become of order 1 within
just one period of the elliptic function is $ge^{4\omega} \approx 1 $ which
leads to the conclusion that for $\eta(0) < g^{1/4}$ the quantum fluctuations
grow very large before the zero mode can actually execute a single
oscillation. In such a situation an analysis in terms of Floquet
(quasi-periodic) solutions is not correct because the back reaction
prevents the 
zero mode from oscillating enough times for periodicity to be a reasonable
approximation. 

We now analyze the behavior of the pressure for the zero mode to compare to the
previous case.  In the approximation where eqs.(\ref{etacR}), (\ref{etac2})
and (\ref{etac3}) hold and adjusting the constant ${\cal{C}}$ in the definition
of the energy, we have
\begin{eqnarray}\label{e0p0R}
\epsilon_0 &=& \frac14  \left(\eta_0^2 - 1 \right)^2  \; , \cr  \cr
p_0(\tau) &=&   -\epsilon_0 +{\dot \eta}(\tau)^2
\end{eqnarray}

Inserting eqs.(\ref{etacR}), (\ref{etac2}) and (\ref{etac3}) in
eq.(\ref{e0p0R}) yields
\begin{equation}\label{presR}
\begin{array}{lrl}
0 \leq  \eta_0 \leq 1 ~:
&\quad p_0(\tau) &= -
\epsilon_0\left[ 1 - 8 \; 
\mbox{sn}^2 \mbox{cn}^2\left((\tau+K)\sqrt{1-{{\eta_0^2}\over 2}}
, k\right) \right]  \; ,\cr \cr    
1 \leq  \eta_0 \leq \sqrt2  ~: &\quad p_0(\tau) &= -
\epsilon_0\left[ 1 - 8 \; 
\mbox{sn}^2 \mbox{cn}^2\left(\tau\eta_0/\sqrt{2} , k\right) \right] \;
, \cr \cr 
\eta_0\geq \sqrt2 ~: &\quad p_0(\tau) &= -
\epsilon_0\left[ 1 - 8 \; k^2 \;
\mbox{sn}^2 \mbox{dn}^2\left(\tau\sqrt{ \eta_0^2 - 1} , k\right) \right] \; .
\end{array} 
\end{equation}
Notice that the functional form of the elliptic modulus $ k $ as a
function of $ \eta_0 $ is different in each interval [see
eqs.(\ref{etacR}), (\ref{etac2}) and  (\ref{etac3})].

Let us now average the pressure over a period as in eq.(\ref{promp}).
We find
\begin{eqnarray}\label{prompB}
<p_0> &=& \epsilon_0 \left\{ \frac83\left[ { {k^2 -2} \over {k^4}}\;\left( 
 1 - {{E(k)}\over {K(k)}}\right) + {1 \over {k^2}}\right] -1 \right\}
\cr \cr
& & {\rm for~} 0 \leq  \eta_0 \leq \sqrt2 \; ,\cr \cr
<p_0> &=& \epsilon_0 \left\{ \frac83\left[  (1 - 2 k^2 )\;\left( 
 1 - {{E(k)}\over {K(k)}}\right) + k^2 \right] -1 \right\}
\cr \cr
& & {\rm for~}   \eta_0 \geq \sqrt2 \; .
\end{eqnarray}

The dimensionless energy $ \epsilon_0 $ tends to $1/4$ both as $\eta_0
\rightarrow 0$ and $\eta_0 \rightarrow \sqrt{2}$ in both cases we find using
eq.(\ref{prompB})
\begin{equation}
{{<p_0>}\over{ \epsilon_0}} \rightarrow
 -1 - { 16 \over {3 \; \log|\frac{1}{4}-\epsilon_0|}} + O(\epsilon_0-\frac{1}{4})\; .
\end{equation}
This is result is recognized as vacuum behaviour in this limit.

For $ \eta_0 \to 1 $, eq. (\ref{prompB}) yields,
\begin{equation}
{{<p_0>}\over{ \epsilon_0}}\buildrel{  \eta_0 \to 1}\over=
O(  \eta_0 - 1 )^2\; .
\end{equation}
That is a dust type behaviour, which is consistent with the small amplitude
limit of the unbroken symmetry case studied before.

Finally, for $ \eta_0 \to \infty $, when the zero mode is released from high up
the potential hill, we find that the pressure approaches radiation behaviour
(from above)
\begin{eqnarray}
{{<p_0>}\over{ \epsilon_0}}&\buildrel{  \eta_0 \to  \infty}\over=&
\frac13 +\frac43\; \left[ {1 \over {\sqrt2}} - 1 + \left(2 -  
{1 \over {\sqrt2}} \right) {{E({1 \over {\sqrt2}})}\over 
{K({1 \over {\sqrt2}})}} \right] \; {1 \over { \eta_0^2}} + 
O( {1 \over { \eta_0^4}})  \cr \cr \cr
&=& \frac13 + {{0.86526\ldots}\over  {\eta_0^2}} 
+ O( {1 \over { \eta_0^4}}) \; .
\end{eqnarray}
Figure 3 shows
 $<p_0>/\varepsilon_0$ vs. $\varepsilon_0$.

As mentioned before, we expect that for $\eta_0 <<1$ the conclusion will be
modified dramatically by the quantum corrections.

\subsection{Numerical Results}

In the region $\eta_0 \approx 1$ the analytic estimates are a good
approximation for large times and weak couplings. We have studied numerically
many different cases with $\eta_0 \geq 0.5$ and weak coupling and confirmed the
validity of the analytic estimates. These cases are qualitatively similar to
the unbroken symmetry case with almost undamped oscillations for a long time
compatible with the weak coupling approximation and when $g\Sigma(\tau)$ grows
by parametric amplification to be of order one with a consequently large number
of produced particles and the evolution of the zero mode damps out.

However as argued above, for $\eta_0 << 1$ the analytic approximation will not
be very reliable because the quantum fluctuations grow on a time scale of a
period or so (depending on the coupling) and the back-reaction term cannot be
neglected. Thus this region needs to be studied numerically.

We numerically solved equations (\ref{modo0R})-(\ref{modokR}) with the initial
conditions (\ref{sigmafin}), (\ref{conds12}),
(\ref{unsfrequ}) and (\ref{stafrequ}). The numerical routines are the
same as in the unbroken 
symmetry case. Again we tested cutoffs $\Lambda/|M_R|= 100,70,50,20$ and for
very small couplings (which is the case of interest, $g=10^{-6} \cdots g=
10^{-12}$) we found no appreciable cutoff dependence, with results that are
numerically indistinguishable even for cutoffs as small as $q_c \approx 2$
. The typical numerical error both in the differential equations and the
integrals are the same as in the unbroken case, less than one part in $10^9$.

We begin the numerical study by considering first the case of very small
coupling and $\eta_0 <<1$; later we will deal with the case of larger couplings
and initial values of the zero mode.  Figure 4.a shows $\eta(\tau)$ vs. $\tau$
for $\eta_0=10^{-5} , \; g=10^{-12}$. In this case we see that within one
period of the classical evolution of the zero mode, $g\Sigma(\tau)$ becomes of
order one, the quantum fluctuations become non-perturbatively large and the
approximation valid for early times and weak couplings breaks down. Fig. 4.b
shows $g\Sigma(\tau)$ and fig. 4.c shows $ g{\cal{N}}(\tau) $ vs. $
\tau $ for
these parameters. We find that only the wave vectors in the region $0<q<1$ are
important i.e. there is only one unstable band whose width remains constant in
time. This is seen in figs. 4.d-f, which show the particle number (defined with
respect to the initial state) as a function of wave vector for different times,
$ g N_q(\tau=30) \; , g N_q(\tau=90) \; , g N_q(\tau=150) $
respectively. Although the analytic approximation breaks down, the prediction
eq.(\ref{rbandap}) for the band width agrees remarkably well with the numerical
result. As in the unbroken case, the band develops structure but its width is
constant throughout the evolution. As can be seen in these figures the peak of
the distribution becomes higher, narrower and moves towards smaller values of
$q$. The concentration of particles at very low momentum is a consequence of
the excitations being effectively massless in the broken symmetry case. The
features are very distinct from the unbroken symmetry case, in which the peak
approaches $q \approx 0.5$.

We found in all cases that the asymptotic behavior corresponds to
\begin{equation}
{\cal{M}}^2(\tau) = -1+\eta^2(\tau)+g\Sigma(\tau)
 \stackrel{\lim \tau \rightarrow \infty}{\rightarrow} 0 \label{asymass}
\end{equation}
This is a consistent asymptotic solution that describes massless ``pions'' and broken
symmetry in the case $\eta(\infty) \neq 0$. 

For times $\tau \approx 100-150$ the value of the zero mode is somewhat larger
than the initial value: $\eta(\tau=150) \approx 2\times 10^{-5}$. This result,
when combined with the result that the average of the effective mass approaches
zero is clearly an indication that the symmetry is {\em broken}. We found
numerically that the final value of the zero mode depends on the initial value
and the coupling and we will provide numerical evidence for this behavior
below.

Figure 4.a, presents a puzzle. Since the zero mode begins very close to the
origin with zero derivative and {\em ends up} very close to the origin with
zero derivative, the classical energy of the zero mode is conserved. At the
same time, however, the dynamical evolution results in copious particle
production as can be seen from fig. 4.c. We have shown in a previous section
that the total energy is conserved and this was numerically checked within the
numerical error. Thus the puzzle: how is it possible to conserve the {\em
total} energy, conserve the classical zero mode energy and at the same time
create ${\cal{O}}(1/g)$ particles?  The answer is that there is a new term in
the total energy that acts as a `zero point energy' that
diminishes during the evolution and thus maintains total energy conservation
with particle production. The most important contribution to the energy arises
from the zero mode and the unstable modes $0<q<q_u$. The energy and
pressure are 
given by (adjusting the constant ${\cal{C}}$ such that the energy coincides
with the classical value):
\begin{eqnarray}
\varepsilon & = & \frac{2 |M_R|^4}{\lambda_R}\left\{ \varepsilon_{cl}+
 \varepsilon_{N}+\varepsilon_C +{\cal{O}}(g) \right\} \label{enebrok}\\
 \varepsilon_{cl} & = & \frac{\dot{\eta}^2}{2}+\frac{1}{4}\left(\eta^2
 -1\right)^2 \label{enebrokclas}\\ 
\varepsilon_N & = & 2g \int_0^{q_u}q^2dq\;
 \Omega_q \; N_q(\tau) \label{enebroknum} \\ 
\varepsilon_C & = &
 \frac{g}{2}\Sigma(\tau)\left[-1-\eta^2_0+{\cal{M}}^2(\tau)-\frac{g}{2}\Sigma(\tau)\right]
 \label{enebrokcons} \\ p & = & \frac{2 |M_R|^4}{\lambda_R}\left\{g
 \int_0^{q_u}q^2 dq \left[ \frac{q^2}{3} |\varphi_q(\tau)|^2
 +|\dot{\varphi}(\tau)|^2\right] + \dot{\eta}^2 + {\cal{O}}(g) \right\}
 -\varepsilon
\label{presubrok} \\
{\cal{M}}^2(\tau) & = & -1+\eta^2(\tau)+g\Sigma(\tau) \label{massbrok}
\end{eqnarray} 
where $ {\cal{M}}(\tau)^2 $ is the effective squared mass of the
$N-1$ ``pions'' 
and again ${\cal{O}}(g)$ stand for perturbatively small terms of order $g$. The
terms displayed in (\ref{enebrokclas}- \ref{presubrok}) are all of
${\cal{O}}(1)$ during the preheating stage. 

We find that whereas $\varepsilon_N$ grows with time, the term $\varepsilon_C$
becomes negative and decreases. In all the cases that we studied, the effective
mass ${\cal{M}}(\tau)$ approaches zero asymptotically; this is
seen in figure 
4.g for the same values of the parameters as in figs. 4.a-c.  This behavior and
an asymptotic value $\eta_{\infty} \neq 0$ is consistent with broken symmetry
and massless pions by Goldstone's theorem.  The term $ \varepsilon_C $ in
eq.(\ref{enebrokcons})  can
be identified with the `zero' of energy. It contributes to the equation of
state as a vacuum contribution, that is $p_C = - \varepsilon_C$ and becomes
negative in the broken symmetry state. It is this term that compensates for the
contribution to the energy from particle production. 

This situation is generic for the cases of interest for which $\eta_0 <<1$,
such is the case for the slow roll scenario in inflationary cosmology.
Figs. 4.h-k show $\varepsilon_{cl}(\tau)$ vs. $\tau$, $\varepsilon_N(\tau)$
vs. $\tau$, $\varepsilon_C(\tau)$ vs. $\tau$ and $ \frac{\lambda_R}{2
|M_R|^4}p(\tau)$ vs $\tau$ for the same values of parameters as fig. 4.a. The
pressure has a remarkable behavior. It begins with $p = -\varepsilon$
corresponding to vacuum domination and ends asymptotically with a
radiation-like equation of state $p= \varepsilon /3$.  A simple explanation for
radiation-like behavior would be that the equation of state is dominated by the
quantum fluctuations which as argued above correspond to massless pions and
therefore ultrarelativistic. 
It must me noticed that we obtain a radiation-like equation of state
in spite of the fact that the 
the distribution is out of equilibrium and far from thermal as can be seen from
figs. 4.d-f.

An important question to address at this point is: why does the zero mode reach
an asymptotic value {\em different} from the minimum of the effective
potential? The answer to this question is that the effective potential is an
irrelevant quantity to study the dynamics\cite{dis,boyveg}. Once there is
profuse particle production, a feature completely missed by the effective
potential, the zero mode evolves in a non-equilibrium bath of these
excitations. Through the time evolution, more of these particles are produced
and the zero mode evolves in a highly excited, non-equilibrium
state. Furthermore we have seen in detail that this mechanism of particle
production modifies dramatically the zero point origin of energy and
therefore the minimum of the effective action, which is now the appropriate
concept to use. The final value reached by the zero mode in the evolution will
be determined by all of these {\em non-perturbative} processes, and only a full
numerical study will capture the relevant aspects. As we have argued above,
approximations based on Mathieu-type equations or the WKB approximation are
bound to miss important details and will lead to a completely different
evolution.

\section{Symmetry Restoration at Preheating?}

The numerical result depicted by figures 4.a-c, which are very similar to
results obtained previously\cite{dis}, has motivated the suggestion that the
growth of quantum fluctuations is so strong that the non-equilibrium
fluctuations restore the symmetry\cite{lindekov,tkachev}. The argument is that
the non-equilibrium fluctuations given by the term 
$\lambda \langle\psi^2(t)\rangle$ in
eq. \ref{pionmodes} for the mode functions grow exponentially and eventually
this term overcomes the term $-|m^2|$ leading to an effective potential with a
{\em positive} mass squared for the zero mode (see eq. \ref{zeromodeeqn}).

Although this is a very interesting suggestion, it is {\em not} borne out by
our numerical investigation.  The signal for broken or restored symmetry is the
final value of the zero mode when the system reaches an equilibrium
situation. Any argument about symmetry restoration based solely on the dynamics
of the fluctuation term $\lambda \langle \psi^2(t)\rangle$ is incomplete if it does not
address the dynamics of the zero mode. In particular for the case of figures
4.a-c, the initial value of the zero mode $\eta_0 \neq 0$ and the final value
is very close to the initial value but still {\em different from zero}. 

At the same time, the asymptotic effective mass of the `pions' is on average
zero. Clearly this is a signal for symmetry breaking. Because the initial and
final values of the order parameter are so small on the scale depicted in the
figures, one is tempted to conclude that the symmetry originally broken by a
very small value of the order parameter is restored asymptotically by the
growth of non-equilibrium fluctuations. To settle this issue we show a
different set of parameters in figures 5.a,b that unambiguously show that the
final value of the order parameter $\eta_{\infty} \neq 0$, while the
effective mass of the pions ${\cal{M}}(\tau) \rightarrow 0 $. Here $\eta_0=0.01$ and
$g=10^{-5}$, and asymptotically we find $\eta(\tau=150) \approx 0.06$, the
average of the effective mass squared ${\cal{M}}^2(\tau) = 0$ and the symmetry is
broken, despite the fact that the fluctuations have grown exponentially and a
number of particles ${\cal{O}}(1/g)$ has been produced.  

The reason that the symmetry is {\bf  not} restored is that when the effective
mass becomes positive, the instabilities shut off and the quantum fluctuations
become small. When this happens $g\Sigma$ is no longer of order one and the
instabilities appear again, producing the oscillatory behavior that is seen in
the figures for $g\Sigma(\tau)$ at long times, such that the contributions of
the oscillatory terms average to zero. It is rather straightforward to see that
there is a self-consistent solution of the equations of motion for the zero
mode and the fluctuations with constant $ \eta_{\infty} $ and
$ {\cal{M}}^2(\infty) = 0 $. Eq.(\ref{modo0R}) takes the asymptotic
form \cite{dis}
$$
 \eta_{\infty} \left[ -1 +  \eta_{\infty}^2 + g  \;\Sigma(\infty)
\right] = 0 \; .
$$
In addition, eq.(\ref{modokR}) yields when $ {\cal{M}}(\infty)^2 = 0
$,
$$
 \varphi_q(\tau) \buildrel{\tau \to \infty}\over = A_q \; e^{-iq\tau} +
B_q  \; e^{iq\tau} \; ,
$$
where $  A_q $ and $ B_q $ depend on the initial conditions and $ g
$. We get from eqs.(\ref{sigmafin}) and (\ref{partnumber}),
\begin{equation}
 \eta_{\infty}^2 = 1 - 4g \int_0^{\infty} {{q^2 \; \Omega_q}\over {q^2 +
\Omega_q^2 }} N_q(\infty) \; dq - g \; S(\eta_0) \; \label{asyneta},
\end{equation} 
where $  S(\eta_0) \equiv \frac14 (1 - \eta_0^2 ) \left[ \log{{1 -
\eta_0^2} \over 4 } - \frac{\pi}2 \right] + \frac12  (1 + \eta_0^2 )
\left[ {\rm ArgTh}\sqrt{{1 -\eta_0^2} \over 2} - {\rm arctg}\sqrt{{1 -\eta_0^2}
\over 2} \right]$.

We see that the value of $\eta_{\infty}$ depends
on the initial conditions. Whereas the last term  in eq. (\ref{asyneta}) is perturbatively
small, the contribution from the produced particles is non-perturbatively large, as
$N_q(\infty) \approx 1/g$ for the unstable wavevectors. Thus the asymptotic
value of the zero mode is drastically modified from the tree level v.e.v (in terms of
renormalized parameters) because of the profuse particle production due to the
non-equilibrium growth of fluctuations. 

Another way to argue that the symmetry is indeed broken in the final state is
to realize that the distribution of ``pions'' at late times will be different
than the distribution of the quanta generated by the fluctuations in the
$\sigma$ field, if for no other reason than that the pions are asymptotically
massless while the $\sigma$ quanta are massive, as long as
$ \eta_\infty $ is
different from zero. If the symmetry were restored during preheating, these
distributions would have to be identical.

The conclusion is that the final value of the zero mode 
 depends strongly on the initial conditions and couplings, and though 
 symmetry restoration can take place for other situations.

A consistent study of the
evolution of the zero mode and quantum fluctuations  determines
what happens in each case \cite{paris}. 
Thus we emphasize that
the ultimate test for symmetry restoration is the asymptotic in time evolution
of the zero mode, which is the order parameter for symmetry breaking.

\section{The Reheating Temperature}

The arena in which these results become important is that of inflationary
cosmology. In particular, the process of preheating is of vital importance in
understanding how the big-bang cosmology is regained at the end of inflation,
i.e. the reheating mechanism. 

While our analysis has been entirely a Minkowski space one, we can make some
comments concerning the reheating temperature. However, a more detailed
analysis incorporating the expanding universe must eventually be done along the
lines suggested in this work, to get more accurate results.

Since the particles created during the preheating stage are far from
equilibrium, thermalization and equilibration will be achieved via collisional
relaxation. In the approximation that we are studying, however, collisions are
absent and the corresponding contributions are of
${\cal{O}}(1/N)$\cite{mottola}. The difficulty with the next order calculation
and incorporation of scattering terms is that these are non-local in time and
very difficult to implement numerically.

However we can obtain an estimate for the reheating temperature under some
reasonable assumptions: in the cosmological scenario, {\em if} the
equilibration time is shorter than the inverse of the expansion rate $H$, then
there will not be appreciable redshifting of the temperature because of the
expansion and we can use our Minkowski space results. 

The second assumption is that the time scales between particle production and
thermalization and equilibration are well separated. Within the large $N$
approximation this is clearly correct because at large $ N $, scattering
processes are suppressed by $ 1/N $. If these two time scales are widely
separated then we can provide a fairly reliable estimate of the reheating
temperature as follows. 

Equilibration occurs via the {\em redistribution} of energy and momentum via
elastic collisional processes. Assuming that thermalization occurs on time
scales larger than that of particle production and parametric amplification,
then we can assume energy conservation in the scattering processes. 
Although a reliable and quantitative estimate of the reheating temperature can only be obtained
after a detailed study of the collisional processes which depend on the interactions, we
can provide estimates in two important cases. If the scattering processes do not change 
chemical equilibrium, that is conserve particle number, the energy per particle is conserved. Since the energy (density) stored in
the non-equilibrium bath and the total number of particles per unit
volume are, as shown in the previous sections:

\begin{eqnarray}
\varepsilon \approx \frac{|M_R|^4}{\lambda_R}  \\
N \approx \frac{|M_R|^3}{\lambda_R}
\end{eqnarray}
with proportionality constants of order one, we can estimate the reheating
temperature to be:
\begin{equation}\label{primero}
\frac{\varepsilon}{N} \approx T \approx |M_R| \label{trehmassive}
\end{equation}
Here $|M_R|$ is the inflaton mass. This is consistent with previous
results\cite{dis}.

This result seems puzzling, since naively one would expect $\varepsilon \approx T^4_R ; 
N \approx T^3_R$ but the powers of $\lambda$ do not match. 
This puzzle arises from intuition  based on an ultrarelativistic free particle gas. 
However the ``medium'' is
highly excited with a large density of particles and the ``in medium'' properties of the
equilibrated particles may drastically modify this naïve result as is known to happen in most
theories at high temperature, where the medium effects are strong and perturbation theory 
breaks down requiring hard thermal loop resummation. 

In the case in which the collisional processes do not conserve particle number and therefore
change chemical equilibrium, the only conserved quantity is the energy. Such is the case for
massless particles interacting with a  quartic
couplings for example or higher order processes in a quartic theory with massive particles. Processes in which $3 \rightarrow 1$ conserving energy
and momentum can occur. The inverse process $1 \rightarrow 3$ occurs with far
less probability since the high momentum modes are much less populated than the
low momentum modes in the unstable bands.  In this case only energy is
conserved whereas the total number of particles is {\em not conserved} and in
this case an estimate of the reheating temperature compares the energy density
in the bath of produced particles to that of an ultrarelativistic gas in
equilibrium at temperature $T_R$,

\begin{equation}
\varepsilon \approx  \frac{ |M_R|^4}{\lambda_R} \approx T^4_R 
\end{equation} 
leading to the estimate
\begin{equation}
T_R \approx \frac{|M_R|}{{\lambda_R}^{\frac{1}{4}}}
\label{trehmassless}
\end{equation}

Thus we can at least provide a bound for the reheating temperature
\begin{equation}
|M_R| \leq T_R \leq \frac{|M_R|}{\lambda} \label{tempbound}
\end{equation}
and a more quantitative estimate requires a deeper understanding of
the collisional processes involved. 

Within the large $ N $ approximation, scattering terms will appear at
order $ 1/N $ and beyond. 
The leading contribution ${\cal {O}}(1/N)$ to collisional relaxation
conserves particle 
number in the unbroken symmetry state because the product particles
are massive. This 
can be seen from the fact that the self-energy to this order is given
by the same chain of 
bubbles that gives the scattering amplitude but with two external legs
contracted. All cut 
diagrams (that give the imaginary part) correspond to $2 \rightarrow
2$ processes that 
conserve particle number because of kinematic reasons. Certainly at
higher order in $ 1/N $  
there will be processes that change chemical equilibrium, but for
large $ N $ these
are suppressed formally.  This argument based on the leading
collisional contribution in the 
$ 1/N $ expansion allows us to
provide a further consistent estimate in the unbroken symmetry case
(when the produced 
particles are massive). In this approximation and consistently with
energy and particle 
number conservation we can assume that the final equilibrium
temperature is of the order of the typical particle energy before
thermalization. Recalling that the unstable band remains stable
during the evolution with the peak shifting slightly in position we can
estimate the typical energy per particle by the position of the peak in the
distribution at $ q_1 $, and use the analytical estimate for the peak given in
section 2. Restoring the units we then obtain the estimate
\begin{equation}
T_R \approx  |M_R| \; q_1 \approx  |M_R| \;  \frac{\eta_0}{2} \approx
\sqrt{\frac{\lambda_R}{8}}\; \Phi_0
\label{trehest}
\end{equation}
which displays the dependence on $\eta_0$
explicitly. Eq. (\ref{trehest}) is an improvement of the simple
estimate (\ref{primero}).

It must be noticed that the peak in the momentum distribution
decreases with time for $ t > t_{reh} $ (see fig.2f-2h). This drift
follows from the non-linear interaction between the modes. For the
case of fig. 2, one sees that $ T_R $ reduces by approximately a
factor $ 3 $ with respect to the value (\ref{trehest}).

The large $ N $ model studied in this article is not a typical model used in
inflationary cosmology, and since we want to make a quantitative statement for
inflationary scenarios (within the approximation of only considering Minkowski
space) we now study a model that incorporates other scalar fields coupled to
the inflaton.

The simplest model \cite{linde,rocky} contains in addition to the inflaton a
lighter scalar field $ \sigma $ with a $ g \sigma \, \Phi^2 $ coupling.  That
is, we consider the Lagrangian \cite{dis},
\begin{equation}
{\cal{L} }= -{1 \over 2} \Phi \,( \partial^2 +m^2+g \, \sigma)\,  
\Phi -{ \lambda \over 4!}  \, \Phi^4 -{1 \over 2} \sigma \,
(\partial^2 + m_{\sigma}^2)  \, \sigma
     -{ \lambda_{\sigma} \over 4!}  \, \sigma^4\; .
\end{equation}

We will consider again the preheating regime of weak couplings and early times
such that we can neglect the back-reaction of the quantum fluctuations of the
$\sigma$ field as well as the back-reaction of the quantum fluctuations of the
inflation itself, focussing only on the parametric growth of the $\sigma$
fluctuations in the unbroken symmetry case.  The mode equations for the $
\sigma $ field take the form

\begin{equation}
\left[\frac{d^2}{dt^2}+k^2+ m^2_{\sigma}+g\, \phi^2(t)\;\right]
V_k(t)=0\;.
\end{equation}
In dimensionless variables this equation becomes
\begin{equation}
\left[\frac{d^2}{d\tau^2}+q^2+ \left({{m_{\sigma}}\over m}\right)^2
+ {{6g}\over {\lambda}}\, \eta^2(\tau)\;\right] V_k(\tau)=0\;.
\end{equation}
In the short time approximation we can replace $  \eta(\tau) $ by the
classical form  (\ref{etac}). We then find a Lam\'e equation which
admits closed form solutions for\cite{ince} 
\begin{equation}
 {{12 \, g}\over {\lambda}} = n(n+1) \quad  ,  \quad n=1,2,3,\ldots
\end{equation}

Although these are not generic values of the couplings, the solubility of the
model and the possibility of analytic solution for these cases makes this study
worthwhile.  In the simplest case, ($ n = 1, \; 6\, g =\lambda $), there
is only one forbidden band for $ q^2 > 0 $. It goes from $ q^2 = 1 -
\left({{m_{\sigma}}\over m}\right)^2 $ to $ q^2 = 1 - \left({{m_{\sigma}}\over
m}\right)^2 + \eta_0^2/2 $. That is,
\begin{equation}
 {\rm forbidden \; band\; :} \; \;
m^2 -  m^2_{\sigma} < k^2 < m^2 -  m^2_{\sigma} + {{\lambda }\over
{4}}\; \Phi_0^2 \; .
\end{equation}
The Floquet index  in this forbidden band takes the form
\begin{equation}
 F(q) = 2 i \, K(k)\; Z(2  K(k)\,v)  \pm  \pi \; ,
\end{equation}
where now $v$ and $q$ are related by the following equation
\begin{equation}
q^2 = 1-    \left({{m_{\sigma}}\over m}\right)^2 + {{\eta_0^2}\over
2}\; \mbox{cn}^2(2  K(k)\,v,k) \; .  
\end{equation}
and $ 0 \leq v \leq 1/2 $.

The imaginary part of the Floquet index is now maximal at $ q^2_1 = 1-
\left({{m_{\sigma}}\over m}\right)^2 + {{\eta_0^2}\over 4} $ and we can use
this value to provide an estimate for the reheating temperature in this model
in the same manner as for eq.(\ref{trehest}), yielding the following
estimate for the reheating temperature
\begin{equation}\label{Treh}
T_{reh} \simeq \sqrt{ m^2 -  m^2_{\sigma} + {{\lambda} \over 8 }\,
\Phi_0^2 }\; . 
\end{equation}
In the special case $  m_{\sigma} = m = |M_R| $, we recover eq.(\ref{trehest}),
as expected. 

Again, the non-lineal field evolution for $ t > t_{reh} $ decreases $
T_{reh} $. In the third reference under \cite{dis}, we found for
late times a  $ T_{reh} $ ten times smaller than the value
(\ref{Treh}) for $ g = 1.6 \pi^2 $. 

The estimates on the reheating temperature provided above should not
be taken rigorously, 
but as an approximate guide. 
A consistent estimate of the reheating temperature and the thermalization time
scales would in principle involve setting up a Boltzmann equation
\cite{rocky}.  Under the assumption of a
separation between the preheating and thermalization time scales one
 could  try to use
the distribution functions $N_q$ at the end of the preheating stage as input in
the kinetic Boltzmann equation. However, we now argue that such a kinetic
description is {\em not valid} to study thermalization.  A kinetic approach
based on the Boltzmann equation, with binary collisions, for example, would
begin by writing the rate equation for the distribution of particles
\begin{eqnarray}
\dot{N}_k & \propto &
\lambda^2 \int d^3k_1 d^3k_2 d^3k_3 \; \delta^4(k_1+k_2+k_3+k) \; [
 (1+n_k)(1+n_{k_1})n_{k_2}n_{k_3}- \nonumber \\
& & n_k n_{k_1}(1+n_{k_2})(1+n_{k_3})].
\end{eqnarray}
However this equation is only valid in the {\em low density} regime. In particular
for the case under study, the occupation numbers for wavevectors in the
unstable bands are non-perturbatively large $\propto 1/\lambda$ and one would
be erroneously led to conclude that thermalization occurs on the same time
scale or {\em faster than} preheating.

Clearly such a statement would be too premature. Without a separation of time
scales the kinetic approach is unwarranted. The solution of the Boltzmann
equation provides a partial resummation of the perturbative series which is
valid whenever the time scales for relaxation is much longer than the
microscopic time scales\cite{boylawrie}, in this case that of particle
production.

In the case under study there is an expansion parameter, $1/N$ and clearly
these scattering terms are subleading in this formal limit, so that the
separation of time scales is controlled. In the absence of such an expansion
parameter, some resummation scheme must be invoked to correctly incorporate
scattering. In particular when the symmetry is broken, the asymptotic
excitations are Goldstone bosons, the medium is highly excited but with very
long-wavelength Goldstones and these have very small scattering cross
sections. Such a resummation is also necessary in the large temperature limit
of field theories in equilibrium. In this case the perturbative expansion of
the scattering cross section involves powers of $\lambda T/m$ with $m$ being
the mass. A correct resummation of the (infrared) divergent terms leads to
$\lambda(T) \rightarrow m/T$ in the large $T/m$ limit\cite{tetradis}. In
particular the $1/N$ corrections in the formal large $ N$ limit involve such a
resummation, but in the non-equilibrium situation,  the numerical implementation of
this resummation remains a formidable problem.

\section{Conclusions}

It is clear that preheating is both an extremely important process in a variety
of settings, as well as one involving very delicate analysis. In particular,
its non-perturbative nature renders any treatment that does not take into
account effects such as the quantum back-reaction  due to the produced particles,
consistent conservation (or covariant conservation) of the relevant quantities and
Ward identities,  
incapable of correctly describing the important physical phenomena during the
preheating stage.

In this work, we dealt with these issues by using the $O(N)$ vector model
in the large $N$ limit. This allows for a  controlled non-perturbative 
approximation scheme that conserves energy and the proper Ward identities, 
to study  the non-equilibrium dynamics of scalar fields. 
Using this model we were able
to perform a full analysis of the evolution of the zero mode as well as of the
particle production during this evolution.

Our results are rather striking. We were able to provide analytic results for
the field evolution as well as the particle production and the equation of
state for all these components in the weak coupling regime and for times for
which the quantum fluctuations, which account for back-reaction effects, are
small. What we found is that, in the unbroken symmetry situation, the field
modes satisfy a Lam\'e equation that corresponds to a Schr\"odinger equation
with a two-zone potential. There are two allowed and two forbidden bands, which
is {\em decidedly} unlike the Mathieu equation used in previous
analysis\cite{branden,lindekov,yoshimura}. The difference between an
equation with two 
forbidden bands and one with an infinite number is profound. We were also able
to estimate analytically
the time scale at which preheating would occur by asking when the
quantum fluctuations as calculated in the absence of back-reaction would become
comparable to the tree level terms in the equations of motion. The equations of
state of both the zero mode and ``pions'' were calculated and were found to be
describable as polytropes. These results were then confirmed by numerical
integration of the equations, and we found that the analytic results were in
great agreement with the numerical ones in their common domain of validity.

When the $O(N)$ symmetry is spontaneously broken, more subtle effects can
arise, again in the weak coupling regime. If the zero mode starts off  very near the
origin, then the quantum back-reaction grows to be comparable to the tree level
terms within one or at most a few oscillations even for very weak coupling. In this
case the periodic approximation for the dynamics of the zero mode breaks down
very early on and the full dynamics must be studied numerically. 

When numerical tools are brought to bear on this case we
find some extremely interesting behavior. In particular, there are situations
in which the zero mode starts near the origin (the initial value depends on
the coupling) and then in one oscillation,
comes back to almost the same location. However, during the evolution it has
produced $1\slash g$ particles. Given that the total energy is conserved, the
puzzle is to find where the energy came from to produce the particles. We found
the answer in a term in the energy density that has the interpretation 
of a ``vacuuum energy'' that becomes {\em negative} during
the evolution of the zero mode and whose contribution to the equation of state
is that of ``vacuum''. The energy given up by this term is the energy
used to produce the particles.

This example also allows us to study the possibility of symmetry
restoration during 
preheating\cite{lindekov,tkachev,kolbriotto}. While there have been arguments
to the effect that the produced particles will contribute to the quantum
fluctuations in such a way as to make the effective mass squared of the modes
positive and thus restore the symmetry, we argued that they were unfounded. 
Whereas the effective squared mass oscillates, taking positive values during the
early stages of the evolution, its asymptotic value is zero, compatible with Goldstone
bosons as the asymptotic states. 

Furthermore, this says nothing about whether the symmetry is restored or
not. This is signaled by the final value of the zero mode. In all the
situations examined here, the zero mode is driven to a {\em non-zero} final
value. At this late time, the ``pions'' become massless, i.e. they
truly are the Goldstone modes required by Goldstone's theorem. 

The arguments presented in favor of symmetry restoration rely heavily on the
effective potential. We have made the point of showing explicitly why such a 
concept is completely irrelevant for the non-equilibrium dynamics when profuse
particle production occurs and the evolution occurs in a highly excited, out of
equilibrium state. 

Finally, we dealt with the issue of how to use our results to calculate the
reheating temperature due to preheating in an inflationary universe
scenario. Since our results are particular to Minkowski space, we need to
assume that preheating and thermalization occur on time scales shorter than the
expansion time, i.e. $H^{-1}$. We also need to assume that there is a
separation of time scale between preheating and thermalization. Under these
assumptions we can estimate the reheating temperature as 
$T_{reh} \propto |M_R|$ in the case
where the produced particles are massive and $T_{reh} \propto |M_R|\slash
\lambda^{\frac{1}{4}}$ in the massless case. 

We have made the important observation that
due to the large number of long-wavelength 
particles  in the forbidden bands, a kinetic or
Boltzmann equation approach to thermalization 
is {\em inconsistent} here. A resummation akin to that of hard thermal loops, that
consistently arises in the next order in $1/N$  must be employed. In equilibrium
such a resummation shows that the scattering cross section for soft modes is
perturbatively small despite their large occupation numbers.

There is a great deal left to explore. Clearly the first step from this
work would be to generalize what we have done to include the expansion of the
Universe. This should be done both in order to understand preheating in more
detail, as well as to understand the evolution of the scalar field during the
inflationary period. Further steps should certainly include trying to
incorporate thermalization effects systematically within the $1\slash N$
expansion. 

As we were finishing the write-up of this work, we learned about complementary 
work by Cooper, Kluger, Habib and  Mottola\cite{newcooper} who studied
similar issues in the broken symmetry phase with results that are
consistent with those found by us. 

\acknowledgements 

D.B. thanks the N.S.F. for support under grant awards: PHY-9302534 and
INT-9216755. R. H. was supported by DOE grant DE-FG02-91-ER40682. D.B. and R.H.
would also like LPTHE at the Universit\'e de Paris VI for its hospitality
during part of this work. D. B. and H. J. d. V. would like to thank
S. Habib and   E. Mottola for stimulating conversations and for
sharing with them results of their work prior to distribution. D. B. thanks the
Pittsburgh Supercomputer Center for  grant award No: PHY950011P.  

\appendix
\section{Some Results on Elliptic Functions}

Here we detail some of the derivations used in the main text for the analytic
results. 

\subsection{The Unbroken Symmetry Case}

We derive here the solutions of the mode equations (\ref{modsn}),

\begin{equation}\label{modsnA}
\left[\;\frac{d^2}{d\tau^2}+q^2+1+ \eta_0^2 - \eta_0^2\;
\mbox{sn}^2\left(\tau\sqrt{1+\eta_0^2},k\right) \;\right] \varphi_q(\tau) =0
\end{equation}
where $ k =  {{\eta_0}\over{\sqrt{2( 1 +  \eta_0^2)}}} $.

It is convenient to express the Jacobi sine in terms of the Weierstrass ${\cal
P}$ function through \cite{gr}
\begin{eqnarray}
& &\mbox{sn}^2\left(\tau\sqrt{1+\eta_0^2},k\right) =
{ 1 \over {k^2\;
\mbox{sn}^2\left(\tau\sqrt{1+\eta_0^2}+iK'(k),k\right)}} \nonumber \\
&& = {1 \over {k^2(e_1-e_3)}}\;\left[ {\cal P}
\left({{\tau\sqrt{1+\eta_0^2}+iK'(k)}\over {\sqrt{e_1-e_3}}}\right) -
e_3 \right] \\
& & 
k^2 = {{e_2-e_3}\over{e_1-e_3}} \; .
\end{eqnarray}
We assume that the discriminant $\Delta$ of the function ${\cal P}$ is here
positive and that the roots $e_1,e_2,e_3$ of the cubic equation obeying
$e_1+e_2+e_3= 0$ are ordered as
\begin{equation}\label{raices}
e_3<e_2<0<e_1. 
\end{equation}
In addition, without any loss of generality we choose $e_1-e_3 = 1+\eta_0^2$.
With such a choice we find the roots to be: $ 3 e_1 = \frac{3}{2}\eta_0^2 + 2 , 3
e_2 = -1 $ and $ 3 e_3 = - \frac{3}{2}\eta_0^2 - 1 $.

Collecting all factors the mode equations (\ref{modsnA}) become
\begin{equation}\label{lame2}
 \left[\;\frac{d^2}{d\tau^2}+q^2+\frac{1}{3}
-2\,  {\cal P}(\tau + \omega')\;\right]
 \varphi_q(\tau) =0
\end{equation}
where $ \omega' \equiv i {{K'(k)}\over {\sqrt{1+\eta_0^2}}} $.

Eq.(\ref{lame2}) is a Lam\'e equation that can be solved in closed form in
terms of Weierstrass functions. The solutions is\cite{ince}
\begin{equation}\label{solame} 
U_q(\tau) = {{\sigma(\tau + \omega'+z)\sigma( \omega')}\over
{\sigma(\tau + \omega') \;\sigma( \omega'+z)}}\; e^{-\tau \zeta(z)}
\end{equation}
where $z$ is defined through
\begin{equation}\label{mapping}
 {\cal P}(z)= -\frac13 - q^2\; ,
\end{equation}
$\sigma(x)$ and $\zeta(x)$ are Weierstrass functions.
We normalize the functions $ U_q(\tau) $ according to eq.(\ref{ucero}).

Changing $\tau$ by $-\tau$ in eq.(\ref{solame}) provides an 
independent solution of eq.(\ref{lame2}).

Eq.(\ref{mapping}) maps the real $q^2$ axis into the sides of the fundamental
square in the $z$-plane\cite{abr,pbm}. That is:
\begin{eqnarray}
z &=& \beta, \; 0\leq \beta \leq \omega \; , \; \;
-\infty \leq q^2 \leq  -1 - {{\eta_0^2}\over 2}\, , \cr \cr
z &=& \omega + i \alpha\;  , \; 0 \leq \alpha \leq \omega'/i \, , \; \;
  -1 - {{\eta_0^2}\over 2} \leq q^2 \leq 0\, ,  \\ \cr
z &=&  \omega' + \beta , \;  \omega \geq \beta \geq 0\;  ; \;
0  \leq q^2 \leq  {{\eta_0^2}\over 2} \, , \cr \cr
z &=& i \alpha\;  , \;  \omega'/i \geq \alpha  \geq 0\;  ; \;
 {{\eta_0^2}\over 2} \leq q^2 \leq +\infty \; .\nonumber
\end{eqnarray}
where $ \omega =  {{K(k)}\over {\sqrt{1+\eta_0^2}}} $.
Notice that $ 2  \omega $ is the period in $\tau$ of the potential in
eqs.(\ref{modsn}) and (\ref{lame2}).

The mapping of the real $q^2$-axis into the sides of the fundamental
 square in the $z$-plane is made explicit by  writing the Weierstrass 
 ${\cal P}(z)$ function in terms of Jacobi sn and cn  functions \cite{gr,abr,pbm}:
\begin{eqnarray}
 q^2&=&  -\frac13 -  {\cal P}( \beta) = {{\eta_0^2}\over 2} -
{{\eta_0^2 + 1}\over {\mbox{sn}^2\left( \beta\sqrt{1+\eta_0^2},k\right)}}\,
, \cr \cr
q^2&=&  -\frac13 -  {\cal P}( \omega + i \alpha) =
-1 -  {{\eta_0^2}\over 2} + ( 1 +  {{\eta_0^2}\over
2})\; \mbox{sn}^2\left(\alpha\sqrt{1+\eta_0^2},k'\right) \, ,  \\ \cr
  q^2&=&  -\frac13 -  {\cal P}( \omega' + \beta) =  {{\eta_0^2}\over
2} \;\mbox{cn}^2\left(\beta\sqrt{1+\eta_0^2},k\right) \, , \cr \cr
 q^2&=&  -\frac13 -  {\cal P}(  i \alpha) = -1 -  {{\eta_0^2}\over 2}
+{{\eta_0^2 + 1}\over
{\mbox{sn}^2\left(\alpha\sqrt{1+\eta_0^2},k'\right)}}\, .\nonumber
\end{eqnarray}
Here $ \alpha$ and $\beta$ are real variables. [Recall that $ 0 \leq 
\mbox{sn}(u,k) \leq 1 $ for $ 0 \leq u \leq K(k) $ and that 
$ \mbox{sn}(u,k)^2+\mbox{cn}(u,k)^2 = 1 $ .]

Generically, a periodic potential possesses an {\em infinite} number of allowed
and forbidden (Floquet) bands that alternate for $0 \leq q^2 < +\infty$.  A
detailed study of the Floquet indices reveals that this is not the case for
eq.(\ref{modsnA}).

Let us now compute the Floquet indices for the mode functions
(\ref{solame}). The quasi-periodicity property of the Weierstrass
$\sigma$-functions states that\cite{gr,abr,pbm}:
\begin{equation}\label{sip}
\sigma(x+2 \omega) = - \sigma(x) \; e^{2(x + \omega)\eta}
\end{equation}
where $\eta \equiv  \zeta( \omega)$. Using  eqs.(\ref{sip}) and
(\ref{floq}), we find from  eq.(\ref{solame})  
\begin{eqnarray}
U_q(\tau+2 \omega) &=& U_q(\tau) \; e^{2[z\eta - \omega  \zeta(z)]}\cr\cr
F(q) & =& 2i[   \omega  \zeta(z) - z\eta] \; .
\end{eqnarray}

The quantity $q$ belongs to an allowed band when $2[z\eta - \omega \zeta(z)]$
is purely imaginary and $q$ belongs to a forbidden band when this quantity has
a nonzero real part.

Using the properties of the Weierstrass functions \cite{gr}, we find {\em two}
allowed bands: 
\begin{eqnarray}\label{banda} z = i \alpha \quad \mbox{and}
\quad z = i \alpha + \omega , \cr \mbox{~with~real~} \alpha,\; 0 < \alpha <
\omega'/i \; .
\end{eqnarray}
and {\em two} forbidden bands \begin{eqnarray}\label{banda2} z = \beta \quad
 \mbox{and} \quad z = \beta + \omega' , \cr \mbox{~with~real~} \beta,\; 0 <
 \beta < \omega \; .
\end{eqnarray}

In terms of $q^2$ the allowed bands correspond to
\begin{equation}
 -1 - {{\eta_0^2}\over 2} \leq q^2 \leq 0
 \quad \mbox{and} \quad
 {{\eta_0^2}\over 2} \leq q^2 \leq +\infty \; ,
\end{equation}
and the forbidden bands to
\begin{equation}
-\infty \leq q^2 \leq  -1 - {{\eta_0^2}\over 2}
 \quad \mbox{and} \quad
0  \leq q^2 \leq  {{\eta_0^2}\over 2} \, .
\end{equation}
The last forbidden band is for {\em positive} $q^2$ and hence {\em does}
contribute to the fluctuation function $\Sigma(\tau)$.  $\Sigma(\tau)$
will grow exponentially in time due to the presence of such unstable modes.

Let us investigate the Floquet indices for the forbidden band $ 0 \leq q^2 \leq
{{\eta_0^2}\over 2} $. Setting $z = \beta + \omega'$ we find
\begin{eqnarray}
2[z\eta - \omega  \zeta(z)] = -
{{\vartheta_1'(\frac{z}{2\omega})}\over{\vartheta_1(\frac{z}{2\omega})}}\cr
=i\pi - 
{{\vartheta_4'(\frac{\beta}{2\omega})}\over{\vartheta_4
(\frac{\beta}{2\omega})}}\; .
\end{eqnarray}
where we used the relation between the Weierstrass $\zeta$-function
and the Jacobi $\vartheta$-functions \cite{erd}
\begin{equation}\label{zetaW}
\zeta(z) = \frac{\eta}{\omega}\, z +
\frac{1}{2\omega}\;{{\vartheta_1'(\frac{z}{2\omega})} 
\over{\vartheta_1(\frac{z}{2\omega})}} \; .
\end{equation}
Here $ \eta \equiv \zeta(\omega) $.

In summary, we have on the forbidden band $ 0 \leq q^2 \leq {{\eta_0^2}\over 2}
$ ,
\begin{equation}
F(q) = \pm \pi + i  {{\vartheta_4'(v)}\over{\vartheta_4(v)}} =
  \pm \pi + 2 i \, K(k)\; Z(2  K(k)\,v)  \; ,
\end{equation}
where $ v \equiv \frac{\beta}{2\omega} \, , 0 \leq v \leq 1/2 $ and $ Z(u)
\equiv E(u,k) - u \; {{E(k)}\over { K(k)}} $ is the Jacobi zeta function
\cite{erd}. Eq.(\ref{Zjacob}) gives its series expansion.
We see that the solution $ U_q(\tau) $ decreases with $\tau$. The other
independent solution $U_q(-\tau) $ grows with $\tau$.

Notice that $ U_q(\tau) $ becomes just an antiperiodic function on the borders
of this forbidden band, $ v = 0 $ and $ v = 1/2 $.

It is useful to write the solution $ U_q(\tau) $ in terms of Jacobi
$\vartheta$-functions. Using eqs. (\ref{solame}), (\ref{zetaW}) and \cite{erd}
\begin{equation}\label{sigmaW}
\sigma(z) = {{2\, \omega}\over{\vartheta_1'(0)}}\; 
e^{{\eta z^2}\over {2 \omega}}\; \vartheta_1(\frac{z}{2\omega})\; ,
\end{equation}
we found  eq.(\ref{uqproh})  after some calculation.

For the allowed band $ {{\eta_0}\over {\sqrt2}} \leq q \leq \infty $, we find
eq.(\ref{Uqperm}) using eqs.(\ref{solame}), eq.(\ref{zetaW}) and
eq.(\ref{sigmaW}).

\subsection{Broken Symmetry Case}

We derive here the solutions of the mode equations (\ref{modsnR}),

\begin{equation}\label{modrA}
 \left[\;\frac{d^2}{d\tau^2}+q^2-1+ {{\eta_0^2} \over 
{\mbox{dn}^2\left(\tau\sqrt{1-{{\eta_0^2}\over 2}},k\right)}}
\;\right] \varphi_q(\tau) =0
\end{equation}

It is convenient to express the Jacobi function dn in terms of the Weierstrass
${\cal P}$ function through \cite{gr}
\begin{equation}
 {1 \over {\mbox{dn}^2}(u,k)}= {1 \over {e_1-e_2}}\;\left[ e_1 -
{\cal P}\left({{u+K(k)+iK'(k)}\over {\sqrt{e_1-e_3}}}\right)
\right]\; .
\end{equation}
Where now for convenience and without loss of generality we choose $e_1-e_3 =
 1- {{\eta_0^2}\over 2}$. Then, we find $ 3 e_1 = 1 \; , \; 3 e_2 = 1
 -\frac{3}{2}\eta_0^2 $ and $ 3 e_3 = \frac{3}{2}\eta_0^2 - 2 $.

Collecting all factors the mode equations (\ref{modrA}) become
\begin{equation}\label{lame2R}
 \left[\;\frac{d^2}{d\tau^2}+q^2-\frac{1}{3}
-2\,  {\cal P}(\tau  + \omega + \omega')\;\right]
 \varphi_q(\tau) =0
\end{equation}
where $ \omega' \equiv i {{K'(k)}\over {\sqrt{ 1 -  {\eta_0^2}}}} $.
This equation is equivalent to eq.(\ref{lame2}) up to the sign in
front of the $\frac13$ and a shift $\tau  \to\tau + \omega $.

The solutions of eq.(\ref{lame2R}) are  given by 
\begin{equation}\label{solrot} 
U_q(\tau) = {{\sigma(\tau + \omega + \omega'+z)\sigma( \omega'+ \omega)}\over
{\sigma(\tau + \omega + \omega') \;
\sigma(\omega'+ \omega +z)}}\; e^{-\tau \zeta(z)}\; .
\end{equation}
$ z $ and $ q^2 $ are now related  by
\begin{equation}\label{mapR}
 {\cal P}(z)= +\frac13 - q^2\; .
\end{equation}
We continue to use the normalization (\ref{ucero}).
Changing $\tau$ by $-\tau$ in eq.(\ref{solrot}) provides in general an
independent solution of eq.(\ref{modrA}). 

Eq.(\ref{mapR}) maps the real $q^2$ axis into the four sides of the
fundamental square in the $z$-plane. This mapping is better seen
 writing the Weierstrass 
 ${\cal P}(z)$ function in terms of Jacobi sn and cn  functions \cite{gr}
in the four cases. That is:
\begin{eqnarray}
z &=& \beta, \; 0\leq \beta \leq \omega \; , \; \;
-\infty \leq q^2 \leq  0 \, , \cr
 q^2&=&  \frac13 -  {\cal P}( \beta) = \left(1 - {{\eta_0^2}\over 2}\right)
{{\mbox{cn}^2\left( \beta\sqrt{1-{{\eta_0^2}\over 2}},k\right)}
\over {{\mbox{sn}^2\left( \beta\sqrt{1 -{{\eta_0^2}\over 2}},k\right)}}}\;
. \cr \cr
z &=& \omega + i \alpha\;  , \; 0 \leq \alpha \leq \omega'/i \, , \; \;
  0 \leq q^2 \leq  {{\eta_0^2}\over 2} \, , \cr
 q^2&=&  \frac13 -  {\cal P}( \omega + i \alpha) = {{\eta_0^2}\over 2}\;
\mbox{sn}^2\left( \alpha\sqrt{1 -{{\eta_0^2}\over 2}},k'\right) \; . \cr \cr
z &=&  \omega' + \beta , \;  \omega  \geq \beta \geq 0\;  ; \;
 {{\eta_0^2}\over 2} \leq q^2 \leq  1-{{\eta_0^2}\over 2} \; , \cr
 q^2&=&  \frac13 -  {\cal P}( \omega' + \beta) =    1-{{\eta_0^2}\over 2}
-( 1 - \eta_0^2 )
\mbox{sn}^2\left( \beta\sqrt{1 -{{\eta_0^2}\over 2}},k\right) \; . \cr \cr
z &=& i \alpha\;  , \;  \omega'/i \geq \alpha  \geq 0\;  ; \;
 1- {{\eta_0^2}\over 2} \leq q^2 \leq +\infty \; , \cr
 q^2&=&  \frac13 -  {\cal P}(  i \alpha) =
{{1 -{{\eta_0^2}\over 2}}\over{ \mbox{sn}^2
\left( \alpha\sqrt{1 -{{\eta_0^2}\over 2}},k'\right)}} .
\end{eqnarray}

where $ \omega \equiv  {{K(k)}\over {\sqrt{1-{{\eta_0^2}\over 2}}}} $.
Notice that $ 2  \omega $ is the period in $\tau$ of the potential in
eqs.(\ref{modrA}) and (\ref{lame2R}).

Eq.(\ref{floq}) holds for the mode functions (\ref{solrot}) too. Hence,
there are {\em two} allowed bands:
 \begin{eqnarray}\label{rbanda}
z = i \alpha \quad \mbox{and} \quad z = i  \alpha + \omega , \cr
\mbox{~with~real~} \alpha,\;  0 <  \alpha <  \omega'/i \; .
\end{eqnarray}
and {\em two} forbidden bands
 \begin{eqnarray}\label{rbanda2}
z =  \beta \quad \mbox{and} \quad z =   \beta + \omega' , \cr
\mbox{~with~real~} \beta,\;  0 <  \beta <  \omega \; .
\end{eqnarray}

In terms of $q^2$ the allowed bands correspond to
\begin{equation}
 0 \leq q^2 \leq  {{\eta_0^2}\over 2}
 \quad \mbox{and} \quad
 1-{{\eta_0^2}\over 2} \leq q^2 \leq +\infty \; ,
\end{equation}
and the forbidden bands to
\begin{equation}
-\infty \leq q^2 \leq  0
 \quad \mbox{and} \quad
 {{\eta_0^2}\over 2}  \leq q^2 \leq  1-{{\eta_0^2}\over 2} \, .
\end{equation}
The last forbidden band is for {\em positive} $q^2$ and hence {\em does}
contribute to the fluctuation function $\Sigma(\tau)$.  $\Sigma(\tau)$ will
grow exponentially in time due to the presence of such unstable modes.

It is useful to write the solution $ U_q(\tau) $ in terms of Jacobi
$\vartheta$-functions. For the forbidden band $ {{\eta_0^2}\over 2} \leq q^2
\leq 1-{{\eta_0^2}\over 2}$, we obtain eq. (\ref{urotaJ}), using
eqs. (\ref{solrot}), (\ref{zetaW}) and (\ref{sigmaW}), after some calculation.

For the allowed band $ 1- {{\eta_0^2}\over 2} \leq q^2 \leq +\infty \; ,$ we
find eq. (\ref{urotaP1}) from eqs.(\ref{solrot}), (\ref{zetaW}) and
(\ref{sigmaW}).  The analogous expression (\ref{urotaP2}) hold in the allowed
band $ 0 \leq q^2 \leq {{\eta_0^2}\over 2} $.

\figure{ {\bf Figure 1:} The ratio $<p_0>/ \varepsilon_0$ for zero
mode vs. $ \lambda_R 
\varepsilon_0 /2|M_R|^4$ for the unbroken symmetry case.\label{fig1}}

\figure{{\bf Figure 2a:} 
$\eta(\tau)$ vs. $\tau$ for the unbroken symmetry case with
$\eta_0=4$, $g=10^{-12}$.\label{fig2a}}

\figure{{\bf Figure 2b:} 
$g\Sigma(\tau)$ for the same values of the parameters as in
Fig. 2(a). The agreement with the analytic prediction is to within $5 \%$ for
$0< \tau \leq 30$.\label{fig2b}}

\figure{{\bf Figure 2c:}
$g{\cal{N}}(\tau)$ for the same parameters as in
fig. 2a.\label{fig2c}}

\figure{{\bf Figure 2d:}
$gN_q(\tau)$ vs. $q$ for $\tau=40$ for 
the same values of
parameters as fig. 2a.\label{fig2d}}

\figure{{\bf Figure 2e:}$gN_q(\tau)$ vs. $q$ for $\tau=120$ for
 the same values of
parameters as fig. 2(a).\label{fig2e}}

\figure{{\bf Figure 2f:}
$gN_q(\tau)$ vs. $q$ for $\tau=200$ for the same 
values of
parameters as fig. 2(a).\label{fig2f}}

\figure{{\bf Figure 2g:}
$\left(\frac{\lambda_R}{2 |M_R|^4}\right) p(\tau)$
 for the same values
of the parameters as in Fig. 2(a). Asymptotically the average over a period
gives $p_{\infty} \approx \varepsilon /3$.\label{fig2g}}

\figure{{\bf Figure 3:}
The ratio $<p_0>/ \varepsilon_0$ for zero 
mode vs. $ \lambda_R
\varepsilon_0 /2|M_R|^4$ for the broken symmetry case.\label{fig3}}

\figure{{\bf Figure 4a:}
$\eta(\tau)$ vs. $\tau$ for the broken symmetry 
case with
$\eta_0=10^{-5}$, $g=10^{-12}$.\label{fig4a}}

\figure{{\bf Figure 4b:}
$g\Sigma(\tau)$ for the same values of the 
parameters as in
fig. 4a.\label{fig4b}}

\figure{{\bf Figure 4c:}
$g{\cal{N}}(\tau)$ for the same parameters as in
fig. 4a.\label{fig4c}}

\figure{{\bf Figure 4d:}
$gN_q(\tau)$ vs. $q$ for $\tau=30$ for the same values of
parameters as Fig. 4a.\label{fig4d}}

\figure{{\bf Figure 4e:}
$gN_q(\tau)$ vs. $q$ for $\tau=90$ for the same values of
parameters as Fig. 4a.\label{fig4e}}

\figure{{\bf Figure 4f:}
$gN_q(\tau)$ vs. $q$ for $\tau=150$ for the same values of
parameters as Fig. 4a.\label{fig4f}}

\figure{{\bf Figure 4g:}
${\cal{M}}^2(\tau)$ vs. $\tau$ for the same parameters as
Fig. 4a.\label{fig4g}}

\figure{{\bf Figure 4h:}
$ \varepsilon_{cl}(\tau)$ vs. $\tau$ for the same 
parameters as
Fig. 4a.\label{fig4h}}

\figure{{\bf Figure 4i}:
$ \varepsilon_{N}(\tau)$ vs. $\tau$ for the same 
parameters as
Fig. 4a.\label{fig4i}}

\figure{{\bf Figure 4j:}
$ \varepsilon_{C}(\tau)$ vs. $\tau$ for the same
 parameters as
Fig. 4a.\label{fig4j}}

\figure{{\bf Figure 4k:}
$\left(\frac{\lambda_R}{2 |M_R|^4}\right) \; p(\tau)$ for the
same values of the parameters as in Fig. 4a. Asymptotically the average over
a period gives $p_{\infty} = \varepsilon /3$.\label{fig4k}}

\figure{{\bf Figure 5a:}
$\eta(\tau)$ vs. $\tau$ for the broken symmetry case with
$\eta_0=10^{-2}$, $g=10^{-5}$.\label{fig5a}}

\figure{{\bf Figure 5b:}
${\cal{M}}^2(\tau)$ vs. $\tau$ for the same parameters as
fig. 5(a).\label{fig5b}}

\begin{table} \centering
\begin{tabular}{|l|l|l|l|}\hline
$ \eta_0 $ &  $ {\hat q} $  &  $ B $ &  $ N $ \\ \hline
$ $&  $ $ & $ $ & $ $ \\
1 & $ 0.017972387\ldots $ & $ 0.1887167\ldots $ & $ 3.778\ldots $\\
$ $&  $ $ & $ $ & $ $ \\ \hline
$ $&  $ $ & $ $  & $ $ \\
$ 3 $ & $ 0.037295557\ldots $ & $ 0.8027561\ldots $ & $ 0.623\ldots $\\
$ $&  $ $ & $ $ & $ $ \\ \hline
$ $&  $ $ & $ $  & $ $ \\
$ 4 $ & $ 0.03966577\ldots $ & $ 1.1007794\ldots $ & $ 0.4\ldots $\\
$ $&  $ $ & $ $ & $ $ \\ \hline
$ $&  $ $ & $ $  & $ $ \\
 $ \eta_0 \to \infty $ &  $ 0.043213918\ldots $ & $ 0.28595318\,
\eta_0 + O(\eta_0^{-1}) $ & $ 3.147 \, \eta_0^{-3/2}[1+O(\eta_0^{-2})] $
\\ $ $&  $ $ & $ $ & $ $ \\ \hline
\end{tabular}

\bigskip

\label{table1}
\caption{Quantum Fluctuations $\Sigma(\tau) \approx  { 1 \over N \,
 \protect\sqrt\tau} 
 \; e^{B\,\tau}$ during the preheating period.}
\end{table}

\hbox{\epsfxsize 14cm\epsffile{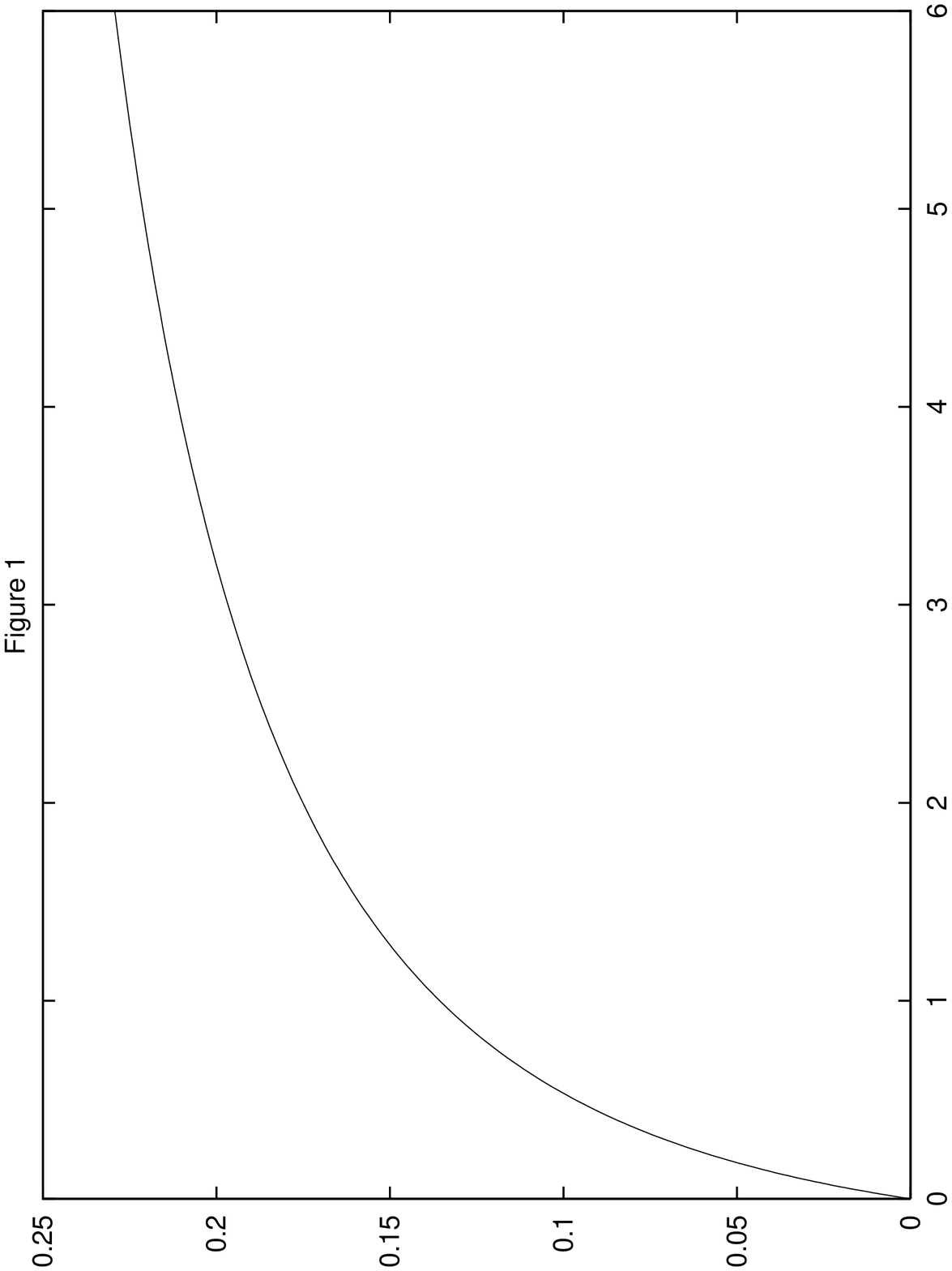}}
\hbox{\epsfxsize 14cm\epsffile{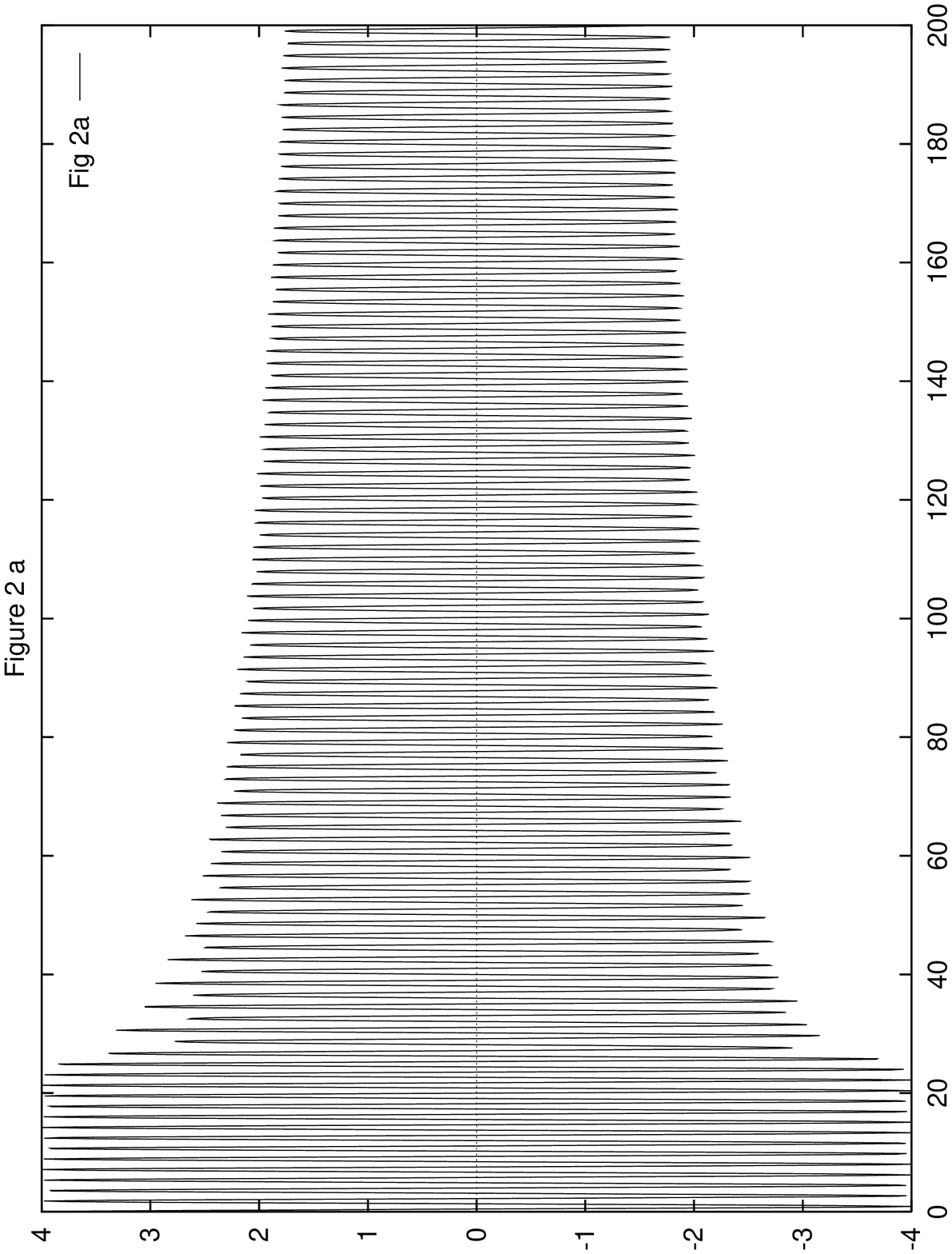}}
\hbox{\epsfxsize 14cm\epsffile{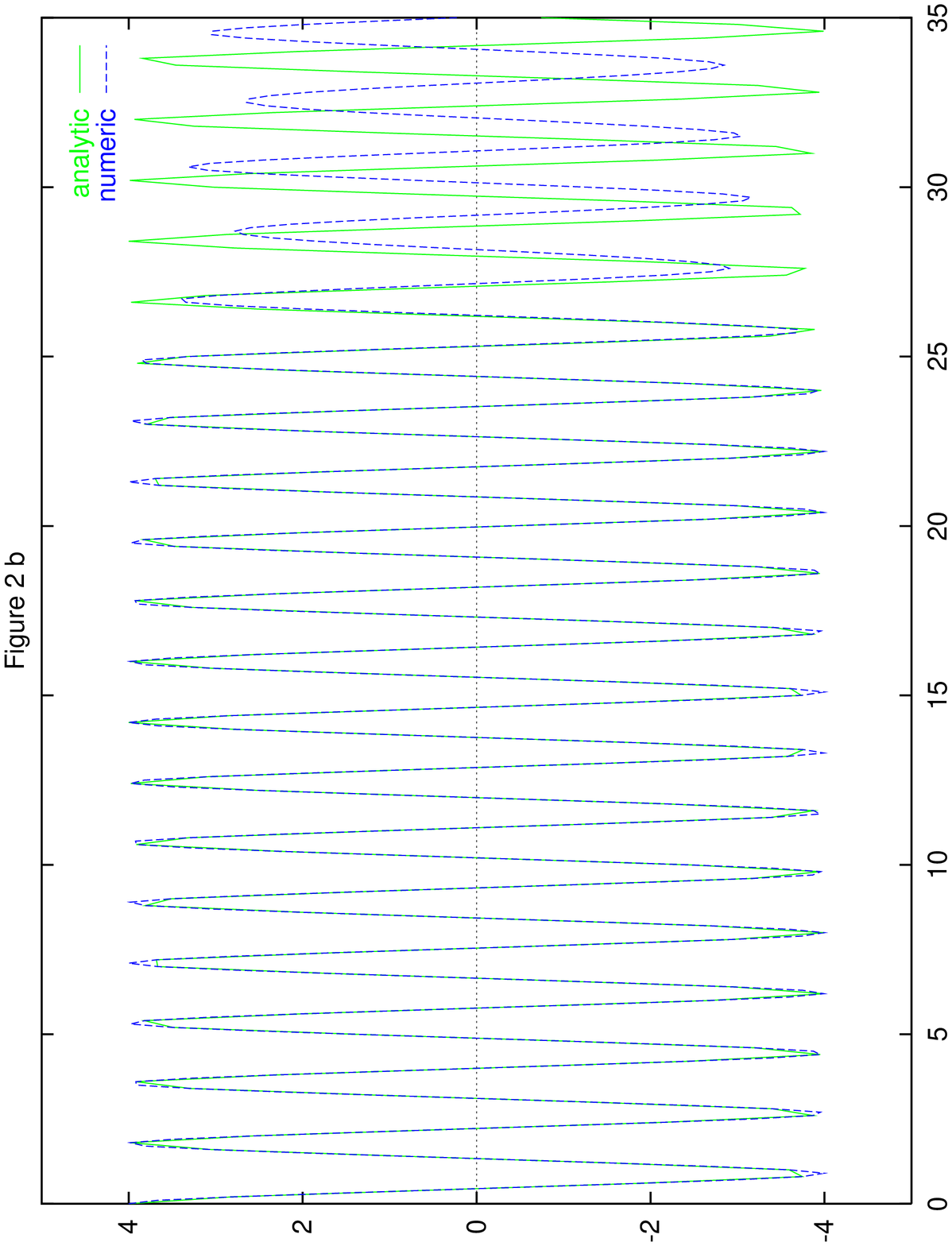}}
\hbox{\epsfxsize 14cm\epsffile{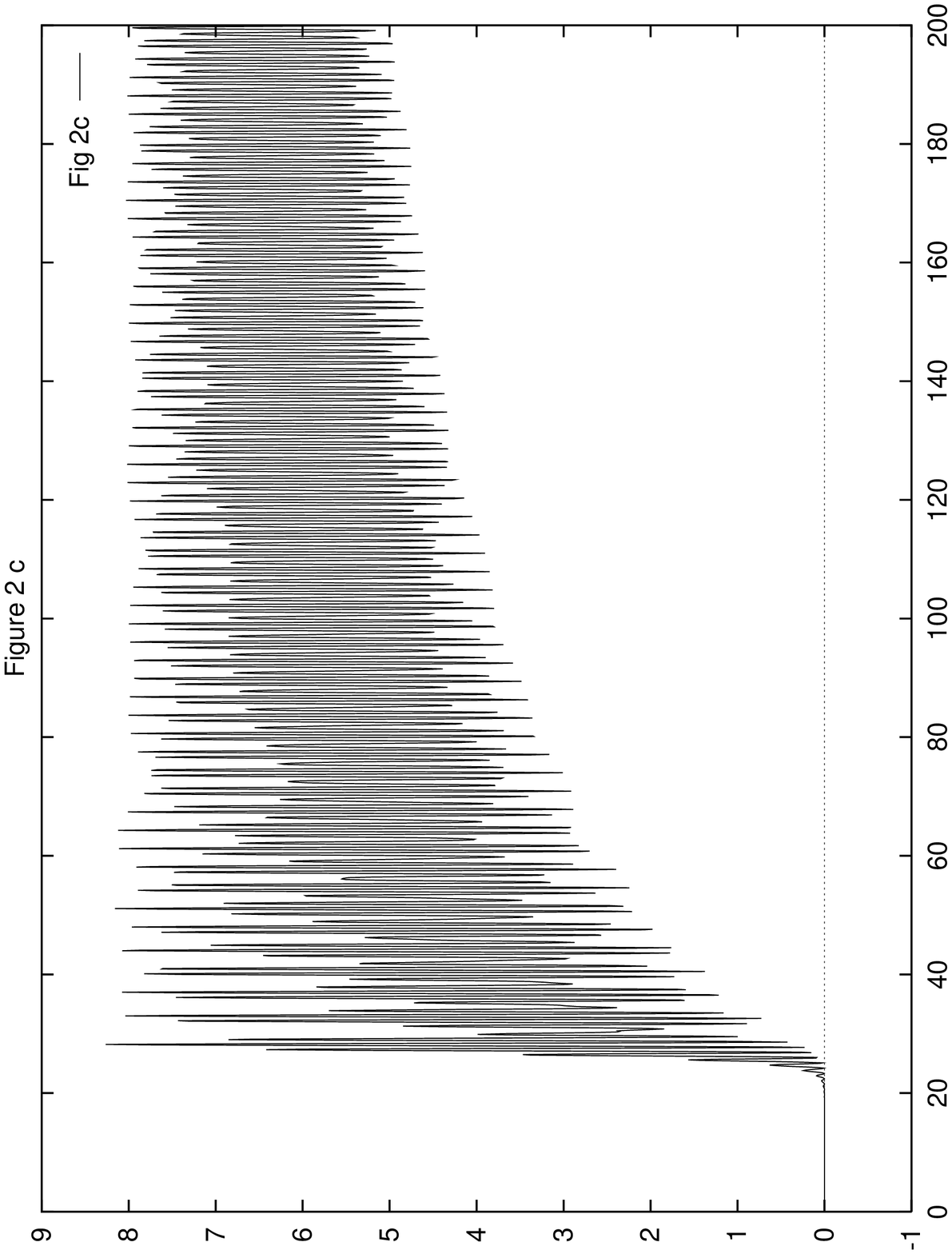}}
\hbox{\epsfxsize 14cm\epsffile{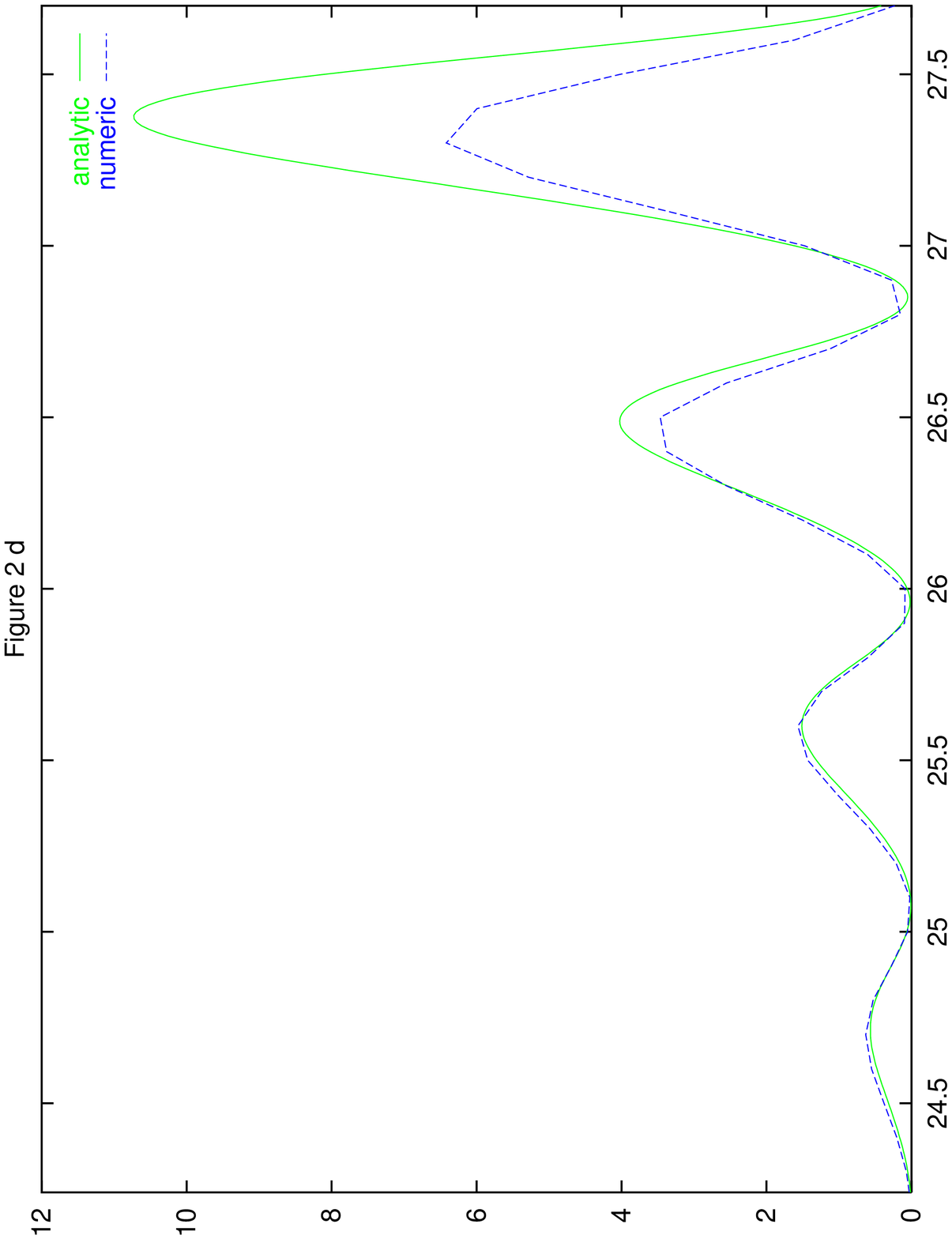}}
\hbox{\epsfxsize 14cm\epsffile{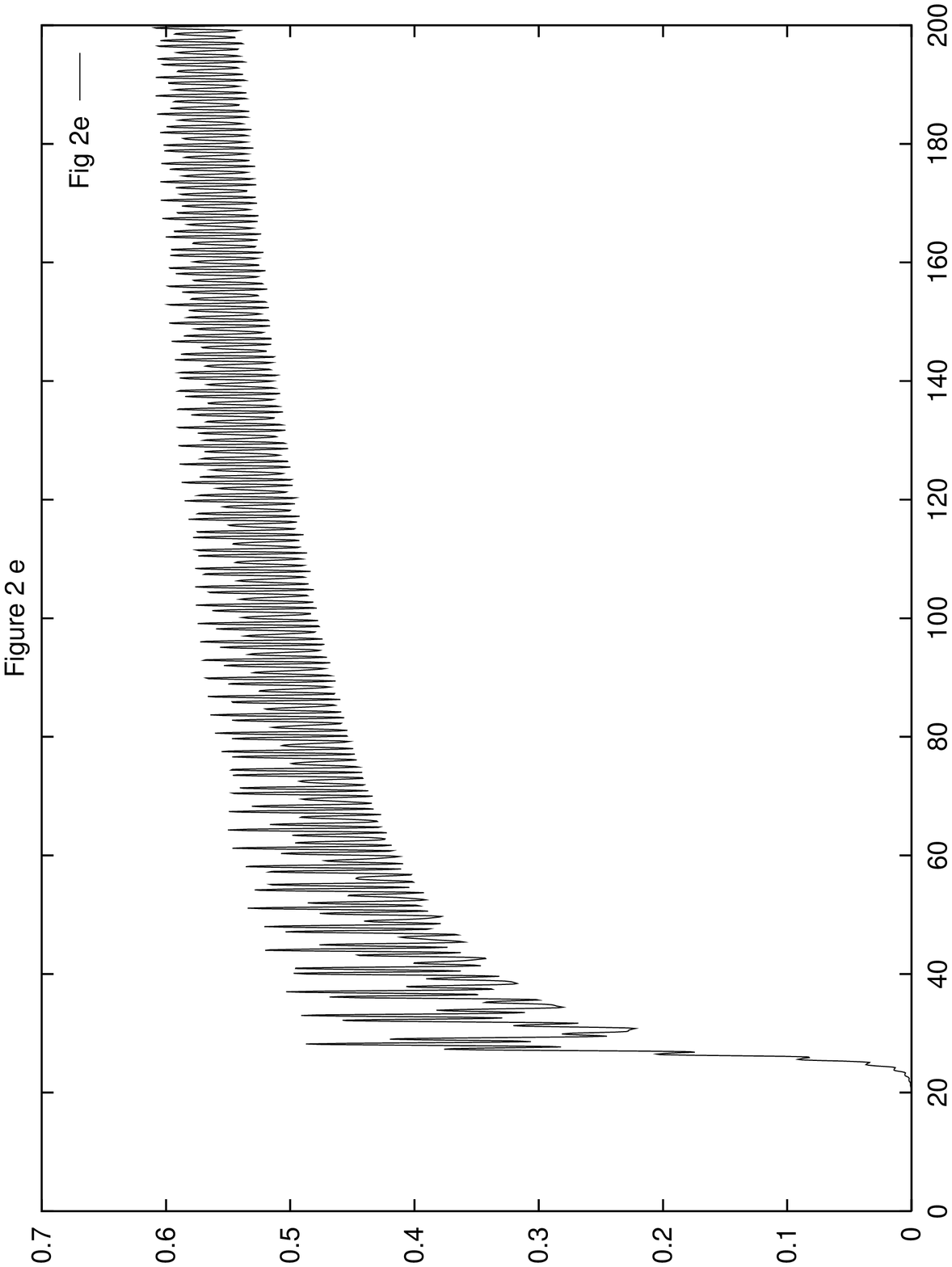}}
\hbox{\epsfxsize 14cm\epsffile{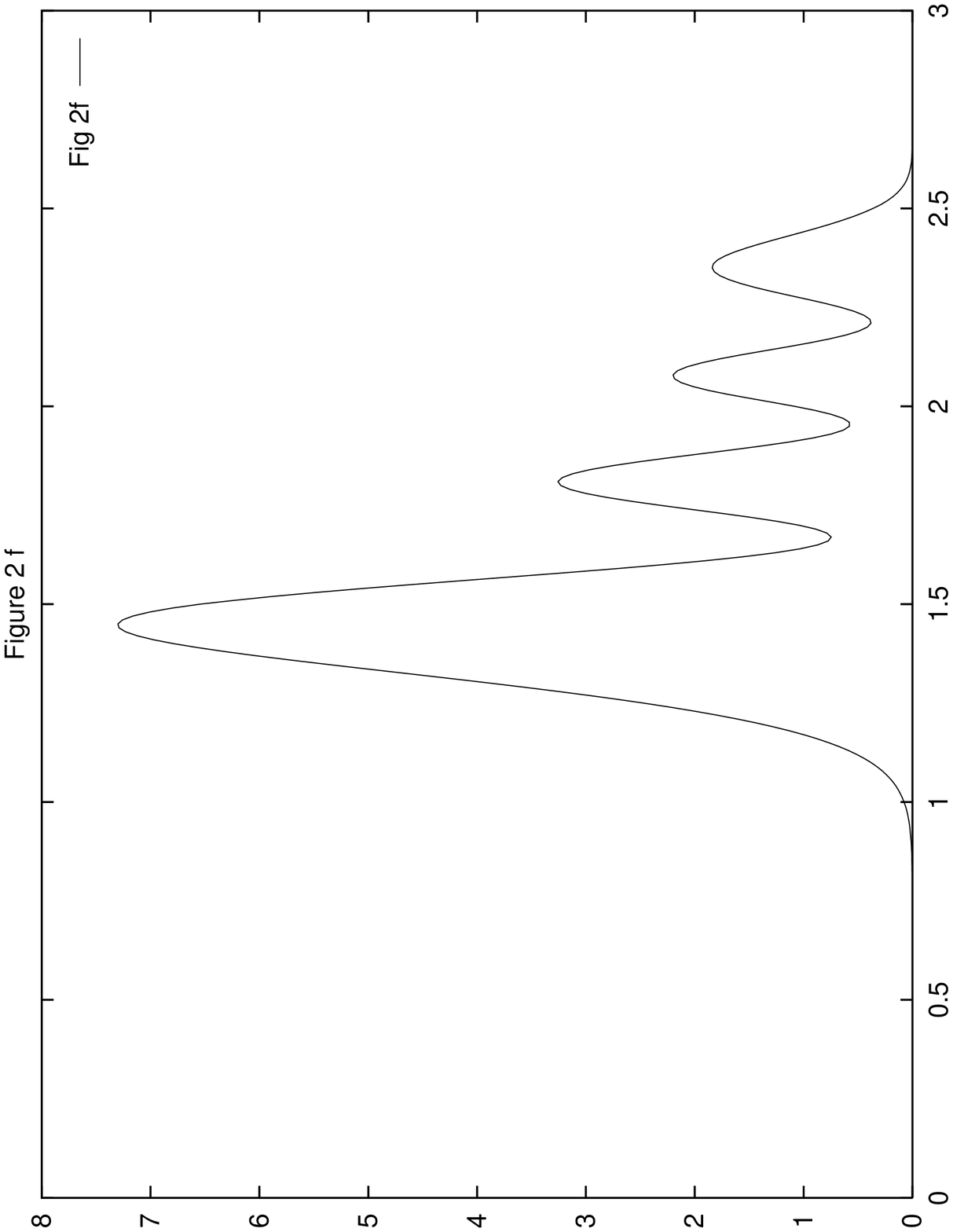}}
\hbox{\epsfxsize 14cm\epsffile{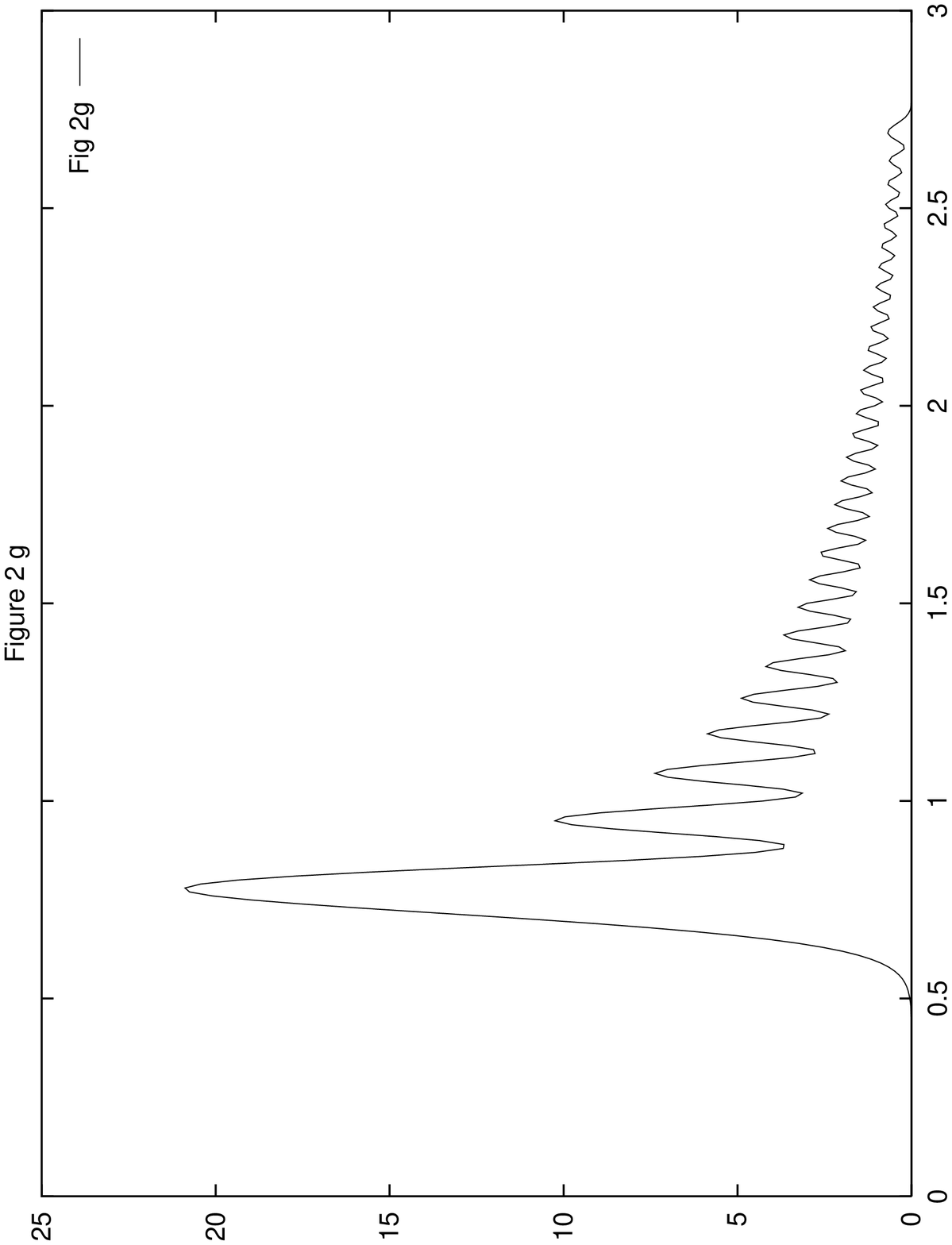}}
\hbox{\epsfxsize 14cm\epsffile{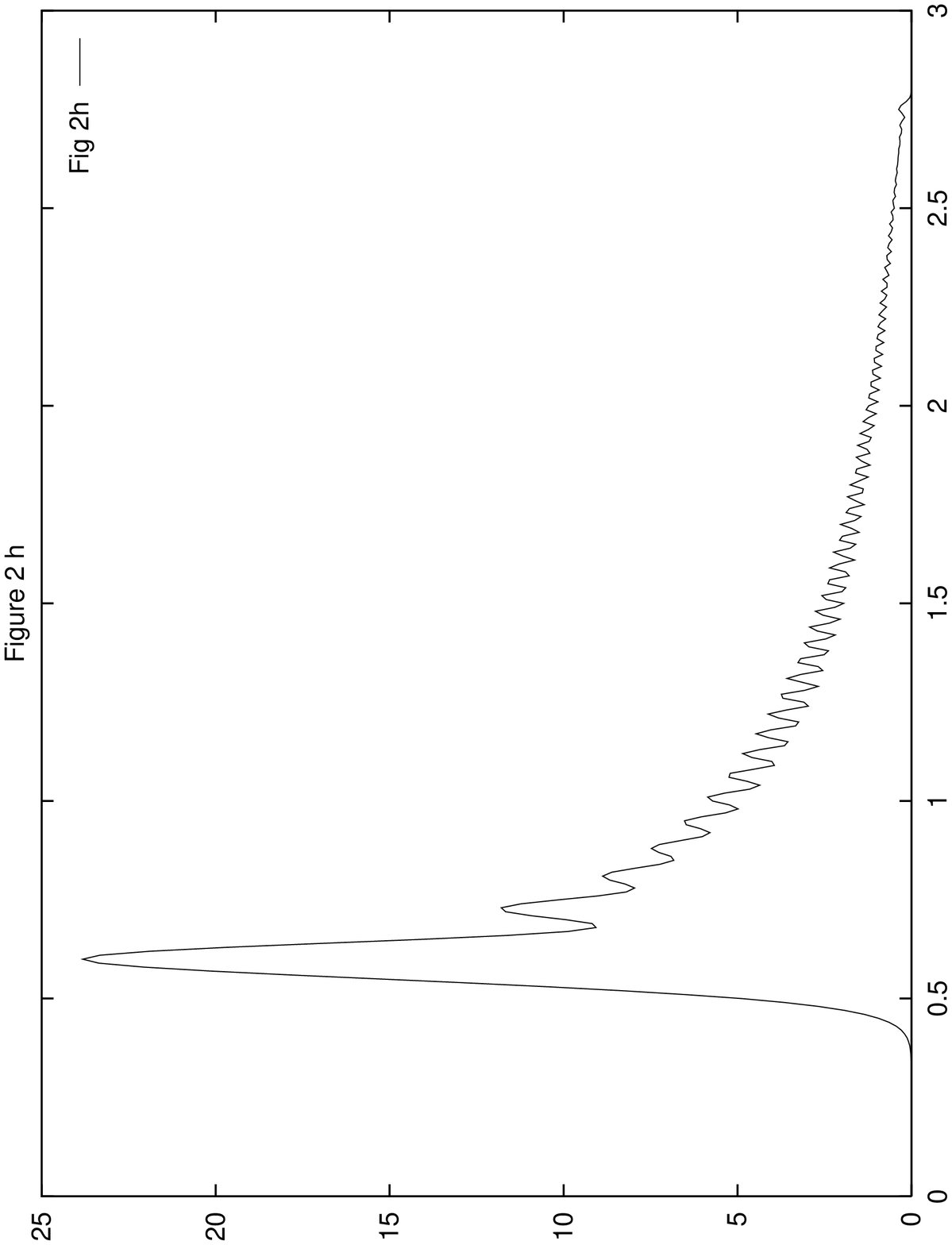}}
\hbox{\epsfxsize 14cm\epsffile{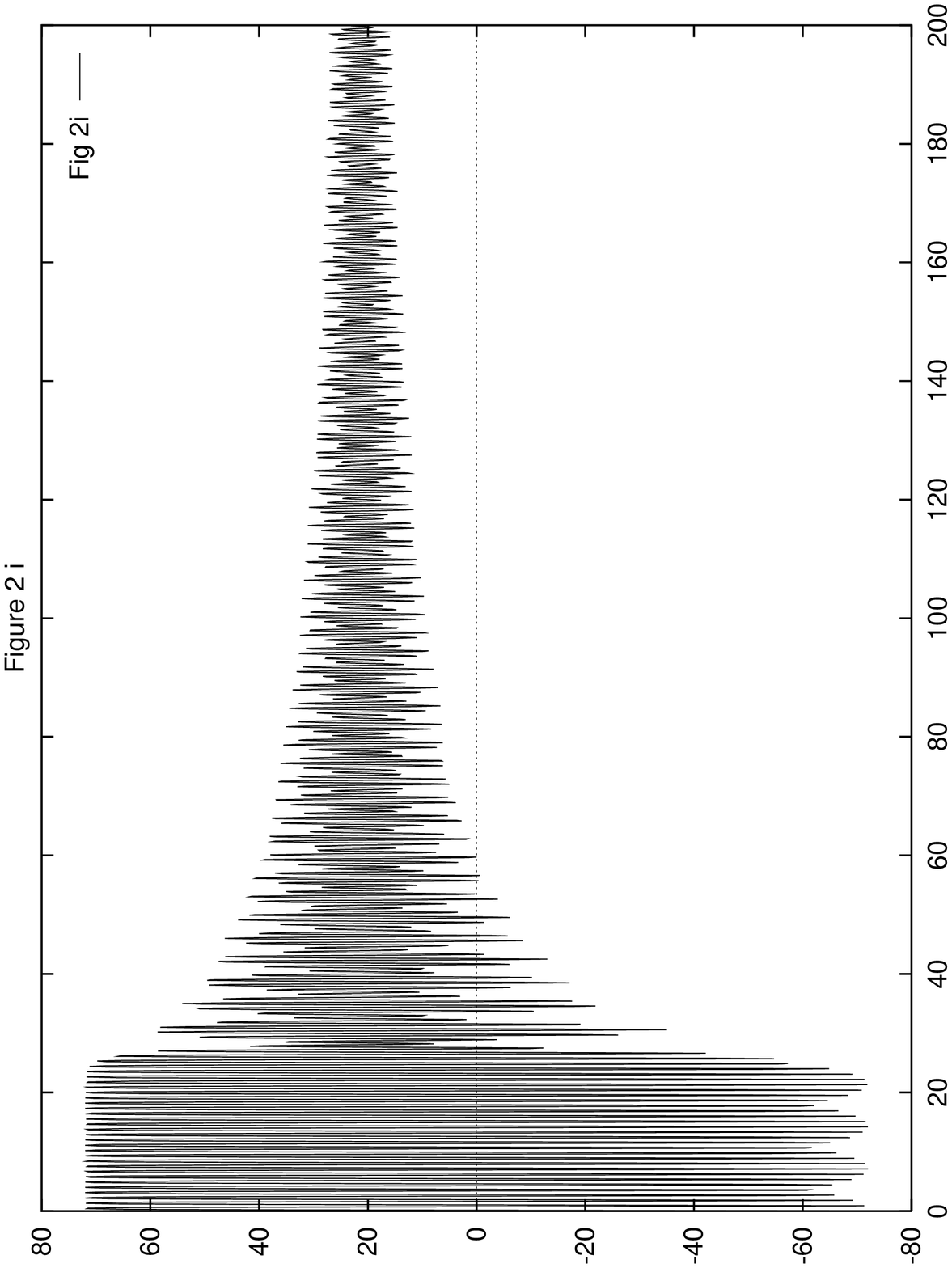}}
\hbox{\epsfxsize 14cm\epsffile{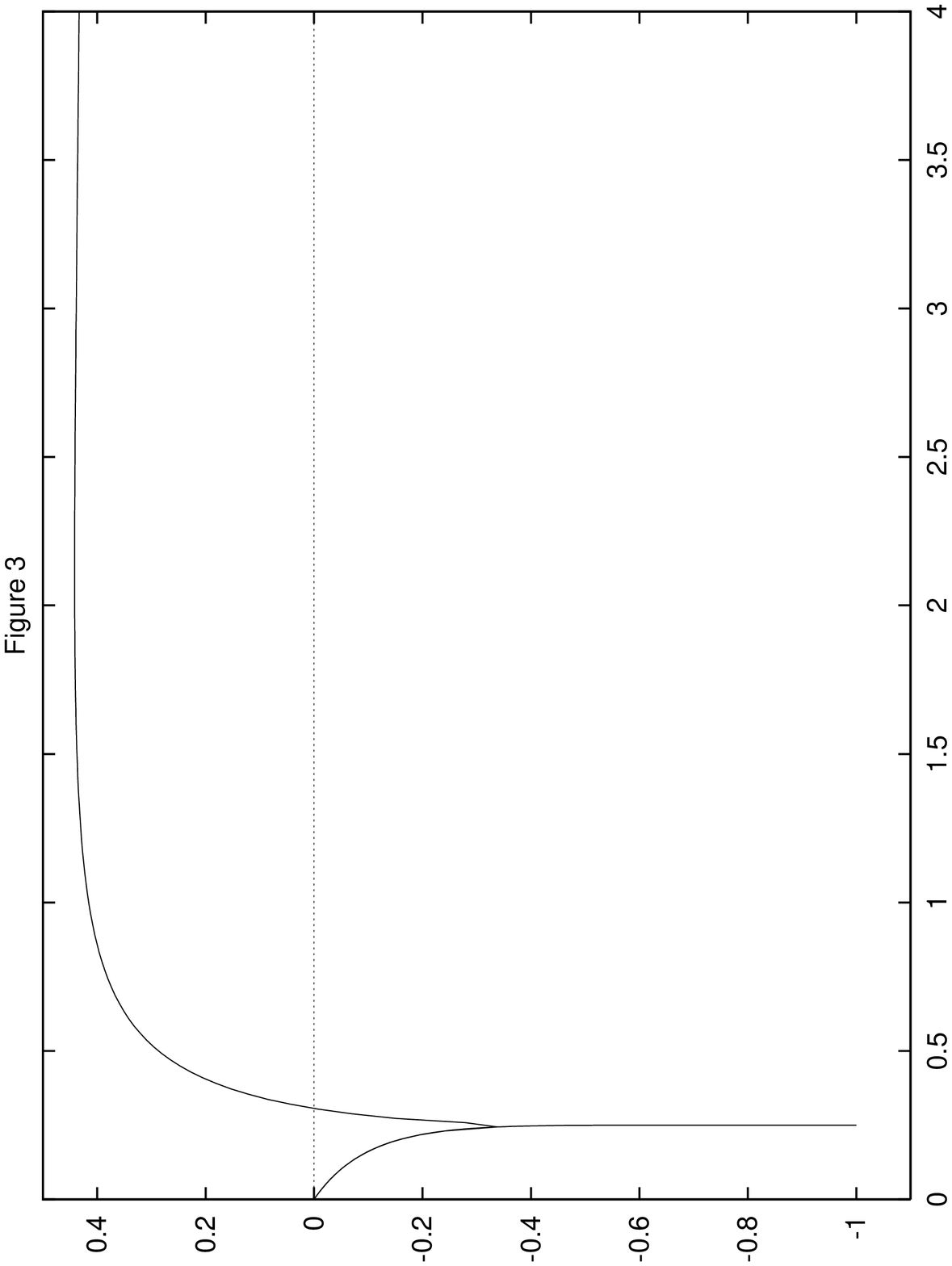}}
\hbox{\epsfxsize 14cm\epsffile{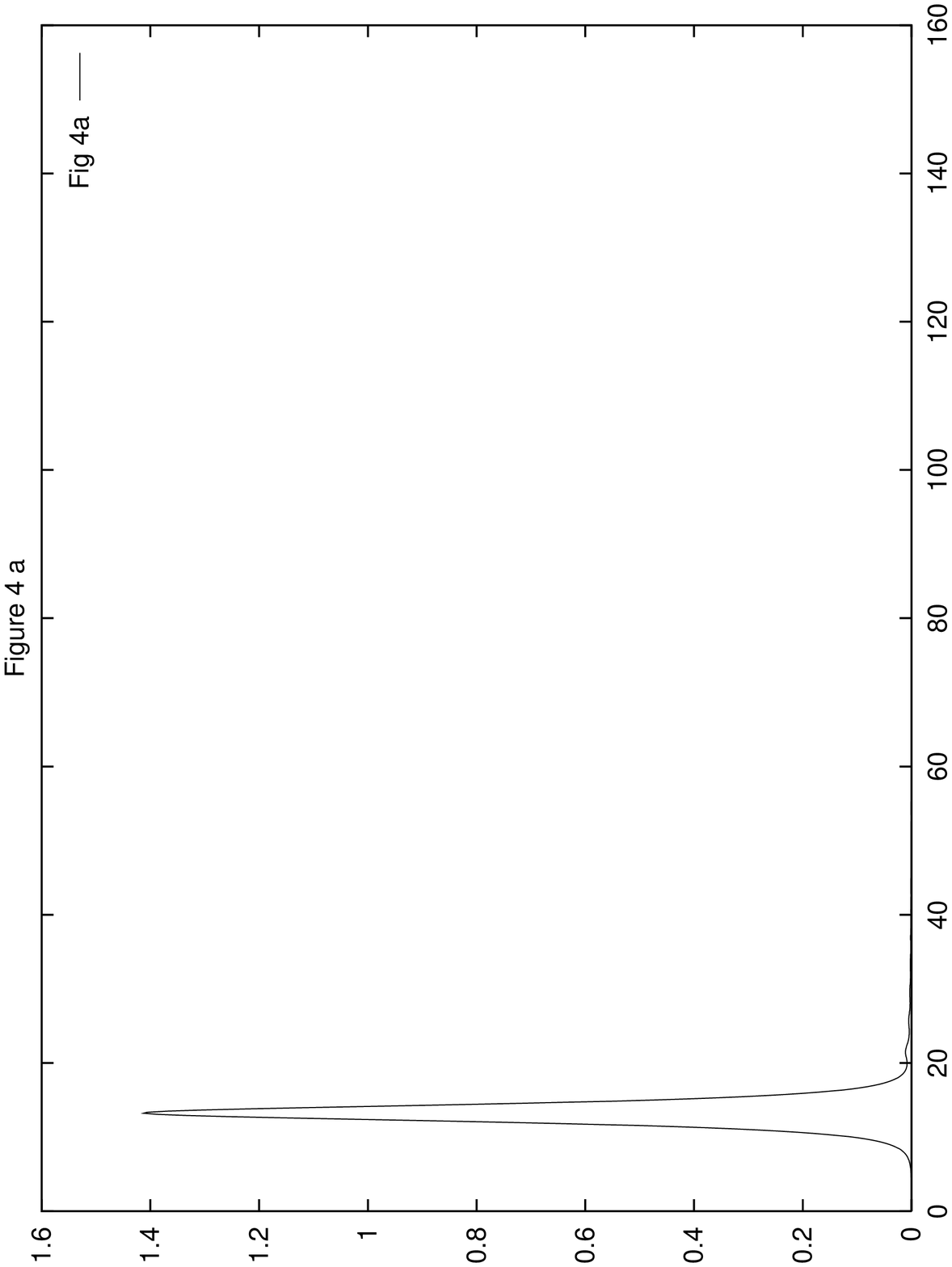}}
\hbox{\epsfxsize 14cm\epsffile{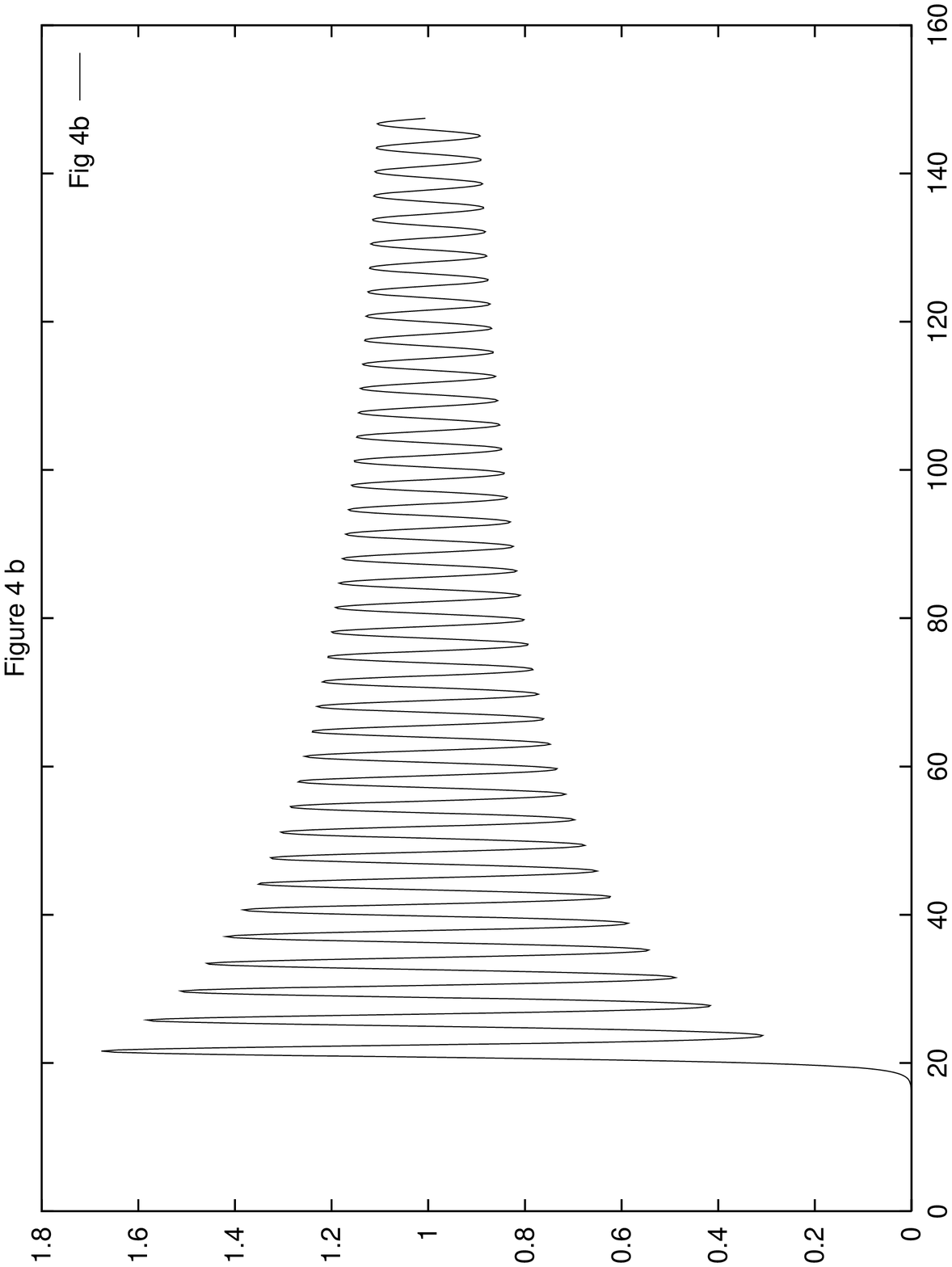}}
\hbox{\epsfxsize 14cm\epsffile{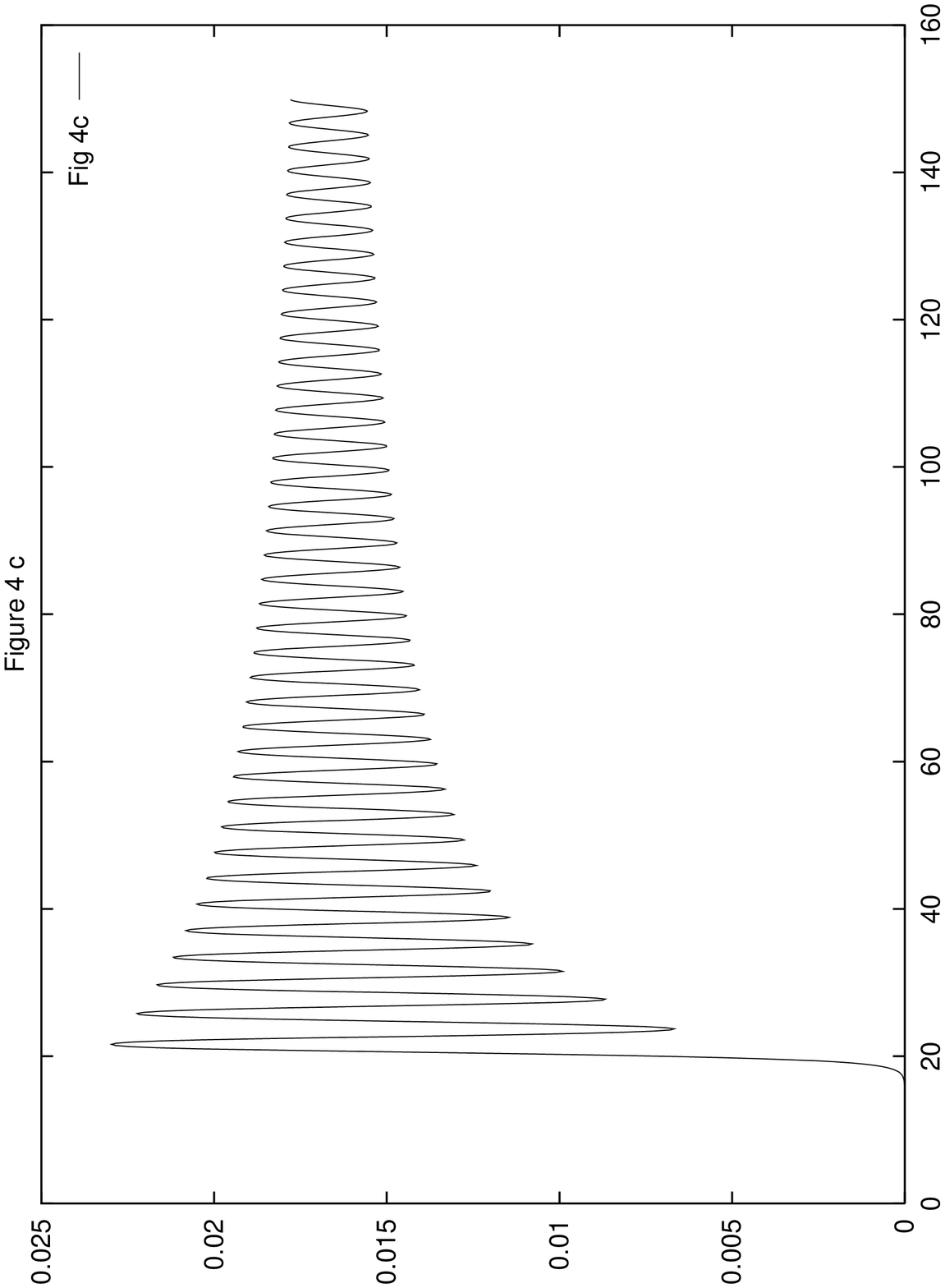}}
\hbox{\epsfxsize 14cm\epsffile{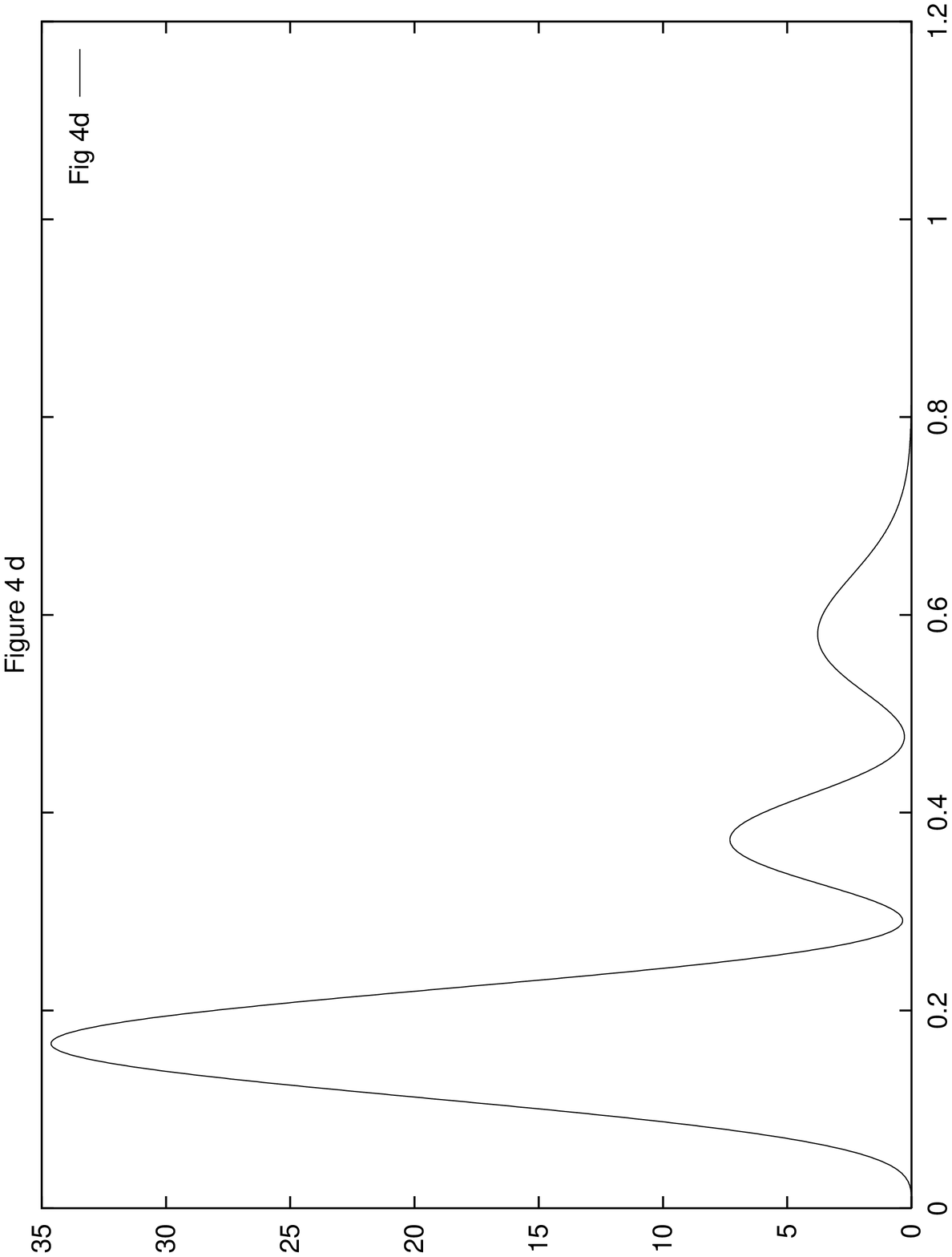}}
\hbox{\epsfxsize 14cm\epsffile{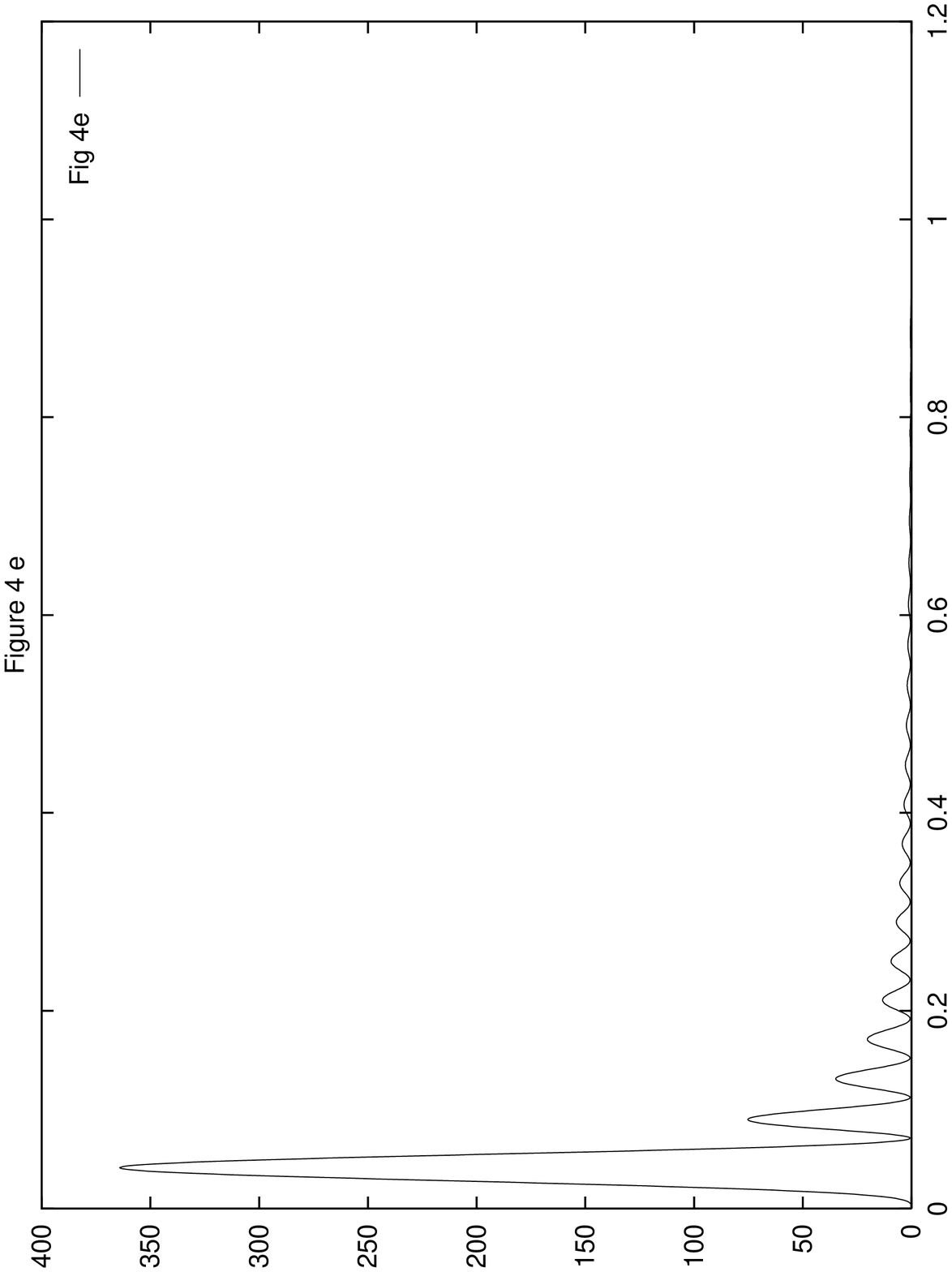}}
\hbox{\epsfxsize 14cm\epsffile{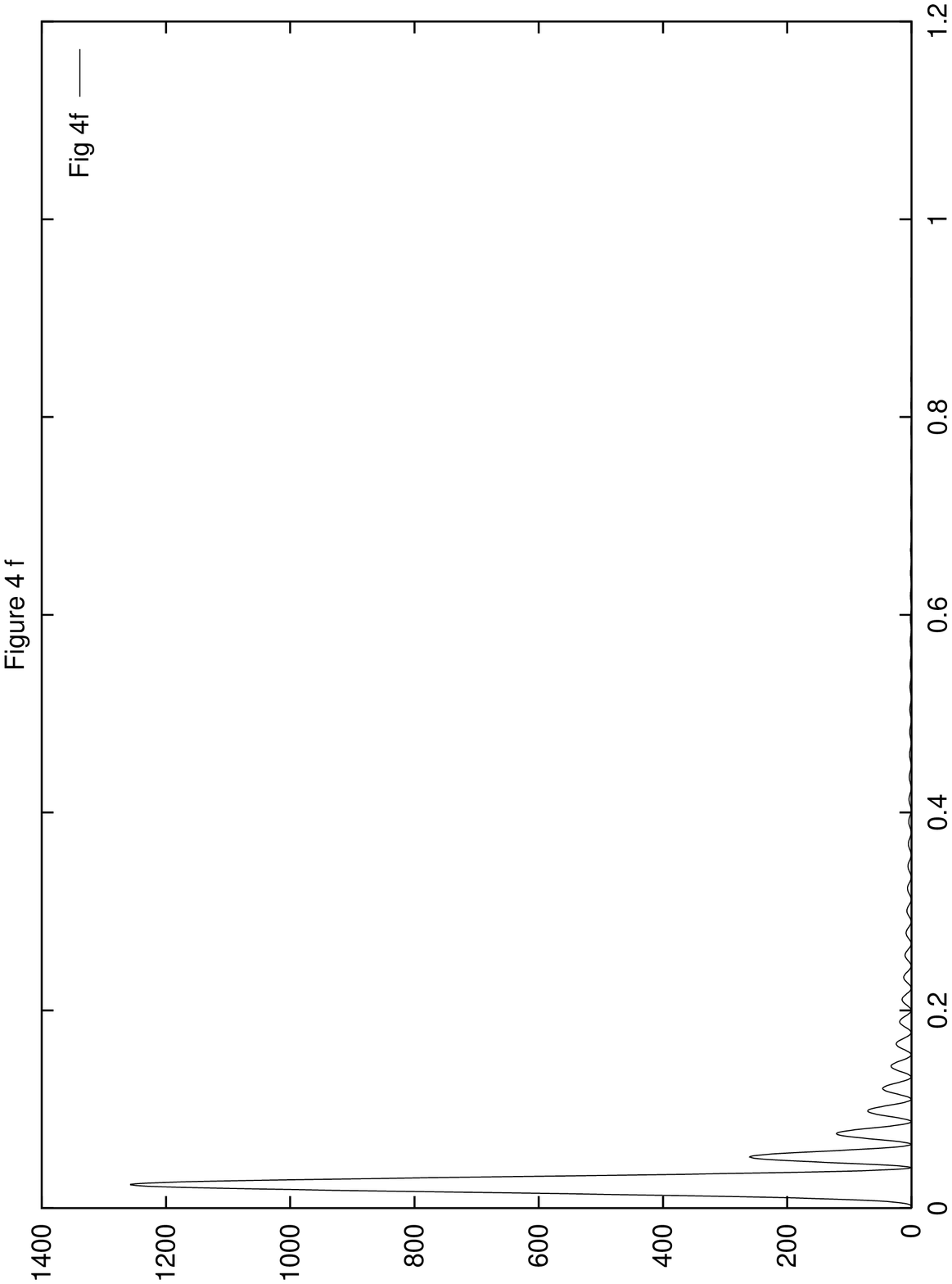}}
\hbox{\epsfxsize 14cm\epsffile{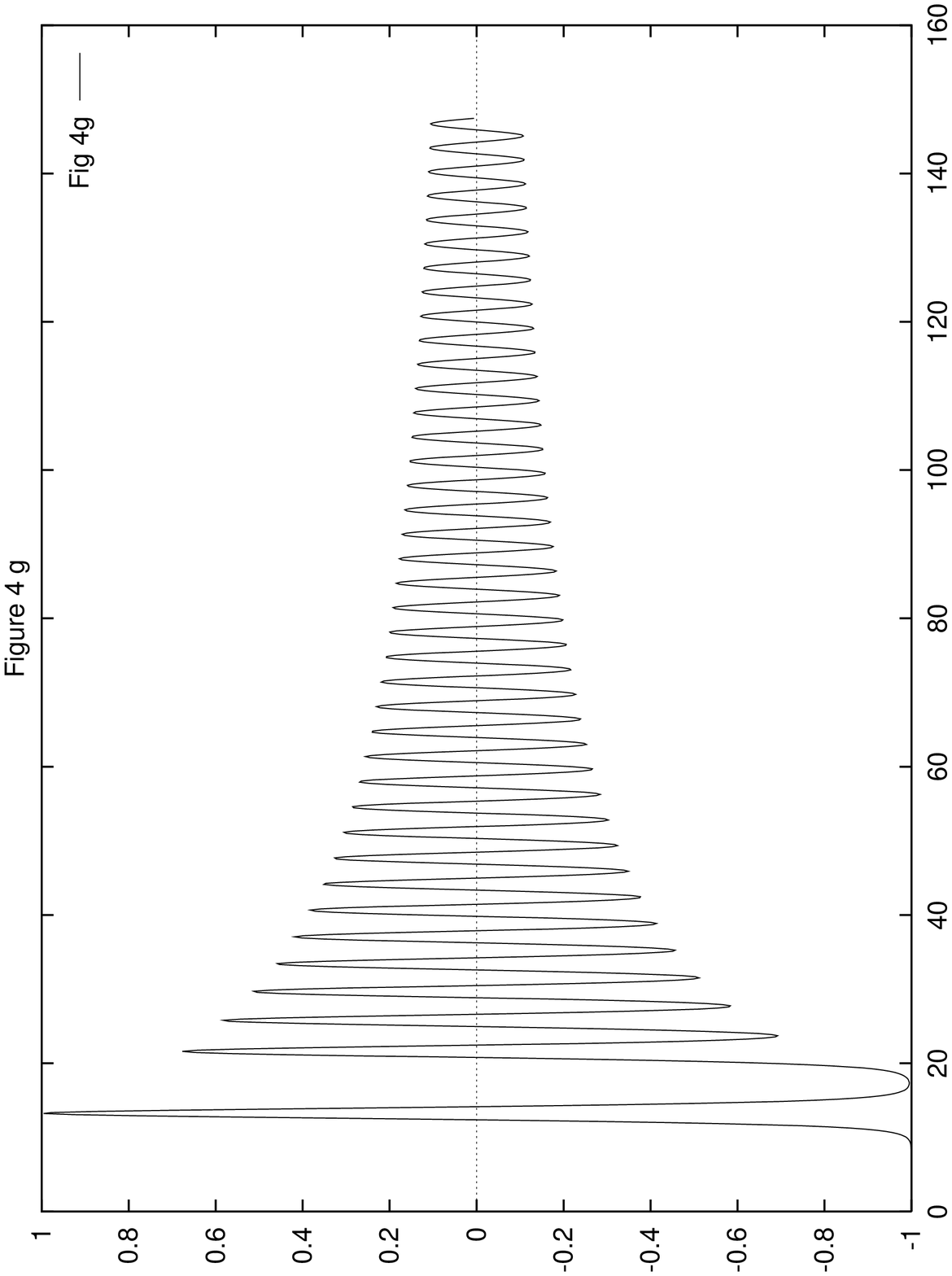}}
\hbox{\epsfxsize 14cm\epsffile{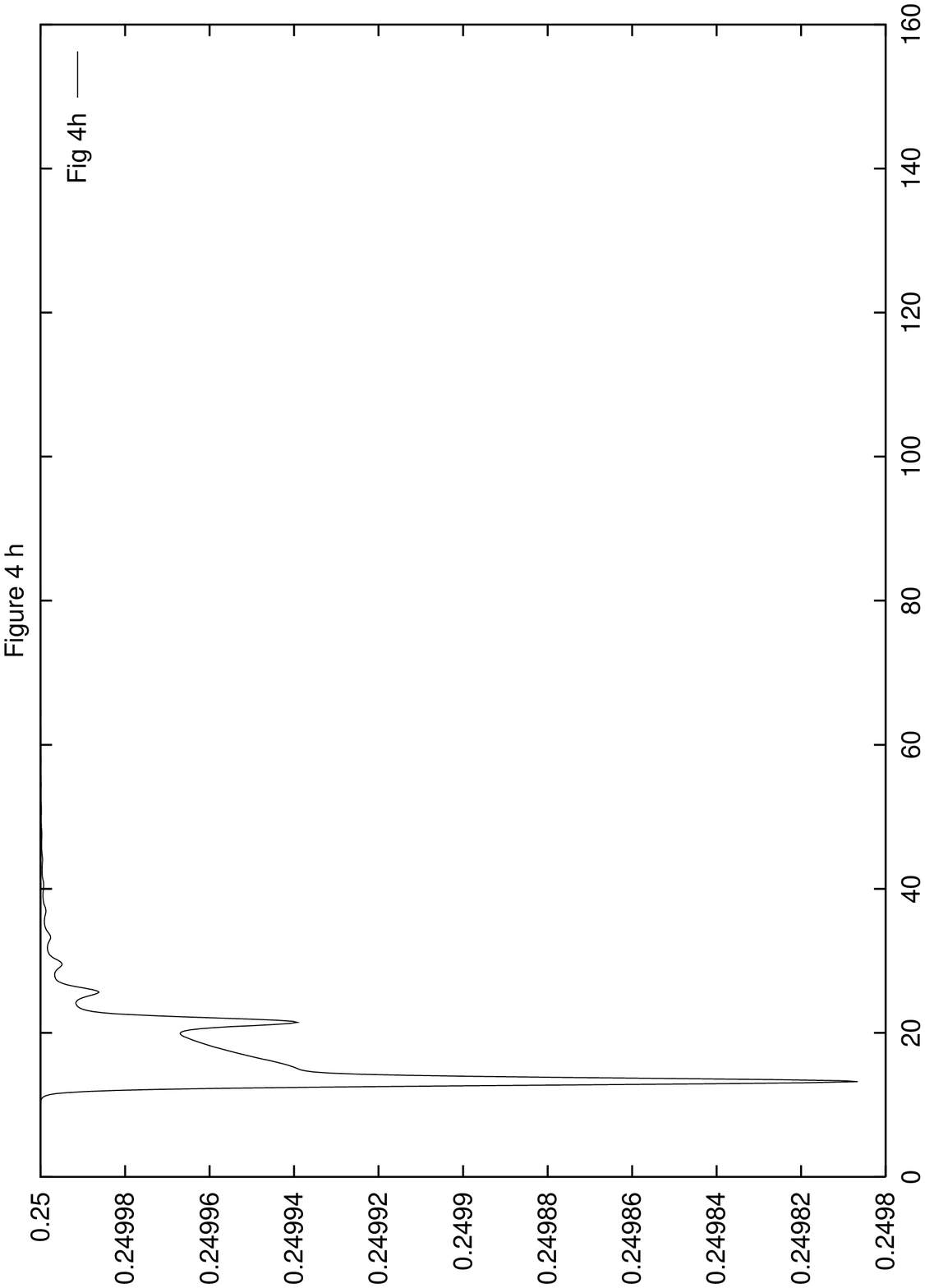}}
\hbox{\epsfxsize 14cm\epsffile{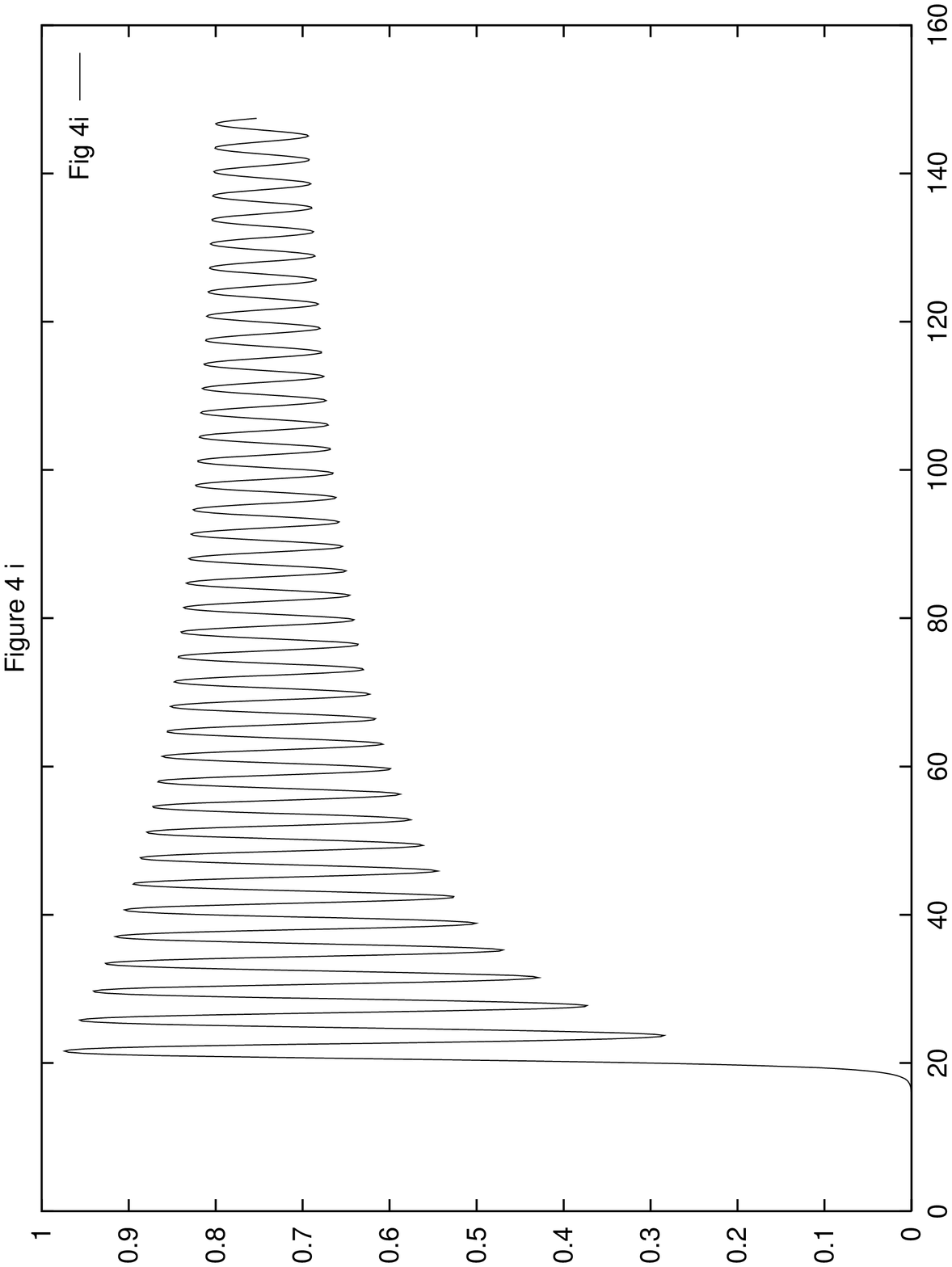}}
\hbox{\epsfxsize 14cm\epsffile{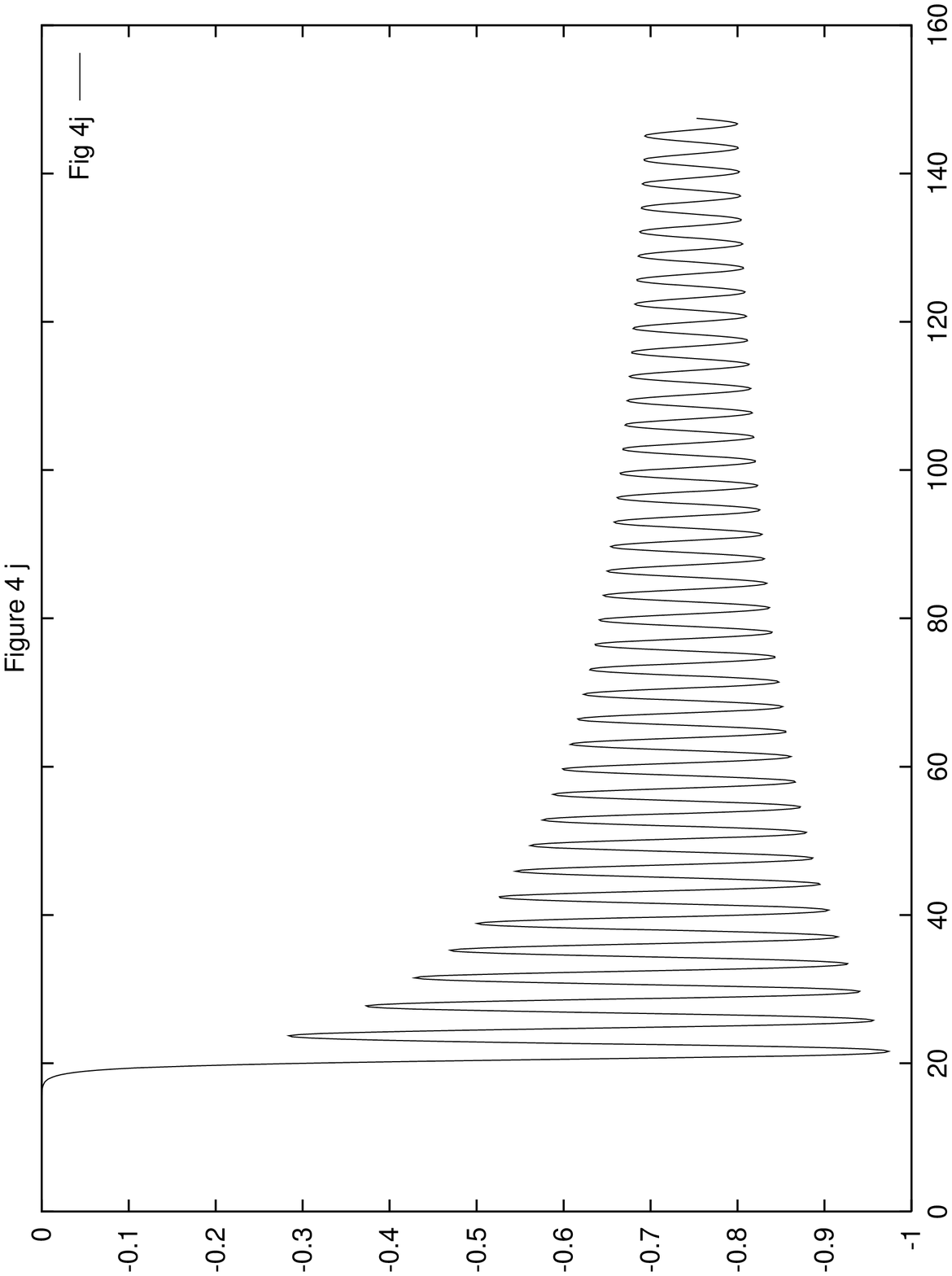}}
\hbox{\epsfxsize 14cm\epsffile{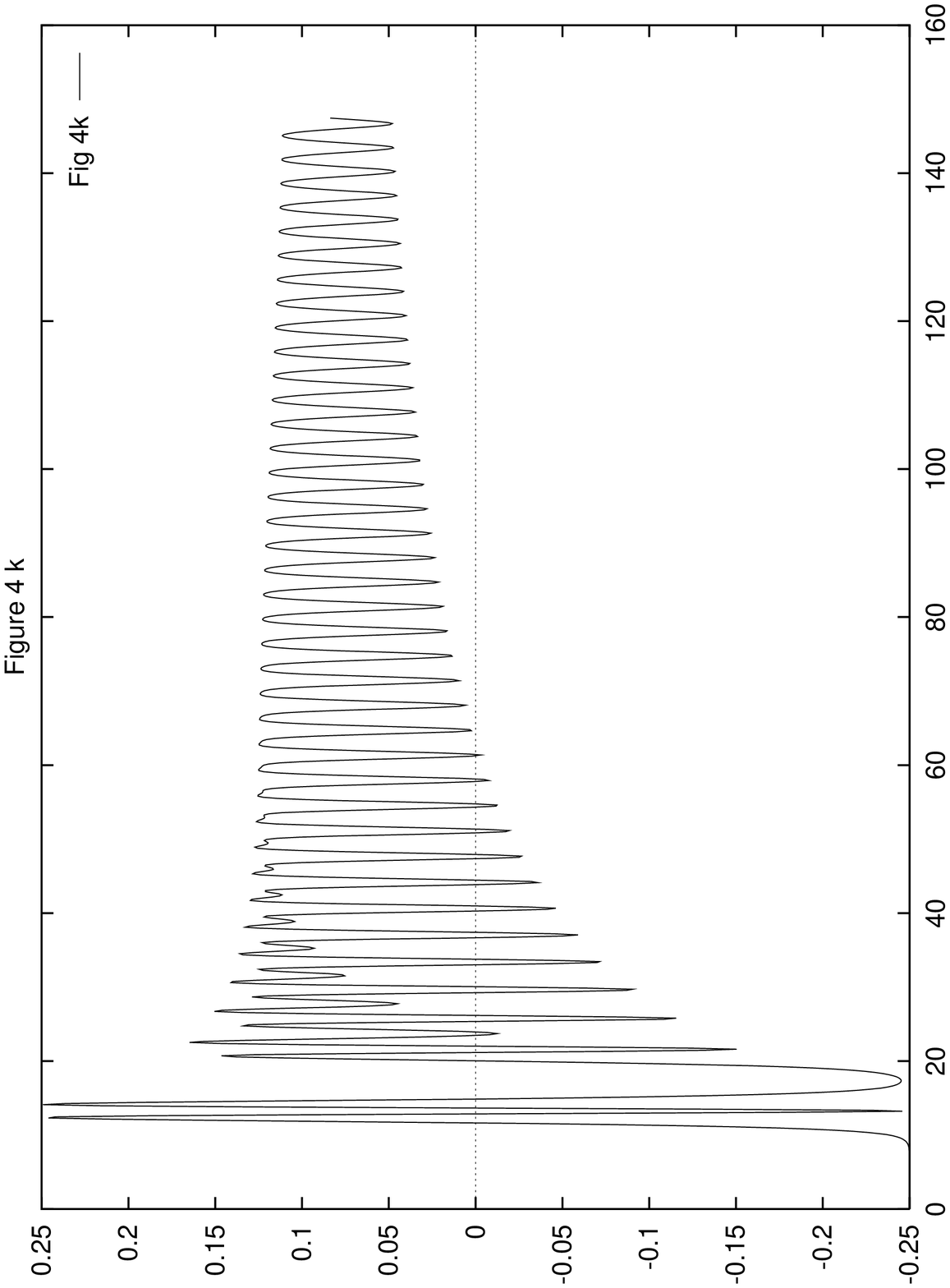}}
\hbox{\epsfxsize 14cm\epsffile{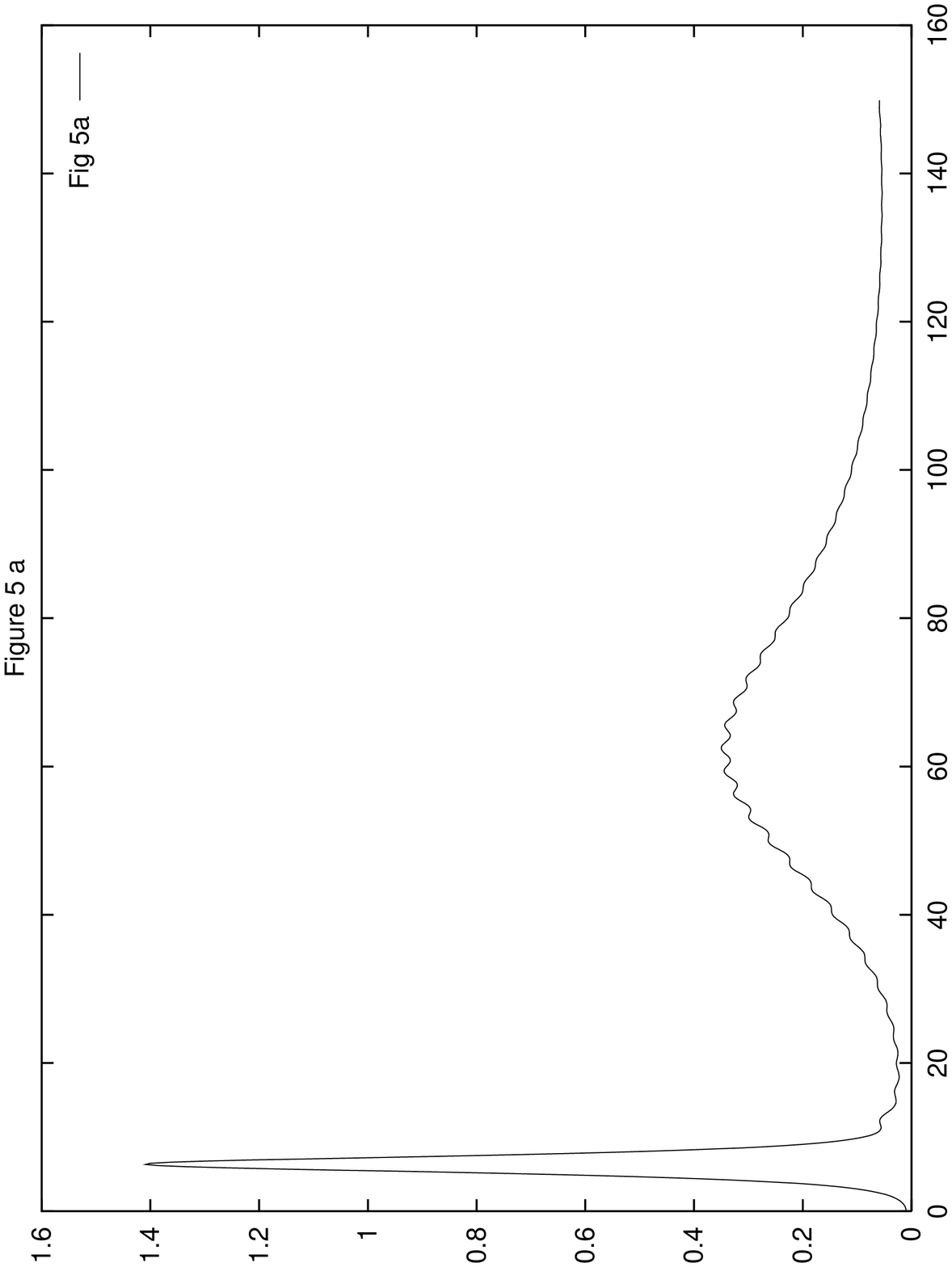}}
\hbox{\epsfxsize 14cm\epsffile{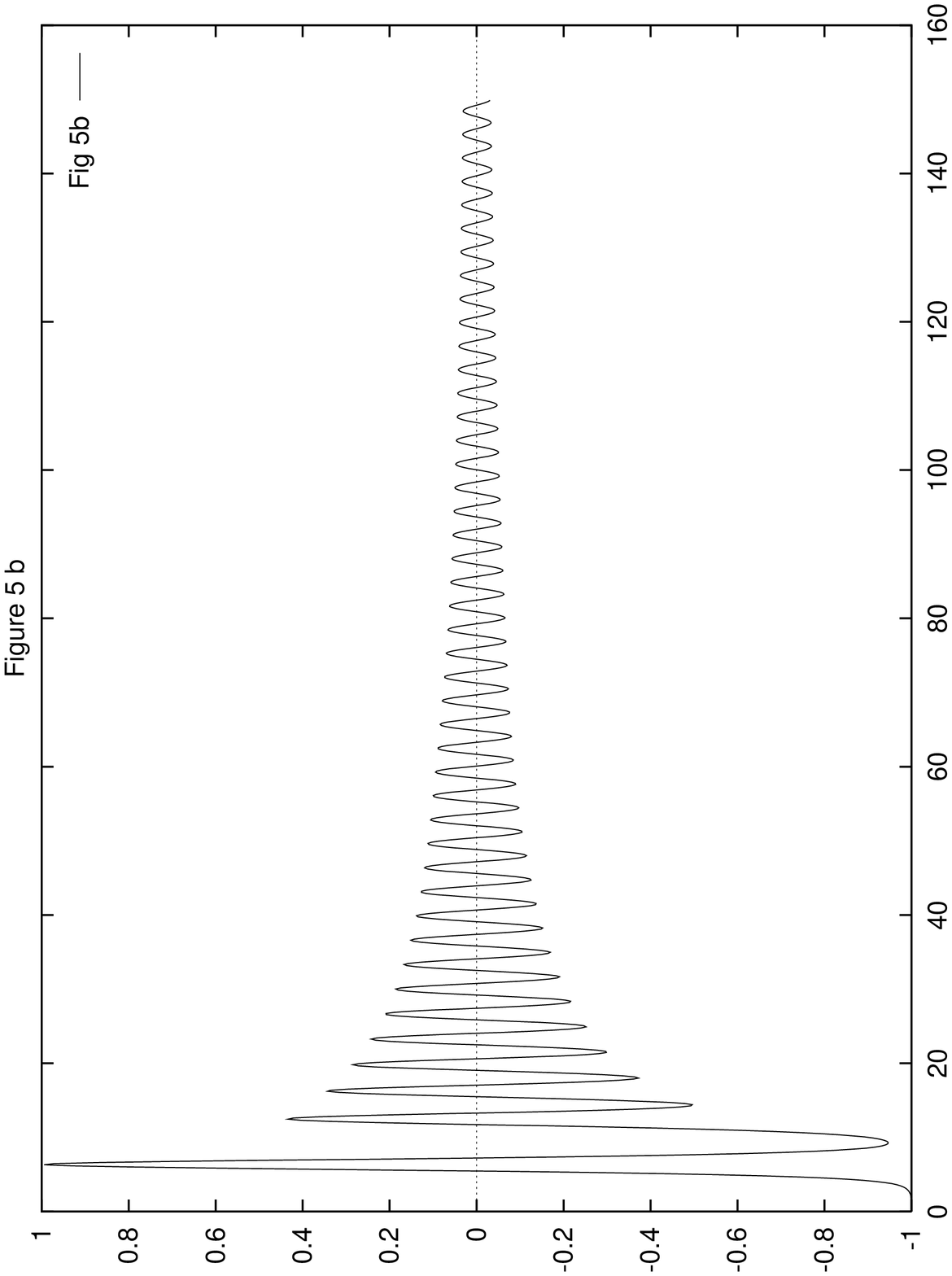}}

\end{document}